\documentclass[preprint]{aastex631}
\pdfoutput=1

\received{October 20, 2024}

\submitjournal{Astrophysical Journal}

\usepackage{longtable}
\usepackage{subfigure}
\usepackage{comment}
\begin{document}

\title{From Stability to Instability: Characterizing the Eccentricities of Multi-planet Systems in the California Kepler Survey as a Means of Studying Stability}

\author[0009-0003-7321-5627]{Matthew Doty}
\affiliation{University of Notre Dame,
Holy Cross Dr,
Notre Dame, Indiana 46556, USA}

\author[0000-0002-3725-3058]{Lauren M. Weiss}
\affiliation{Department of Physics and Astronomy, University of Notre Dame, Notre Dame, IN 46556, USA}

\author[0000-0002-5223-7945]{Matthias Y. He}
\affiliation{NASA Ames Research Center, Moffett Field, CA 94035, USA}

\author[0000-0003-1970-1790]{Antoine C. Petit}
\affiliation{Universit\'e C\^ote d'Azur, Observatoire de la C\^ote d'Azur, CNRS, Laboratoire Lagrange, Nice, France}



\begin{abstract}
Understanding the stability of exoplanet systems is crucial for constraining planetary formation and evolution theories. We use the machine-learning stability indicator, SPOCK, to characterize the stability of 126 high-multiplicity systems from the California Kepler Survey (CKS). We constrain the range of stable eccentricities for each system, adopting the value associated with a 50\% chance of stability as the characteristic eccentricity. We confirm characteristic eccentricities via a small suite of N-body integrations. In studying correlations between characteristic eccentricity and various planet-pair and system-level metrics we find that minimum period ratio correlates most strongly with characteristic eccentricity. These characteristic eccentricities are approximately 20\% of the eccentricities necessary for two-body mean-motion resonance overlap, suggesting three-body dynamics are needed to drive future instabilities.  Systems in which the eccentricities would need to be high ($>0.15$) to drive instability are likely dynamically relaxed and might be the fossils of a previous epoch of giant impacts that increased the typical planet-planet spacing.

\end{abstract}



\section{Introduction}\label{sec:intro}
Since 1992, roughly 5,600 exoplanet candidates have been discovered. At least a quarter of these systems contain two or more planets. While the 2-body problem and the two-planet + star problem are well understood (\citealt{1993Icar..106..247G}, \citealt{2018AJ....156...95H}), a full understanding of the dynamics of high multiplicity systems remains elusive despite rich observational evidence of their widespread existence.

One architectural theme of higher multiplicity systems is the ‘peas-in-a-pod’ configuration of evenly sized, evenly spaced multi-planet systems (\citealt{2023ASPC..534..863W}
and references therein). This striking pattern among multi-planet systems is a clear departure from the architecture of our own solar system. Since final system architecture is impacted by dynamics both during  (\citealt{2024ApJ...960...89B})  and after gas disk dispersal (\citealt{2024ApJ...971....5W}), understanding the specific dynamics surrounding in the stability of peas-in-a-pod systems could provide a key insight into the processes by which peas in a pod are formed.

If a system's instability timescale is shorter than the time between system formation and observation, then the system is unlikely to be observed. Consequently, the age of the observed systems puts a lower limit on the instability timescales of observed systems. The average stellar age (and therefore system age) of the Kepler field is roughly a few Gyr \citep{2020AJ....160..108B}. The few Gyr age of the Kepler field is comparable with the age of the universe suggesting  that architectural elements of Kepler systems lead to long timescales. Therefore, by probing the stability of Kepler systems  in relation to their architectures we can find what elements of the peas-in-a-pod pattern are most sensitive to stability. This defines the main question of this work: what properties of the peas-in-a-pod Kepler multi-planet systems correlate with stability and by what mechanism?

A physical parameter that is both fairly important in determining system stability and poorly measured in individual systems is planetary eccentricity. Based on statistical measurements of the transit durations, multi-planet systems appear to generally have lower eccentricities than the planets in systems with just one transiting planet (\citealt{2016PNAS..11311431X}, \citealt{2019AJ....157..198M}, \citealt{2020AJ....160..276H}).  
Ideally, the eccentricities found via stability limits would be compared to the actual measured values in the system. Some systems do contain measurable eccentricities. These systems have significant transit timing varitiations (TTVs), which allow the eccentricities of the individual planets to be measured.  \citealt{2021AJ....162...55Y} investigated the relationship between SPOCK characterized eccentricities and the TTV measured values for such systems, finding that the eccentricities of planets in systems with TTVs are much lower than is required for stability. While such a comparison between observed values and stability limits would be ideal, most systems contain small, nonmeasurable eccentricities unable to be constrained via TTV measurements. While these systems cannot be compared to observed values, calculated eccentricities for these systems can be compared to other theoretical limits to better understand what is the dominant factor in deciding system stability.  Many authors have used N-body experiments to explore the relationship between stability and eccentricity in idealized multi-planet systems (\citealt{2007IJMPB..21.3981Z}, \citealt{2015ApJ...807...44P}, and \citealt{2024arXiv240317928L}).   In general, eccentricity plays a significant role in system stability as it defines the level of dynamical excitation and can unlock higher order resonances through a variety of mechanisms (\citealt{2017A&A...605A..72L}, \citealt{2020A&A...641A.176P},\citealt{2021AJ....162..220T}). The aim of this study is to determine which eccentricities are possible in the specific architectures of observed multi-planet systems, and how much the allowed eccentricities vary from one system to another. In comparing these values to previous theoretical limits, we will better understand the general dominant factors in system stability. This will set the stage for future work to investigate for specific observed high-multiplicity system architectures which theorized stability mechanisms dominate in sculpting the orbital properties including eccentricities. 

\section{Methods}
We explored the stability of high-multiplicity systems that were characterized as part of the California Kepler Survey (\citealt{2018AJ....155...48W}).  These systems are typically old ($\sim5$ Gyr) and therefore should be stable on long timescales. The CKS multi-planet systems also host a large number of peas-in-a-pod like systems (as the pattern was first discovered in this sample), although systems with diverse planet sizes and spacing are also present and available to test. An advantage of the CKS systems is that their host stars are homogeneously characterized based on Gaia data and high-resolution spectra (\citealt{2018AJ....156..264F}), yielding accurate stellar properties, which propagate to accurate physical radii and semi-major axes for the transiting planets.  However the planet masses are generally not measured.  We estimated planet masses using the mass-radius relationship from \citep{2014ApJ...783L...6W}. We assume the systems are coplanar, which is consistent with the very low inclination dispersion inferred from the transit duration ratios and high multiplicities in these systems (\citealt{2012ApJ...761...92F}, \citealt{2014ApJ...790..146F}, \citealt{2019AJ....157..198M}).

\subsection{Mapping Stability Onto Eccentricity}\label{sec:style}
To find how the stability of different configurations of the CKS high-multiplicity systems varied with eccentricity, we used the machine-learning tool SPOCK \citep{2020PNAS..11718194T}. Trained on near-resonant compact multi- planet systems SPOCK is roughly $10^{5}$ times faster than N-body integration of these systems and gives the probability that the system is stable after roughly $10^{9}$ orbits. SPOCK generates the same result as N-body roughly 95$\%$ of the time and outperforms various analytical metrics \citep{2020PNAS..11718194T}. The speedup of SPOCK compared to N-body allowed us to test a wide range of eccentricities for a much larger sample of planets than have been been explored traditionally (e.g. \citealt{2021AJ....162...55Y}, but see \citealt{2024AJ....167..271V} for broad N-body study).  Because SPOCK was trained on compact multiplanet systems that are near (but not in) resonances, SPOCK works best on these types of systems.  SPOCK is not useful for 2 planet systems as analytical criteria are more efficient. For the purposes of our project we therefore select CKS systems with at least three planets. 
This resulted in a sample of 126 systems with at least 3 transiting planets.

\begin{figure} [hbt!]
    \centering
    \centering
    \begin{subfigure}{}
       \centering
       \includegraphics[width=.4\linewidth]{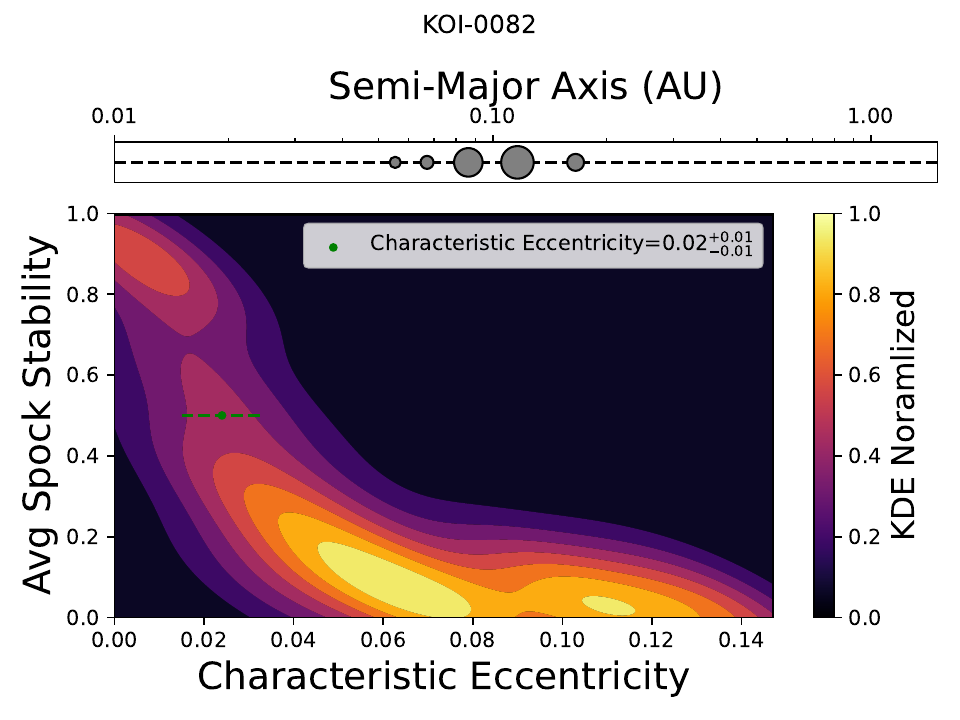}
    \end{subfigure}
    \begin{subfigure}{}
       \centering
       \includegraphics[width=.4\linewidth]{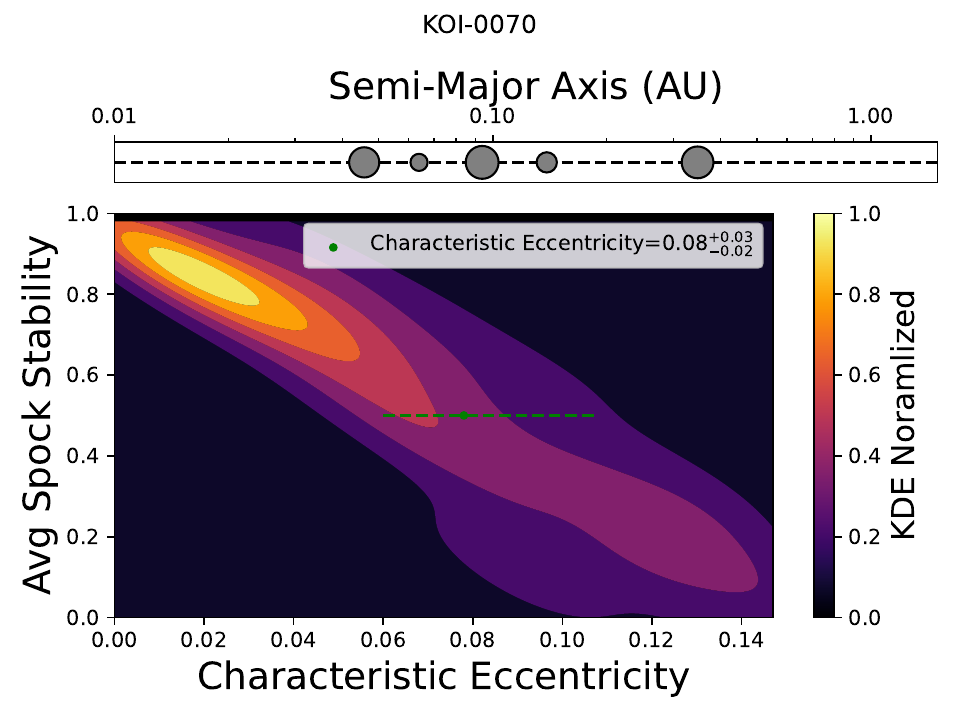}
    \end{subfigure}
    \caption{A KDE plot of SPOCK Stability vs. Eccentricity for KOI-82 on the left and KOI-70 on the right. Color denotes the density of samples. The KDE is normalized with 1 being the highest density of points for the system and 0 being the lowest density of points for the system. A small range of possible eccentricities for a specific stability are denoted by a lighter colored area, while a large range of possible eccentricities for a specific stability are denoted by a darker colored area. As eccentricity increases, stability decreases. Note how there are two well constrained regions of stability and instability with a transition region between the two. The eccentricity that describes the location of this transition region, as marked by a SPOCK Stability of 50$\%$, we called the "characteristic" eccentricity of the system. }
    \label{fig:kde_examples}
\end{figure}

In each system, we fixed the orbital periods and masses of the planets and performed a grid search of 50 linearly spaced eccentricities in [0,0.15]. This range was motivated by the low eccentricities that have been measured so far in compact multiplanet systems (\citealt{2015ApJ...808..126V},\citealt{2016PNAS..11311431X},\citealt{2019AJ....157..198M},\citealt{2021AJ....162...55Y}). For simplicity, we initially set all planet eccentricities to be the same.  However, in Section 4, we allowed the individual eccentricities to vary.  We drew the argument of periastron and initial mean anomaly of each planet from a random uniform distribution on [0, $2\pi$). We ran 250 SPOCK trials at each grid value of eccentricity, with each trial yielding a likelihood of stability on [0,1] based on the initial parameters.  For most systems, there is substantial scatter in stability at a given eccentricity, so we applied a Kernel Density Estimator (KDE) to the results to generate a smooth and continuous distribution of stability as a function of eccentricity.  Examples illustrating different behavior of stability vs. eccentricity are shown in Figure \ref{fig:kde_examples}, and the Appendix contains a plot for each system in our sample.

\section{Results}
As expected, low eccentricities are uniformly stable, and as eccentricity increases, the average stability decreases.  For each system, we identify the value of eccentricity that yields a SPOCK stability prediction of 50$\%$ as the ``characteristic eccentricity,'' $e_\mathrm{char}$. We estimated the error in $e_\mathrm{char}$ by taking the 68\% bounds on the distribution of eccentricities for which the stability prediction ranged from 0.49 to 0.51.  The characteristic eccentricity and its uncertainty are shown as a red dot and dashed line on the KDE plots. 
 We report the characteristic eccentricity of each system in the Appendix. Surprisingly, for some systems, even the highest eccentricities we tested (0.15) yielded stable SPOCK predictions more than 50\% of the time. For these systems, we give a lower limit of $e_\mathrm{char} > 0.15$.  Predominately three planet systems were the ones that exceeded characteristic eccentricities of 0.15. Since other three-planet systems had well-determined values for $e_\mathrm{char}$, the number of transiting planets is not sufficient to predict $e_\mathrm{char}$.

While the KDE plots have several common features, both the characteristic eccentricity and the shape of the eccentricity-stability relation vary substantially from system to system. For example, KOI-82 has a lower characteristic eccentricity and a much steeper slope than KOI-70 (Figure \ref{fig:kde_examples}). The differing shapes and characteristic eccentricities suggest that there is not a single value of eccentricity that an equally good estimate for all compact multi-planet systems.  While previous observational studies (\citealt{2019AJ....157..198M}, \citealt{2021AJ....162...55Y}) have suggested that an eccentricity of 0.05 is a good estimate for compact multi-planet systems, we find characteristic eccentricities that range from 0.02 to $e_\mathrm{char} > 0.15$.

The variety of characteristic eccentricities indicates that other properties of the planetary system architecture play a major role in stability.  The remainder of this paper is dedicated to searching for architectural attributes that are maximally predictive of the characteristic eccentricities for our sample.  The two questions that underpinned our search were: (1) are peas-in-a-pod systems more stable than systems with diverse planet properties, and (2) what metric of planet spacing best predicts the stability (via characteristic eccentricity) of compact multi-planet systems?

\subsection{Are Peas-in-a-Pod More Stable?}
 Since peas-in-a-pod is a common pattern in the CKS multi-planet systems, we tested whether peas-in-a-pod architectures were more or less stable than systems with diverse sizes and spacings. The planetary system size diversity and spacing diversity are well described by two metrics, the gap complexity (spacing) and mass partitioning (size) (\citealt{2020AJ....159..281G}).  We measured the Spearman-R correlation between each of these metrics and the characteristic eccentricity (our proxy for stability), finding a Spearman r-value of 0.04 (p=.627) for gap complexity and a r-value of 0.12 (p=.149) for mass partitioning.  Because neither gap complexity nor mass partitioning correlates strongly with the characteristic eccentricity, the peas-in-a-pod systems do not appear to be significantly more or less stable than systems with diverse planet properties. 
 \begin{table}[hbt!]
    \centering
    \begin{tabular}{||c || c | c||} 
    \hline
    Parameter & p-value & r-value \\ [0.5ex] 
    \hline\hline
    Minimum Period Ratio ($P_{i+1}/P_{i}$) & $\ll$0.001 & 0.727 \\ 
    \hline
    Minimum Separation in Mutual Hill Radii ($\Delta$) & 0.001 & 0.688 \\
    \hline
    Multiplicity ($N_{p}$) & 0.001 & -0.444\\
    \hline
    Mass Partitioning ($Q$) & 0.149 & 0.12 \\
    \hline
    Gap Complexity ($C$) & 0.627 & 0.04 \\
    \hline
    \end{tabular}
    \caption{Spearman r-values and p-values for different metrics tested against characteristic eccentricity.} 
    \label{tab:correlations}
\end{table}
 \subsection{Relationship Between Stability and Various Planet Spacing Metrics}
Various studies have found that the dynamical spacing between the planets is important for stability (\cite{1996Icar..119..261C},\citealt{2015ApJ...807...44P}).  To test how these results apply to the CKS multi-planet systems, we measured the correlation between characteristic eccentricity and several metrics that describe the spacing of adjacent planets. Motivated by the possible role of mean motions resonance overlap (\citealt{2013ApJ...774..129D}), we used the adjacent planet period ratio as one measure of planet spacing. Since planet and stellar mass affect the strength of gravitational interactions, we also tested a metric based  on  the separation of the planets in mutual Hill radii. Mutual Hill radius is defined as:
\begin{equation}
 R_{H} = \left(\frac{\mu_{i}+\mu_{i+1}}{3}\right)^{\frac{1}{3}}\left(\frac{a_{i}+a_{i+1}}{2}\right).
\end{equation} 
Where $\mu$ is the mass ratio between planet and star and $a$ is the semi major axis. From this, the separation can be defined via the canonical $\Delta$:
\begin{equation}
    \Delta_{i,i+1}=(a_{i+1}-a_{i})/R_{H}
\end{equation} 
Note that for this definition of mutual Hill radius and $\Delta$ the mass ratio is raised to the $\frac{1}{3}$. While earlier work such as in \citealt{2016ApJ...823..118M} has suggested 2 and 3+ planet interactions rely more convincingly on dynamical spacing (which takes the same form except with $\mu^{\frac{1}{4}}$), due to the regular masses of CKS systems the same general $\Delta$ results should hold for both metrics. Since general spacing metrics are defined between 2 planets rather than for an entire system, such metrics must be consolidated to be applied to entire systems. Earlier work by \citealt{2020A&A...641A.176P} did this by taking the harmonic average of the spacing. One challenge is that some systems contain a variety of adjacent period ratios and values of $\Delta$.  Because our definition of instability (which is shared by SPOCK) only requires a single ejection or collision for the system to become unstable, we looked for correlations between the minimum values of each of these metrics (period ratio and $\Delta$) in each system, and and the characteristic eccentricity of that system. We found meaningful correlations between characteristic eccentricity and both minimum period ratio and minimum $\Delta$, with an r-value of 0.727 (p$<$0.001) for minimum period ratio and an r-value of 0.688 (p=0.001) for minimum $\Delta$.
\begin{figure} [h]
    \centering
    \includegraphics[width=1\linewidth]{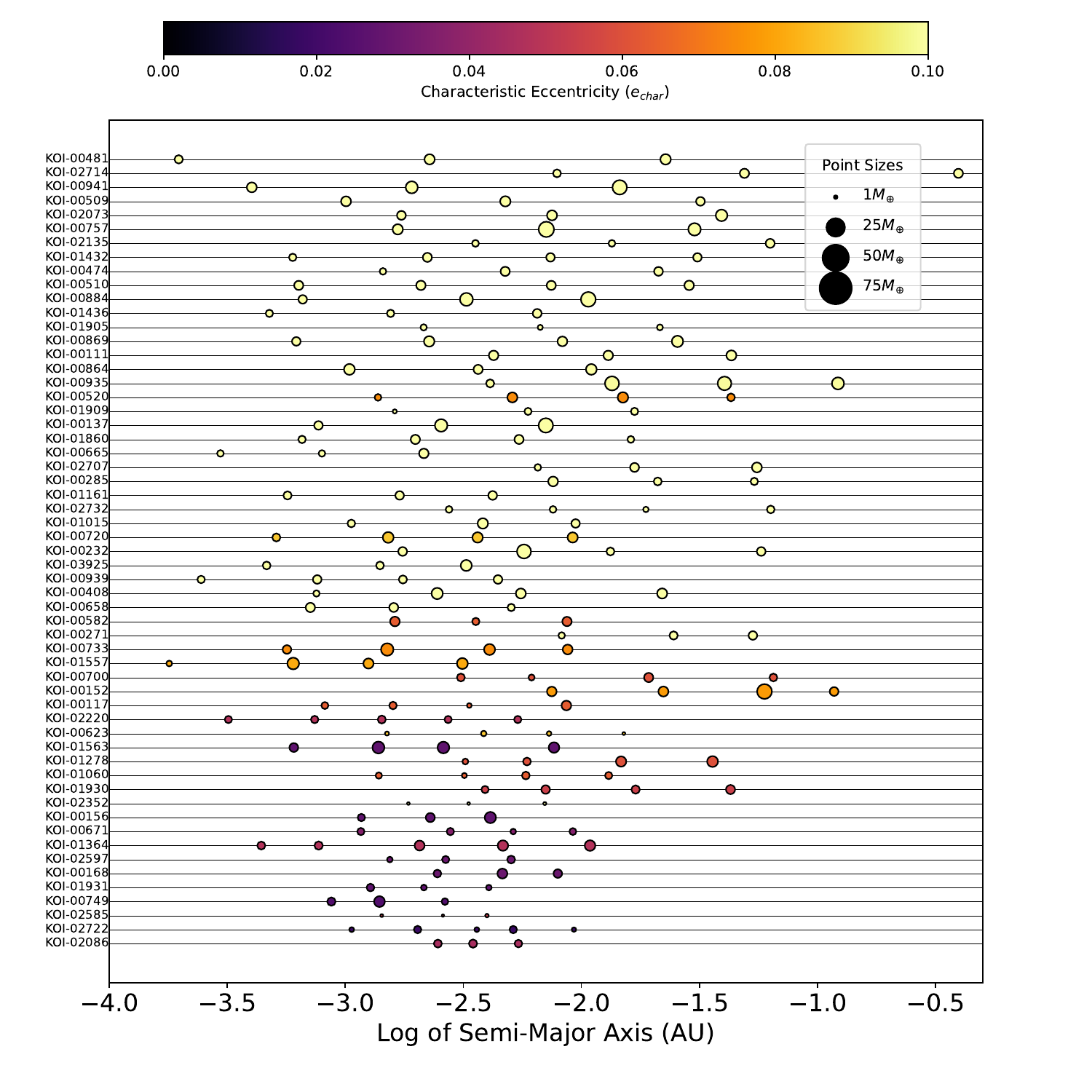}
    \caption{A gallery plot from all peas-in-a-pod (Gap Complexity and Mass Partitioning $<0.1$) SPOCK compatible systems in the CKS database sorted largest (top) to smallest (bottom) minimum period ratio and colored by the system's ``characteristic" eccentricity. As can be seen, systems with a high ``characteristic" eccentricity are more towards the top of the plot and therefore have larger minimum period ratios. This relationship suggests that a single planet pair drives the stability of the system and the dominant factor in stability is the strength of the interaction.}
    \label{fig:enter-label}
\end{figure} 
 
 The results in Table \ref{tab:correlations} show that when these metrics are applied to observed data for real systems, the best predictor of characteristic eccentricity (and thus stability) is the smallest period ratio of adjacent planets, followed by the smallest planet separation in units of mutual Hill radii. While system multiplicity seems to have some relation with characteristic eccentricity, other system-wide metrics of gap complexity and mass partitioning are very poor predictors of characteristic eccentricity.  Figure 2 displays the architectures of planets in our sample ranked by minimum period ratio.  Systems with small minimum period ratios have low eccentricities and systems with larger minimum period ratios have larger eccentricities.  In a sample of 165 Kepler and K2 systems modeled with N-body integrations spanning 5 billion orbits of the innermost planet, \cite{2024AJ....167..271V} also found that period ratio is an excellent predictor of long-term stability and a superior predictor to the separation in mutual Hill radii. These results are consistent with predictions made using idealized systems \citep{2021AJ....162..220T}.

\section{Relaxing Eccentricity to vary Within Systems}\label{sec:style}
The relationship between stability, minimum period ratio, and eccentricity highlights the need to understand these systems at the planet pair level, motivating us to vary the eccentricities of the individual planets. For each planet, we randomly selected eccentricities from a normal distribution centered around the system-wide characteristic eccentricity and its uncertainty, performing 300 trials per system. We then reported which eccentricity for each individual planet best corresponded to 50\% of the trials being stable (Appendix). While values varied, most planetary characteristic eccentricities were very similar to the system wide value.

\begin{figure} [hbt!]
    \centering
    \begin{subfigure}{}
       \centering
       \includegraphics[width=.48\linewidth]{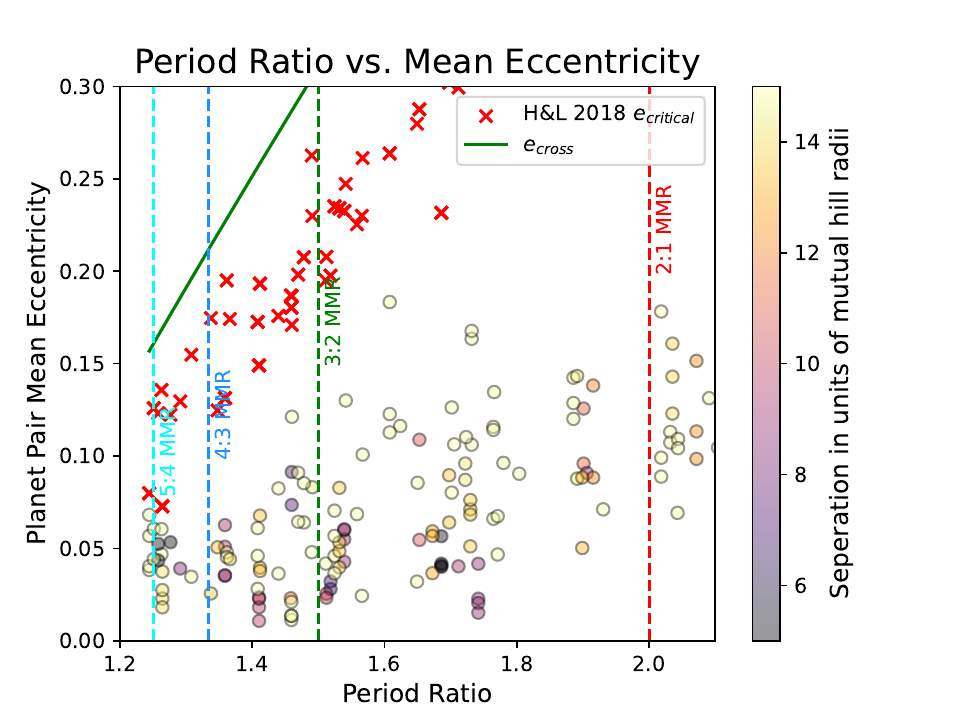}
    \end{subfigure}
    \begin{subfigure}{}
       \centering
       \includegraphics[width=.48\linewidth]{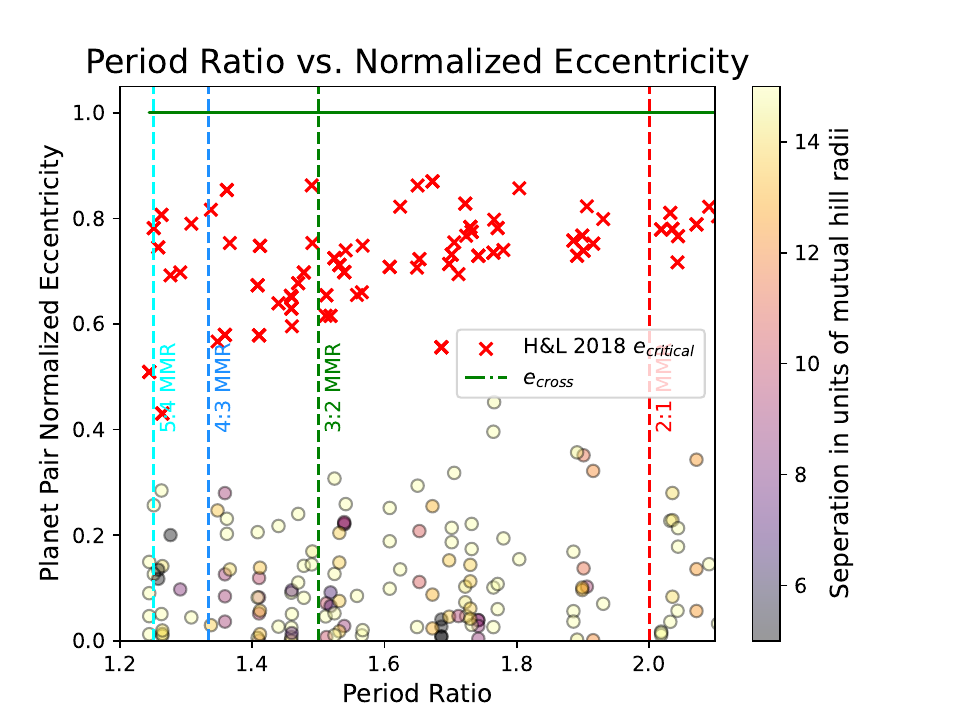}
    \end{subfigure}
    \caption{(Left) Mean planetary characteristic eccentricity of an adjacent planet pair vs. planet pair period ratio. The characteristic eccentricities from SPOCK are circles outlined in black, the critical eccentricities from \cite{2018AJ....156...95H} are red crosses, and the eccentricities are represented by a green line. The vertical lines plotted are common first order 2-body, mean-motion resonances. While the correlation between period ratio and eccentricity still exists there is no relation between eccentricities and the locations of first order mean-motion resonances. (Right) Normalized eccentricity vs. planet-pair period ratio. The color scheme remains the same, however all values are normalized to the crossing eccentricity for the planet pair. When normalized eccentricity is used the relation between period ratio and eccentricity disappears. For both plots the characteristic eccentricities are well below those needed for first order mean-motion resonance overlap. This suggests that first-order mean-motion resonances and their overlap are not the initial drivers of instability.}
    \label{fig:prat-e}
\end{figure}

Figure \ref{fig:prat-e} shows period ratio vs. the pairwise mean characteristic eccentricity for adjacent planets.  There is a strong correlation between adjacent planet period ratio and the mean characteristic eccentricity of the planet pair (Spearman r=0.61, p$\ll$0.001).
 
Figure \ref{fig:prat-e} also shows the eccentricity necessary for orbit crossing, which grows as a function of period ratio.  Note that the characteristic eccentricities for stability are much lower than the orbit crossing eccentricities. However, the natural boundary for multi planet system eccentricities is the orbit crossing eccentricity. Therefore, the eccentricity values must be normalized to the orbit crossing eccentricity to be properly studied in a dynamical sense (\citealt{2018AJ....156...95H},\citealt{2021AJ....162..220T}).  

In the right panel, we plot the normalized eccentricity ($e_{norm}=\frac{(e_{i+1}-e_{i})}{e_{cross}}$) as a function of period ratio.  In this scheme, the pairwise normalized eccentricity is approximately $<0.3$ at all period ratios.  We interpret this to mean that when planet orbits are initialized with 30\% or more of the orbit-crossing eccentricity, they tend toward unstable solutions.  Much of the rest of this manuscript is our attempt to understand which physical mechanisms drive the instability at 30\% of the orbit-crossing eccentricity.

\subsection{Confirming Results with N-body}
To confirm the validity of our results, we took a representative sample of our systems and ran N-body integrations using Rebound. For our sample we chose 10 systems that spanned the range of characteristic eccentricities and were the most ``peas-in-a-pod''-like. We used the same initial orbital parameters as for the SPOCK runs but with the eccentricities set as the characteristic planet eccentricities computed in Section 4. We ran 20 trials per system, varying only the initial orbital angles. Each run was integrated $10^{7}$ years.

\begin{table}[hbt!]
    \centering
    \begin{tabular}{||c || c |c|c|c||} 
    \hline
    System & Multiplicity & $e_{sys}$ & Minimum Period Ratio & Percentage of Stable runs \\ [0.5ex] 
    \hline\hline
    KOI-137 & 3 & $0.11^{+0.03}_{-0.02}$ & 1.558 & $55\%$  \\ 
    \hline
    KOI-2220 & 5 & $0.05^{+0.02}_{-0.01}$ & 1.329 & $55\%$  \\
    \hline
    KOI-1557 & 4 &  $0.08^{+0.03}_{-0.02}$ & 1.353 & $75\%$  \\
    \hline
    *KOI-733 & 4 &  $0.08^{+0.02}_{-0.02}$ & 1.392 & $100\%$  \\
    \hline
    KOI-2722 & 4 &  $0.02^{+0.01}_{-0.01}$ & 1.167 & $90\%$  \\ 
    \hline
    KOI-707 & 5 &  $0.04^{+0.01}_{-0.01}$ & 1.185 & $55\%$  \\ 
    \hline
    KOI-710 & 3 &  $0.07^{+0.03}_{-0.02}$ & 1.241 & $40\%$  \\ 
    \hline
    KOI-1931 & 3 &  $0.04^{+0.02}_{-0.01}$ & 1.254 & $80\%$  \\ 
    \hline
    *KOI-700 & 4 &  $0.08^{+0.04}_{-0.03}$ & 1.349 & $100\%$  \\ 
    \hline
    KOI-939 & 4 &  $0.1^{+0.02}_{-0.02}$ & 1.437 & $40\%$  \\[1ex] 
    \hline 
    \end{tabular}
    \caption{Percentage of stable N-body runs for 5 systems spanning a range of minimum period ratios and characteristic eccentricities. The +/- values denote the range of eccentricities in which a SPOCK stability of 50$\%$ was within a standard deviation of the mean SPOCK stability at that eccentricity. The percentage of stable runs is determined from 20 iterations of the system with randomly varied orbital angles for every iteration. }
    \label{tab:n-body}
\end{table}

For each run we recorded the evolution of each planet's orbital elements. We also noted whether the run resulted in final planetary eccentricities greater than or less than 1. Systems with final eccentricities greater than 1 were considered unstable while systems with eccentricities less than 1 were considered stable. No tested systems that finished with eccentricities less than 1 experienced linear or exponential growth in their eccentricities.
Table \ref{tab:n-body} shows our results. We expect that roughly half of the integrations for each system would be stable on average. This is because the characteristic eccentricity found via SPOCK is the eccentricity that yields a stability percentage of $50\%$. 
Generally,  the percentage of stable N-body runs was between $40\%$ and $100\%$ with an average value of $60\%$ percent. There was no relationship between percentage of stable trials and minimum period ratio (r-value=-0.132, p-value=.718) or system multiplicity (r-value=0.026, p-value=.943). Two systems, KOI-700 and KOI-733 did not go unstable. However, the large range of values leading to a SPOCK stability of $50\%$ for KOI-700 (e=0.05 to 0.12), as well as the larger characteristic eccentricity for KOI-733 (e=0.08) could be the causes for their lack of instability.  While the sample size is limited, the wide range of tested system architectures suggest that these N-body results are at least consistent with the characteristic eccentricities.


\section{Possible Origins of Instability}
Resonances are important drivers in non-linear dynamics that can either promote or hinder stability.  For example, Laplace resonant chains, in which three or more planets librate about the mean-motion resonance exact solutions, are particularly stable (e.g., the Galilean moons, \citealt{2021A&A...649A..26L}).  However, resonant configuration can also create regions of phase-space with resonant overlap.  A test particle in a region of resonant overlap experiences the forcing of two or more distinct resonances, which tend to drive chaotic behavior.  For example, resonant overlap is the origin of the Kirkwood gap in our solar system's asteroid belt.  Given the important role of resonances in sculpting stability, we investigate which forms of resonance and resonance overlap might sculpt the tendency of planets to have $<0.3$ of the orbit-crossing eccentricity.

Figure \ref{fig:prat-e} indicates the first-order mean motion resonances as vertical lines.  There are no clear features at these special period ratios.  This is understandable as the systems in our sample are generally not located exactly in mean motion resonances. Next, we consider the role of 2-body, mean-motion resonance overlap as discussed in \cite{2024AJ....167..271V}. An analytical description of where this resonance overlap occurs is given by \cite{2018AJ....156...95H}:
\begin{equation}
    e_{\mathrm{critical},i,i+1}=0.72e_{\mathrm{cross}}\mathrm{exp}[-1.4\mu_{i+1}^{1/3}\big(\frac{a_{i+1}}{a_{i+1}-a_{i}}\big)^{4/3}]
\end{equation}
where $e_{\mathrm{critical}}$ is the eccentricity necessary for 2-body mean motion resonance overlap, $e_{cross}$ is the eccentricity necessary for the orbits to cross, $a$ is the semi-major axis and $\mu$ is the mass ratio between planet and star. The critical eccentricity for 2-body resonance overlap is indicated in Figure \ref{fig:prat-e} and is generally much higher than the characteristic eccentricities we found.  Therefore, 2-body resonance overlap is not the sole driver of instability for our sample.   
 While 2-body resonant interactions are not the sole contributor to system stability, there seem to be 
 both secular (\citealt{2024AJ....167..271V}) as well as higher-order nonsecular (\citealt{2021CeMDA.133...39P}) effects that could lead to instabilities. Below, we will give a brief description of possible mechanisms and comment on their suitability for sculpting the observed normalized eccentricity ceiling.
 
\subsection{Secular Resonances}
One possible mechanism for promoting instability is secular behavior. The secular approximation can be applied to systems that are far from mean-motion resonance.  In the secular approximation, we take the orbit-averaged dynamics of the planets, and so the important dynamical interactions are determined by the procession rates of the orbital elements.  In secular resonance, two bodies (which do not have to be adjacent) have equal precession rates of their orbital elements, allowing them to interact in a fixed pattern configuration over long timescales.
With the secular approximation, resonance overlap can also occur. Like with 2-body mean-motion resonances, if overlap occurs between two secular resonances, chaotic behavior occurs and instability of the system is assumed \citep{1990CeMDA..49..177S}, with the width of the overlap region being eccentricity dependent. Previously this methodology has been applied to the solar system, specifically Mercury \citep{2011ApJ...739...31L}. While system compactness makes secular resonance overlap likelier, the angular momentum deficit (AMD) of the system \citep{1997A&A...317L..75L}as well as the necessity of long timescales and matching procession ratios limits this method's effectiveness in producing instabilities.

\subsection{Combined Resonant Effects}
However, multiple perturbative effects can be used together to achieve an instability. A possible mechanism promoting instability could be three-body resonances. Due to the addition of a third angular term, 3-body resonances are more numerous than their two-body counterparts \citep{2011MNRAS.418.1043Q}. These three-body resonances can then cause an evolution of the planets semi-major axis leading the planet to cross into a two-body MMR overlap causing system instability to occur. This method of chaotic diffusion to 2-body MMRs was investigated for zeroth-order circular planets by \cite{2020A&A...641A.176P}, who calculated an instability timescale for circular, co-planar systems that were equally massed and spaced. Further computational work by \cite{2024arXiv240317928L} showed that such a methodology also explains the N-body integrative results well. Such three-body resonances, however, require dynamically packed systems and could therefore be too weak to drive instability in widely spaced multi-planet systems. 

Alternatively, secular effects can also evolve the system to a point where mean-motion resonances can cause instability. Secular resonances can redistribute AMD from outer sources, (\citealt{1999ssd..book.....M}) growing inner system eccentricities to the point where 2-body MMR overlap can be achieved, causing an instability \citep{2006ApJ...639..423M}. Specifically, \cite{2020AJ....160...98V} use secular resonance excitation of eccentricity as a way to cause an instability in a Kepler-102. Furthermore, in their broader study of the general population of Kepler and K2 systems, \cite{2024AJ....167..271V} use the spectral fraction index to show that such eccentricity exciting secular resonances can occur in many systems. However, the low observed eccentricities of these systems as well as the long timescales that these effects take suggest that these effects at the very least did not occur on the architectures of the systems we observe today.



\section{High Characteristic Eccentricities - A possible sign of Dynamical Relaxation}
One of the more surprising groups of planets in this study are those systems with characteristic eccentricities greater than 0.1. Such systems are greater than double the average observed exoplanet eccentricity. Of systems with characteristic eccentricities over 0.14, roughly half are three planet systems in the 75th percentile in minimum period ratio. This relationship between architecture and characteristic eccentricity could be a sign of dynamical relaxation. \cite{2016ApJ...822...54D} demonstrated how high-e, dynamically packed systems can become low-e dynamically loose systems through the late giant impacts phase. Such a relaxation process could mean that the primordial versions of these systems were prone to secular, three-body, as well as 2-body effects. Observed systems, rather than being an example of stability, could be the dynamically relaxed remnants of earlier system architectures.  Therefore, widely spaced 3-planet systems should have their eccentricities observed to determine whether these systems have gone through dynamical relaxation or currently have eccentricities that reside at or near the stability limit. Such data could provide further insight into how much dynamical processing has the system undergone.

%

\section{Conclusions}
In this work we have studied the stability of 126 multi-planet systems from the California Kepler Survey (CKS) sample. We note the relationship between eccentricity and SPOCK stability, and denote a characteristic eccentricity that leads to a system stability of 0.5. We compared multiple system level and planet pair based metrics to these characteristic eccentricities. Like theoretical results suggest, we confirm period ratio to be a strong predictor with characteristic eccentricity for real, observed systems. Because of the role individual planet pairs seemed to play in stability, we have also found individual planet characteristic eccentricities for every planet in 103 of the 126 SPOCK compatible systems. We find that the relationship between period ratio and eccentricity holds for individual planet pairs. The relationship disappears, however, when one compares the period ratio and normalized eccentricity. Using the individual characteristic planetary eccentricities we also explored the importance of two-body resonances in system stability. Using the relation derived in \cite{2018AJ....156...95H}, we calculated the critical eccentricities for two-body MMR overlap. In comparing our characteristic eccentricities to the critical eccentricities we note that our characteristic values are significantly lower. For planets in the CKS sample we conclude:
\begin{enumerate}
    \item The stability of a system is better described by the spacing of the closest pair of planets than a system-wide metric. This suggests that it is a perturbation of a single pair of planets (either by only those two planet of another source in the system) that drives (in)stability. 
    \item While the correlation between period ratio and eccentricity is present for individual planet pairs, there is no correlation between period ratio and the normalized eccentricity. 
    \item The SPOCK calculated characteristic eccentricities were much lower than the critical eccentricities necessary for 2-body MMR overlap. This suggests that 2-body MMR overlap is not the sole driving force of system stability.
\end{enumerate}
While we explored the relationship between eccentricity and stability as well as it's relationship with 2-body resonances it is clear that there is more to this story. Other mechanisms such as those outlined above clearly must play a role in determining system stability. While it would be convenient to find one mechanism that defines the stability of every system, in reality that is probably not the case. Furthermore, the possibility of dynamical relaxation could mean that these systems are not primordial, but rather the remnants of previously unstable systems. Therefore, creating a complete picture of planetary systems as well as comparing the constraints these mechanisms might put on system parameters, such as eccentricity, to observed values, could help determine which of these mechanisms dominates. 

\begin{acknowledgments}
   The authors thank Cristobal Petrovich, Katherine Volk, Renu Mahlotra, Caleb Lammers, and Daniel Tamayo for thought-provoking and insightful discussions. MJD acknowledges support from the University of Notre Dame through the College of Science Summer Undergraduate Research Fellowship (COS-SURF).  MJD and LMW acknowledge support from the NASA Exoplanet Research Program (grant no. 80NSSC23K0269). MJD additionally thanks the organizers of the 55th AAS Division of Dynamical Astronomy Meeting, Exoplanets 5, and the Great Lakes Exoplanet Area Meeting (GLEAM) for the oppoturnity to present early versions of this project and to network within the astrodynamics and exoplanet communities. 
\end{acknowledgments}

\clearpage
\section{Appendix}
\subsection{Eccentricity vs. SPOCK Stability Plots}
\begin{figure*}[hbt!]
    \centering
    \begin{subfigure}{}
        \includegraphics[width=0.48\textwidth]{spock_cks_CHAREvsStabilityKDEContourWARCHVALSCbarNTLimGKOI_K00070.pdf}
    \end{subfigure}
    \begin{subfigure}{}
        \includegraphics[width=0.48\textwidth]{spock_cks_CHAREvsStabilityKDEContourWARCHVALSCbarNTLimGKOI_K00082.pdf}
    \end{subfigure}
    \begin{subfigure}{}
        \includegraphics[width=0.48\textwidth]{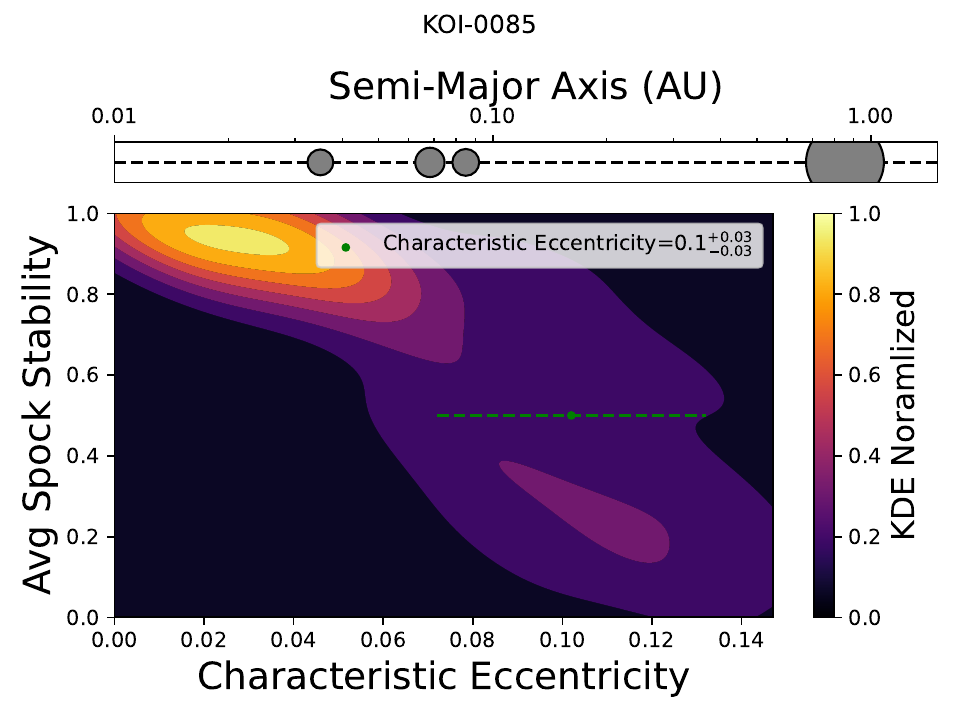}
    \end{subfigure}
    \begin{subfigure}{}
        \includegraphics[width=0.48\textwidth]{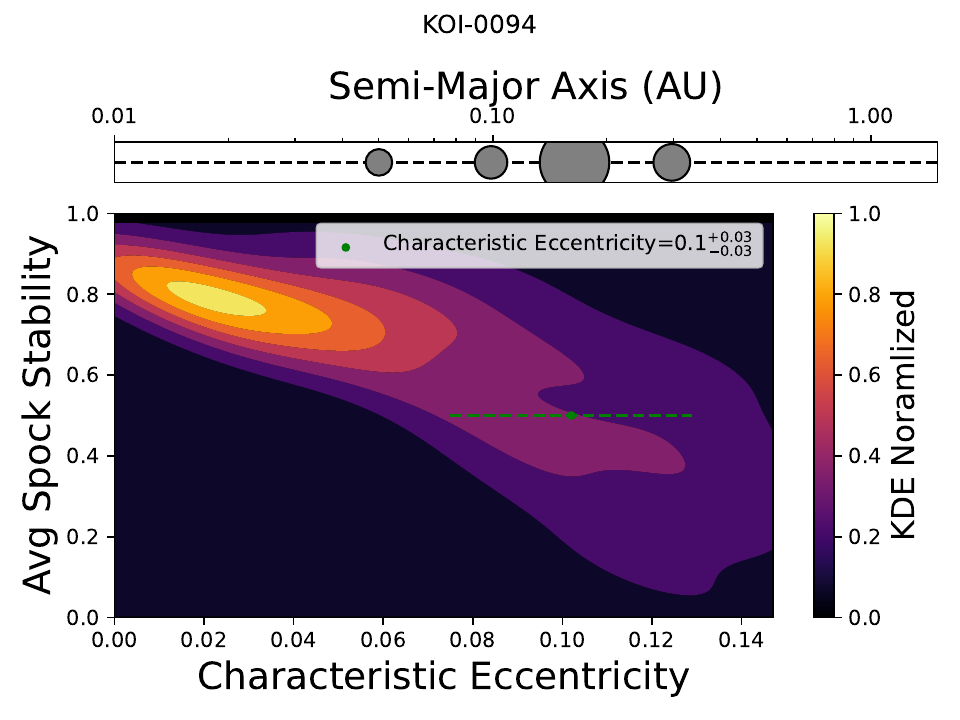}
    \end{subfigure}
    \begin{subfigure}{}
        \includegraphics[width=0.48\textwidth]{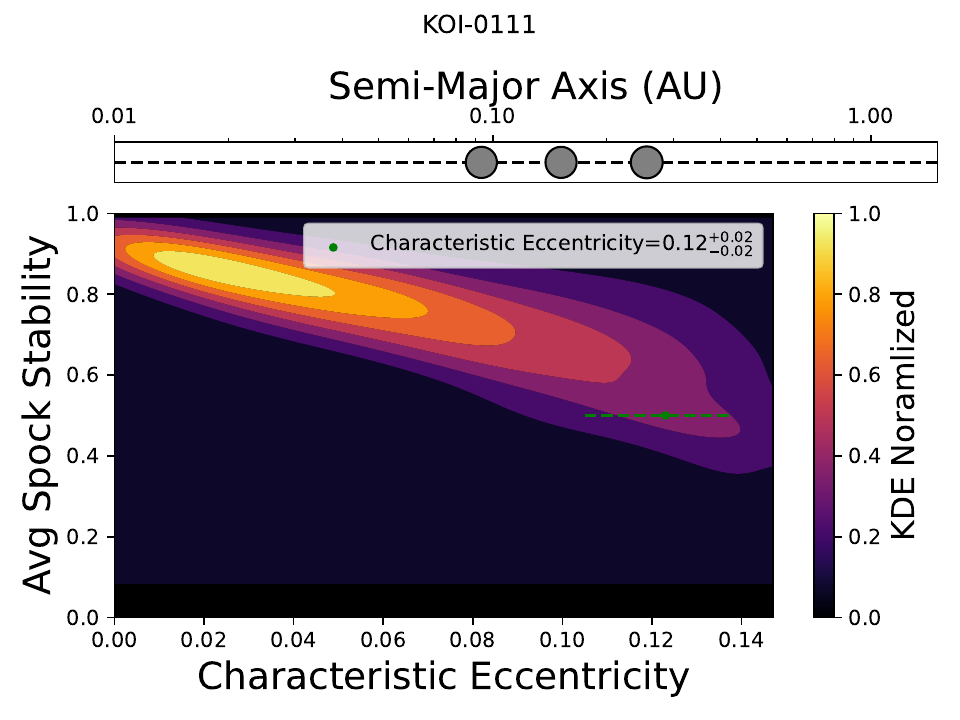}
    \end{subfigure}
    \begin{subfigure}{}
        \includegraphics[width=0.48\textwidth]{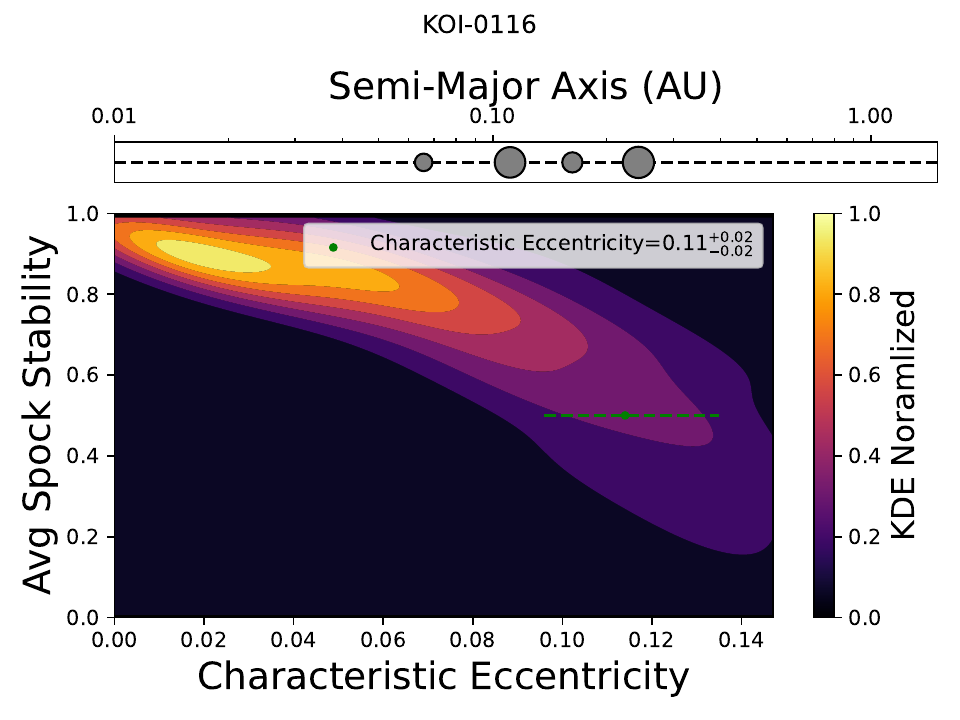}
    \end{subfigure}
\end{figure*}
\begin{figure*}
    \begin{subfigure}{}
        \includegraphics[width=0.48\textwidth]{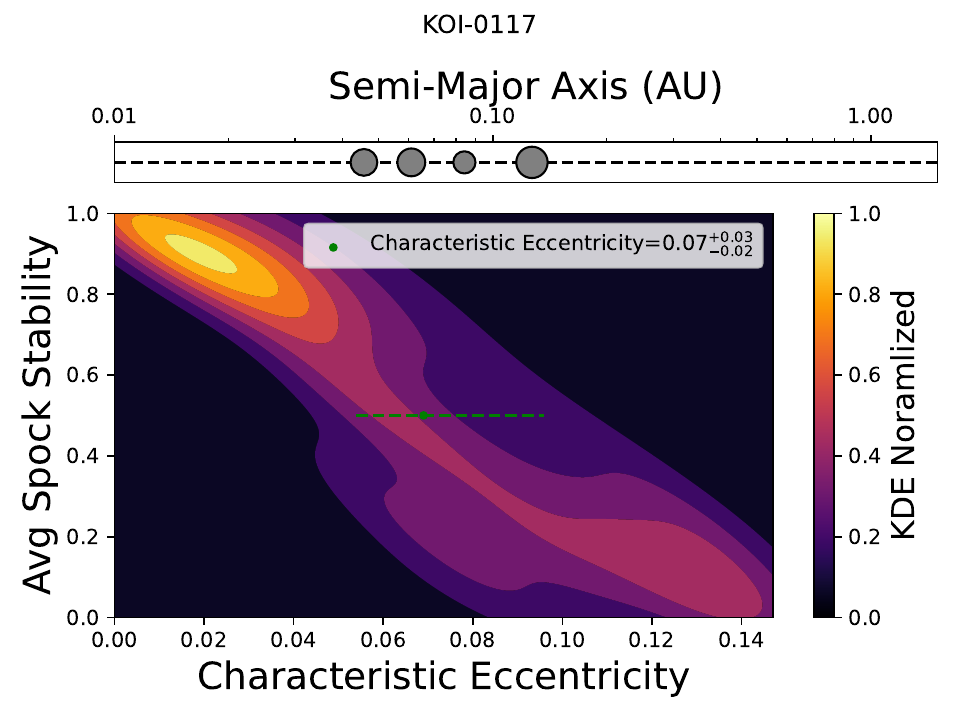}
    \end{subfigure}
    \begin{subfigure}{}
        \includegraphics[width=0.48\textwidth]{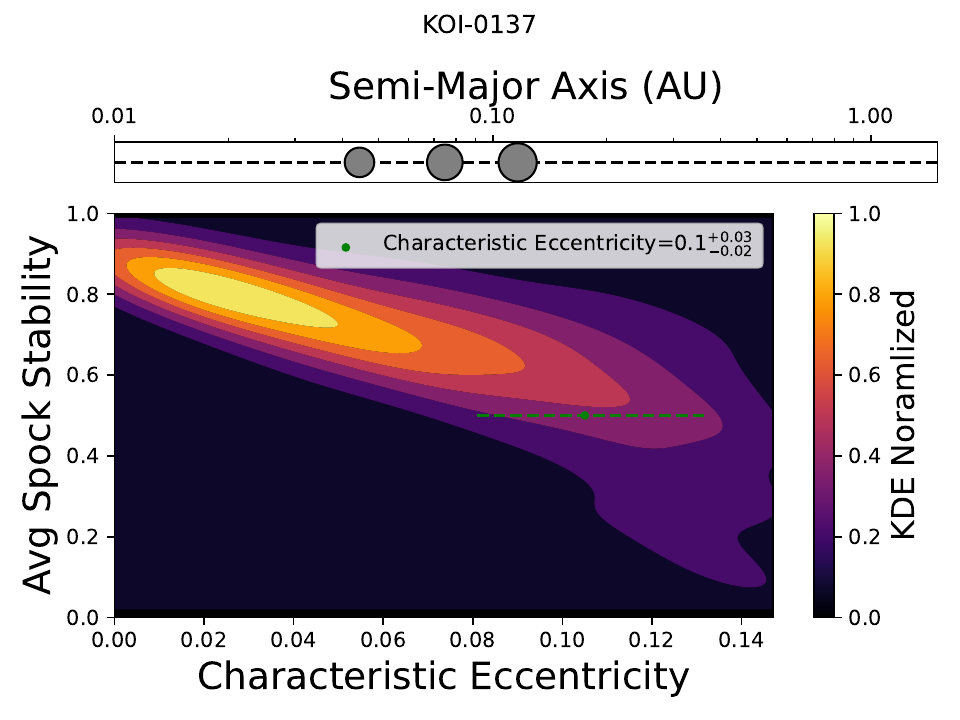}
    \end{subfigure}
    \begin{subfigure}{}
        \includegraphics[width=0.48\textwidth]{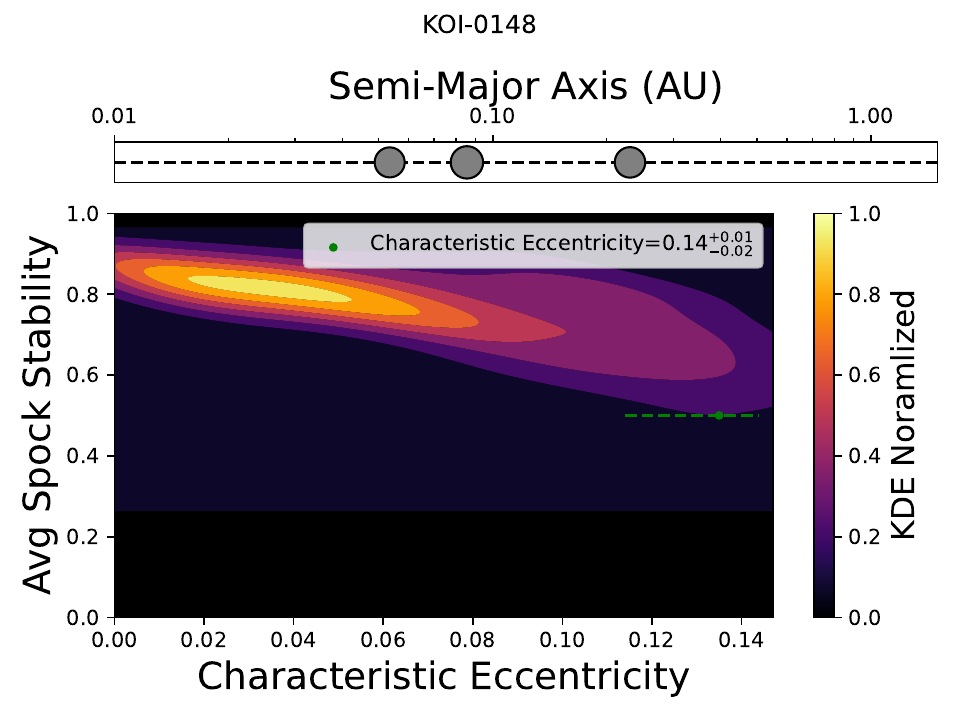}
    \end{subfigure}
    \begin{subfigure}{}
        \includegraphics[width=0.48\textwidth]{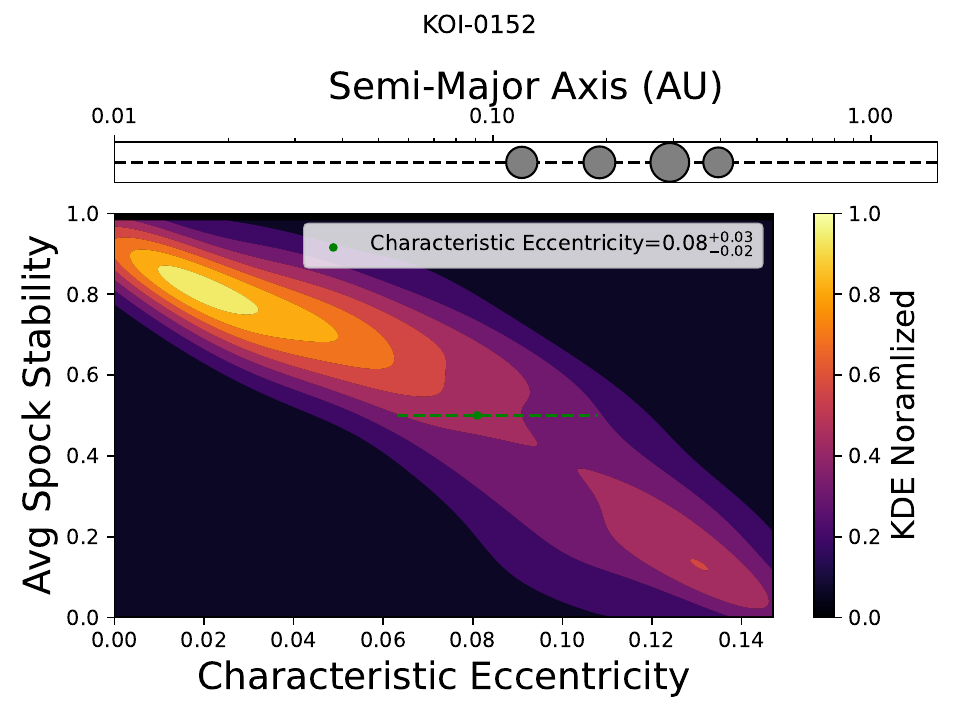}
    \end{subfigure}
    \begin{subfigure}{}
        \includegraphics[width=0.48\textwidth]{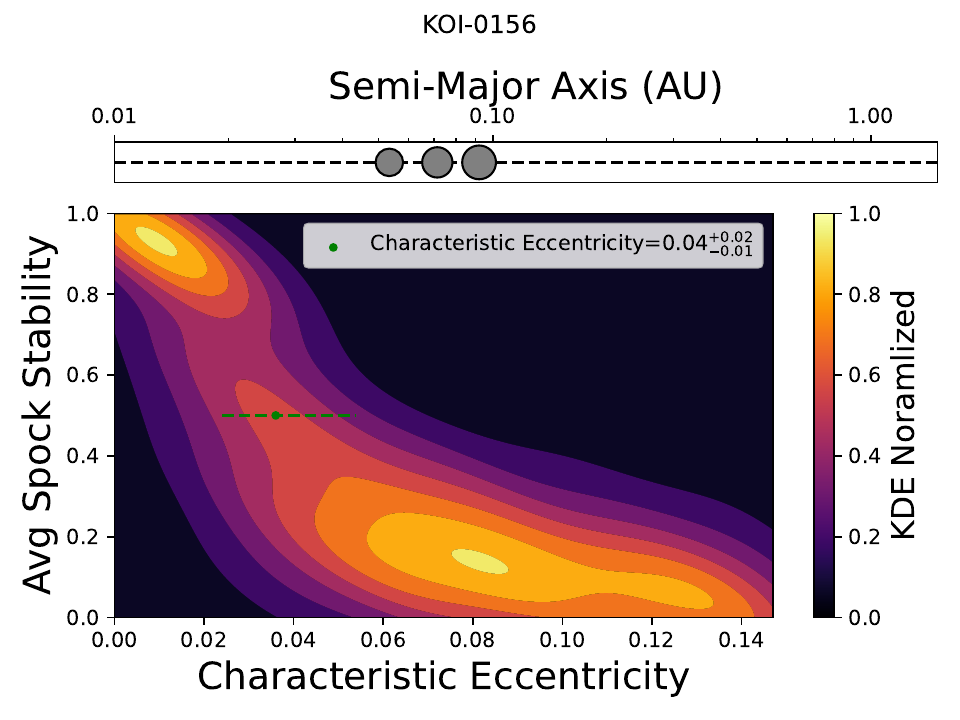}
    \end{subfigure}
    \begin{subfigure}{}
        \includegraphics[width=0.48\textwidth]{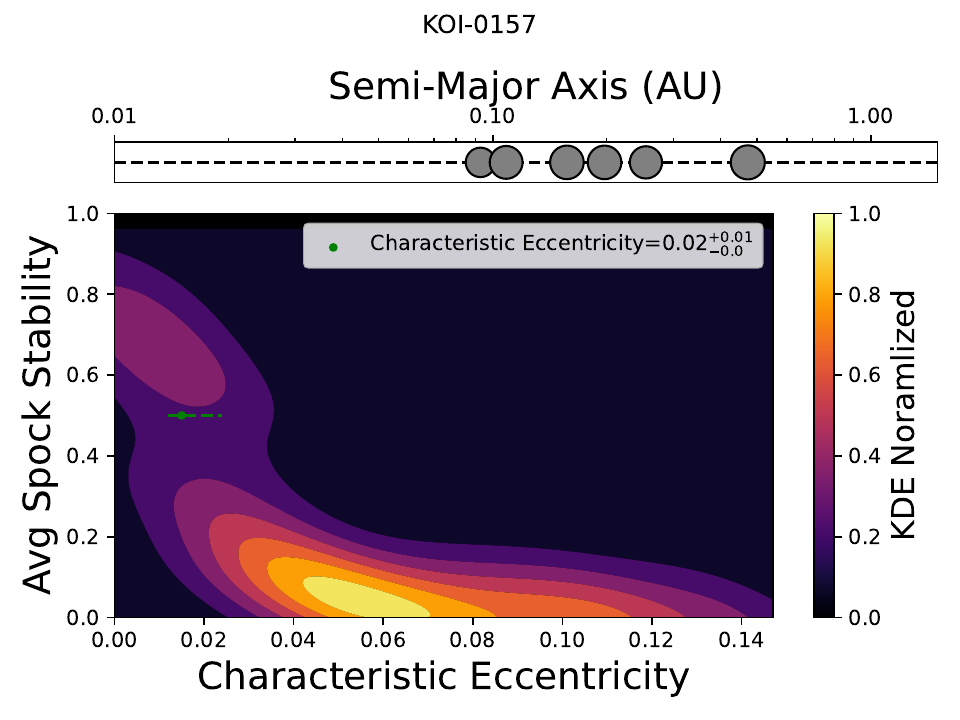}
    \end{subfigure}
\end{figure*}
\begin{figure*}
    \begin{subfigure}{}
        \includegraphics[width=0.48\textwidth]{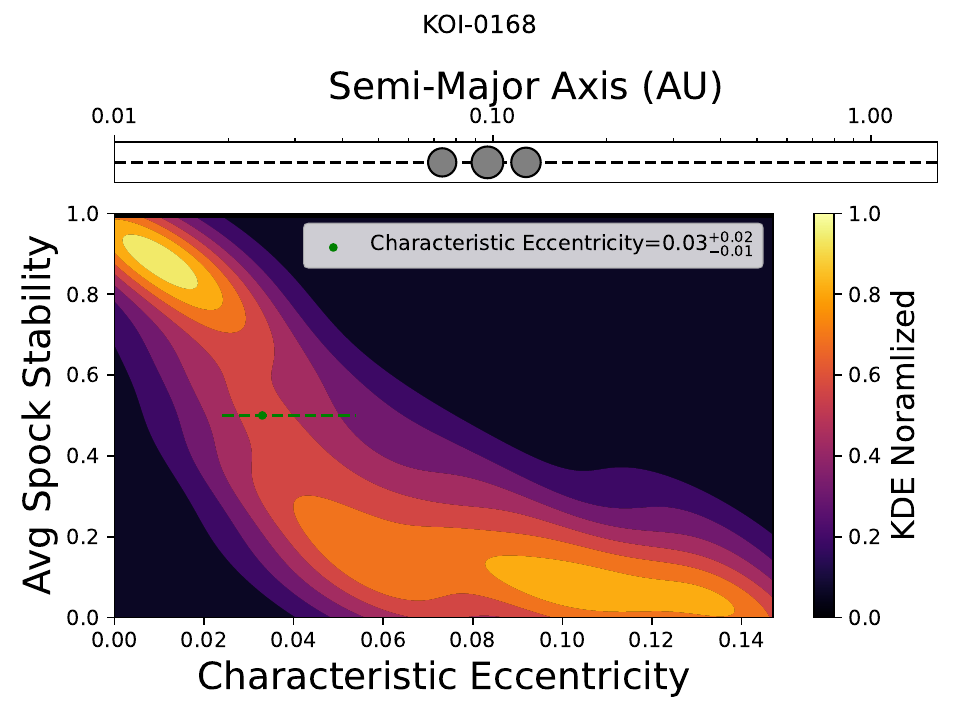}
    \end{subfigure}
    \begin{subfigure}{}
        \includegraphics[width=0.48\textwidth]{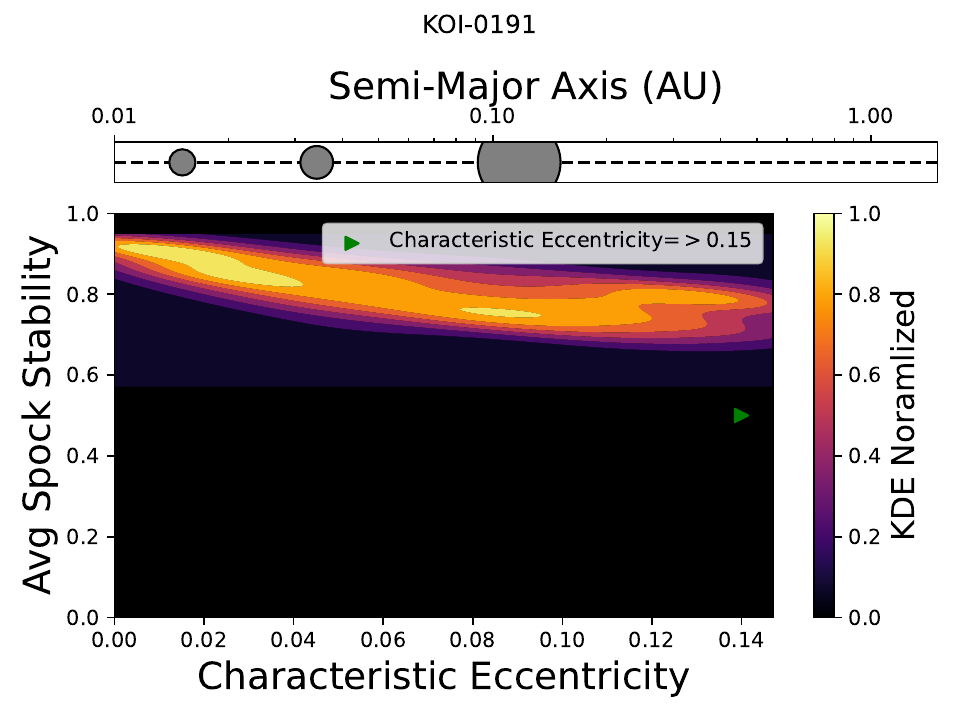}
    \end{subfigure}
    \begin{subfigure}{}
        \includegraphics[width=0.48\textwidth]{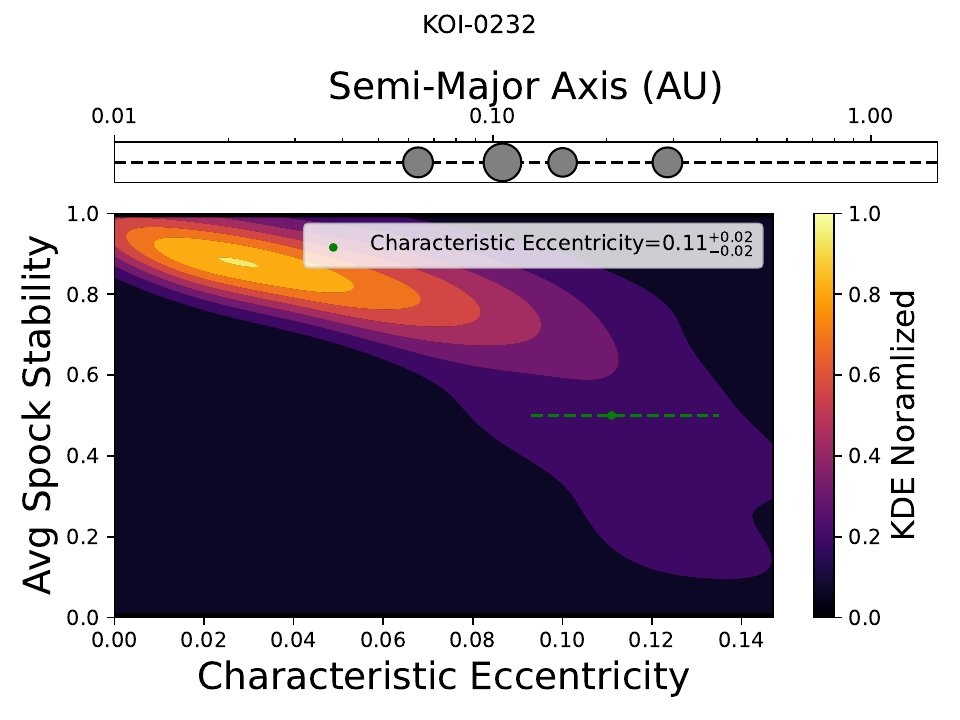}
    \end{subfigure}
    \begin{subfigure}{}
        \includegraphics[width=0.48\textwidth]{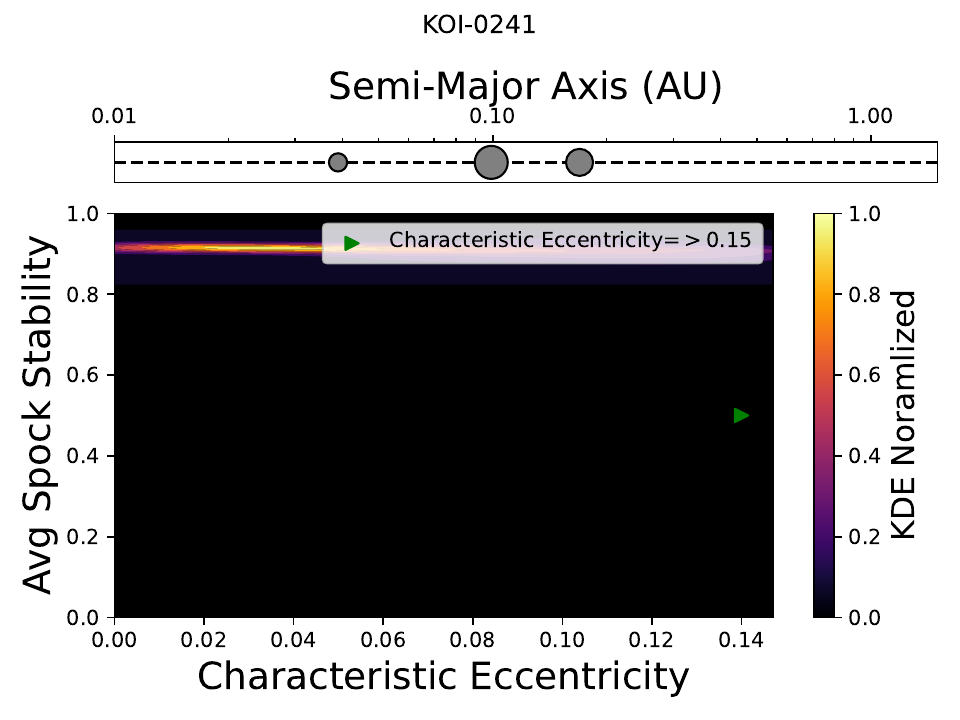}
    \end{subfigure}
    \begin{subfigure}{}
        \includegraphics[width=0.48\textwidth]{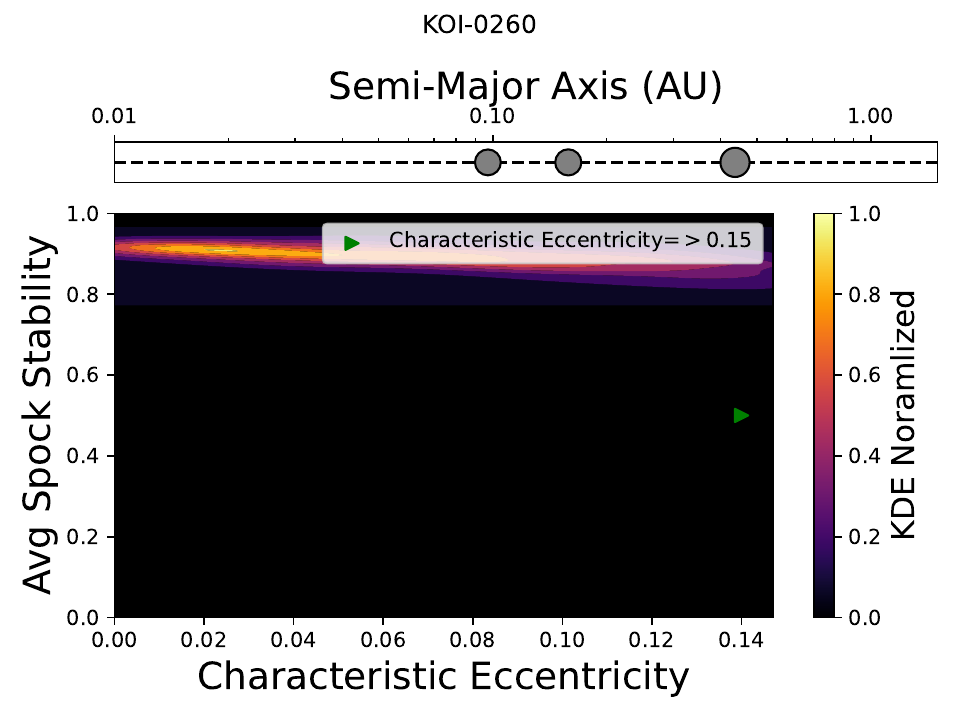}
    \end{subfigure}
    \begin{subfigure}{}
        \includegraphics[width=0.48\textwidth]{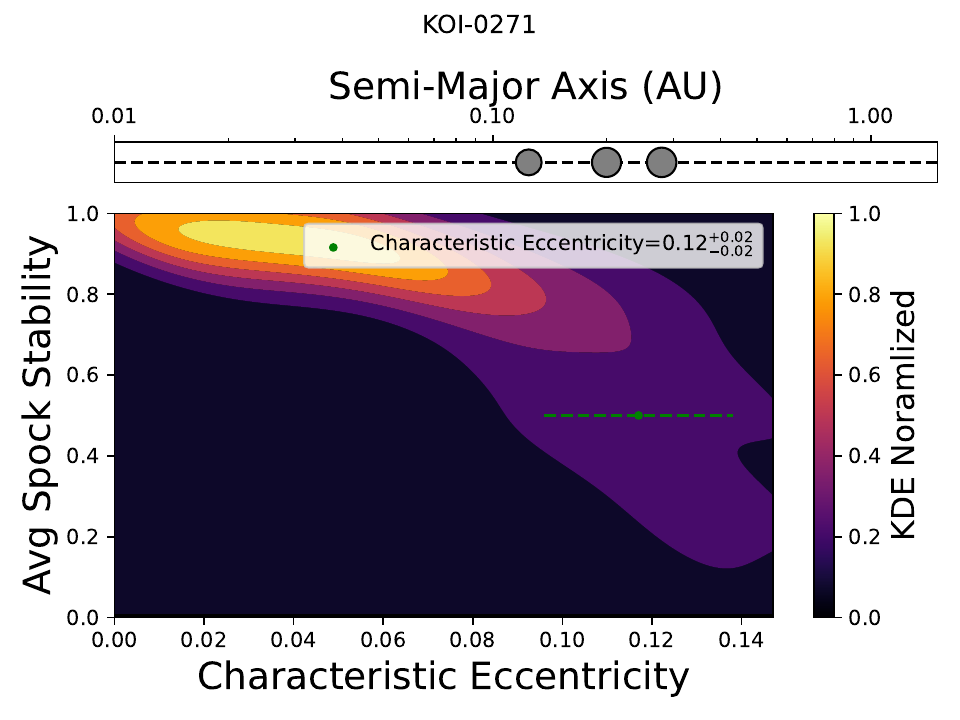}
    \end{subfigure}
\end{figure*}
\begin{figure*}
    \begin{subfigure}{}
        \includegraphics[width=0.48\textwidth]{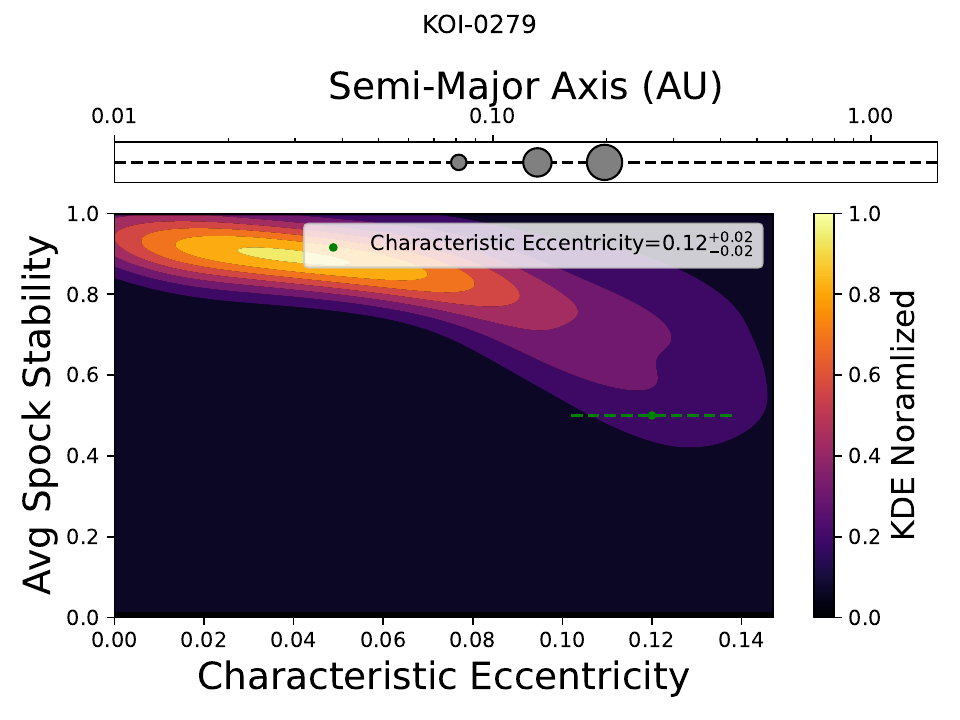}
    \end{subfigure}
    \begin{subfigure}{}
        \includegraphics[width=0.48\textwidth]{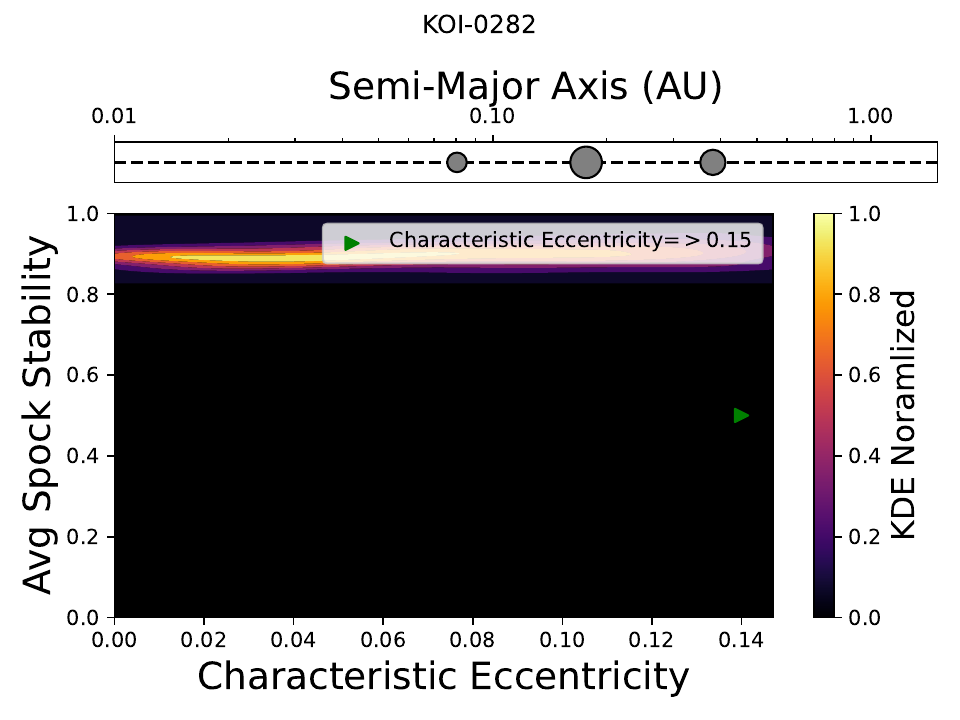}
    \end{subfigure}
    \begin{subfigure}{}
        \includegraphics[width=0.48\textwidth]{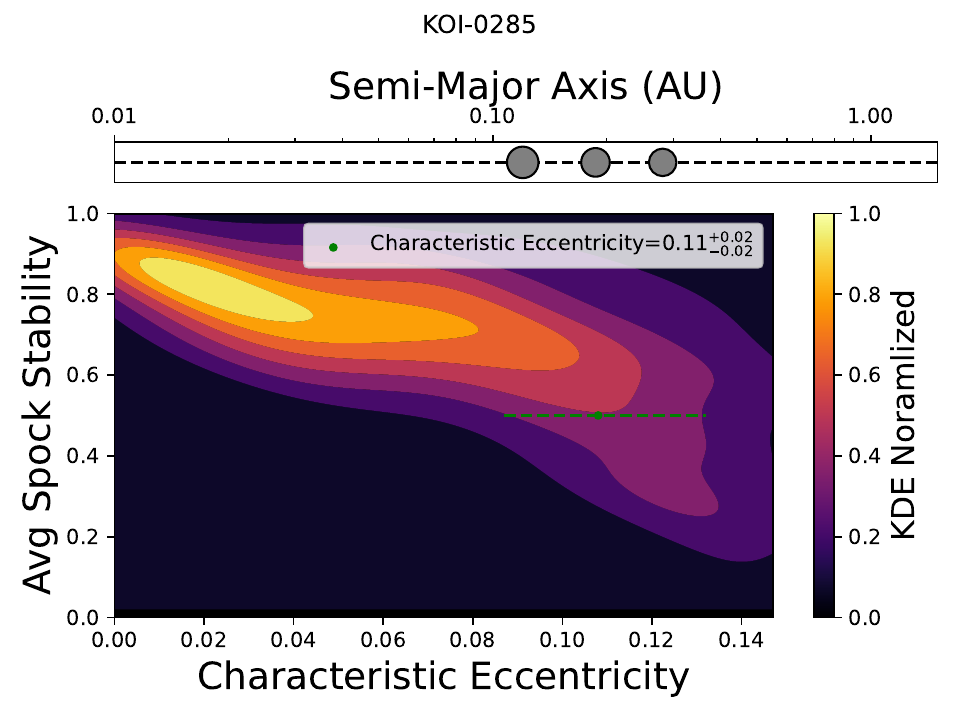}
    \end{subfigure}
    \begin{subfigure}{}
        \includegraphics[width=0.48\textwidth]{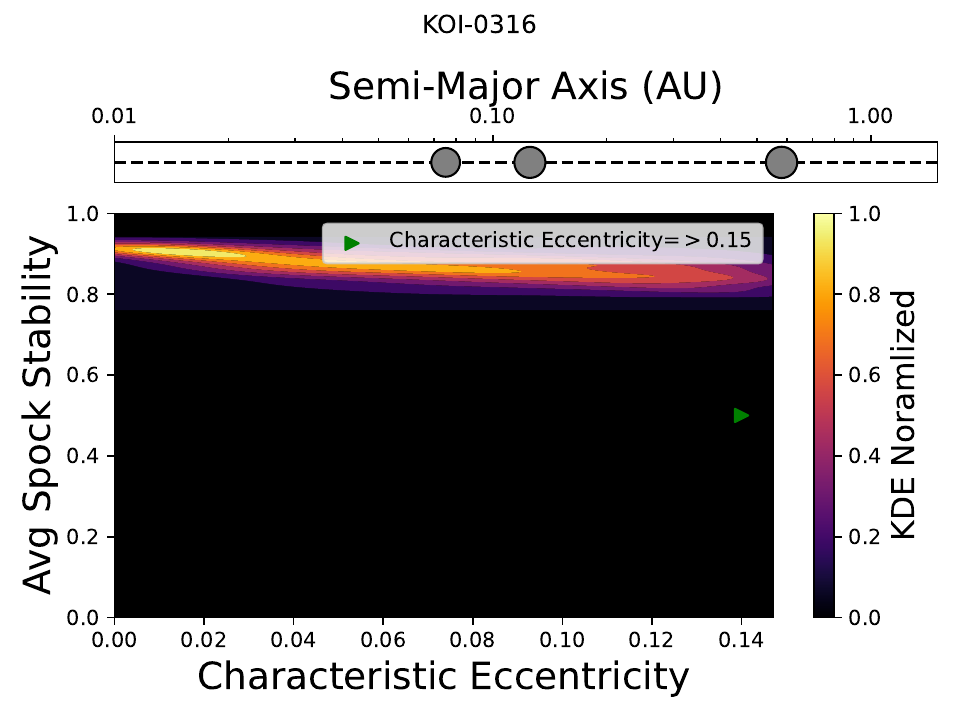}
    \end{subfigure}
    \begin{subfigure}{}
        \includegraphics[width=0.48\textwidth]{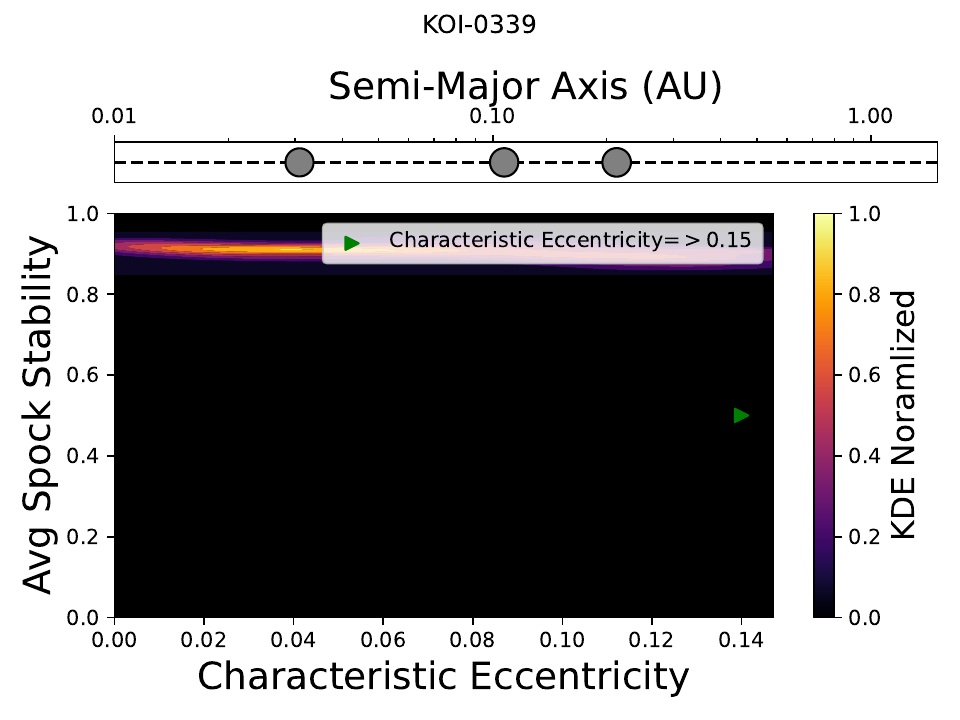}
    \end{subfigure}
    \begin{subfigure}{}
        \includegraphics[width=0.48\textwidth]{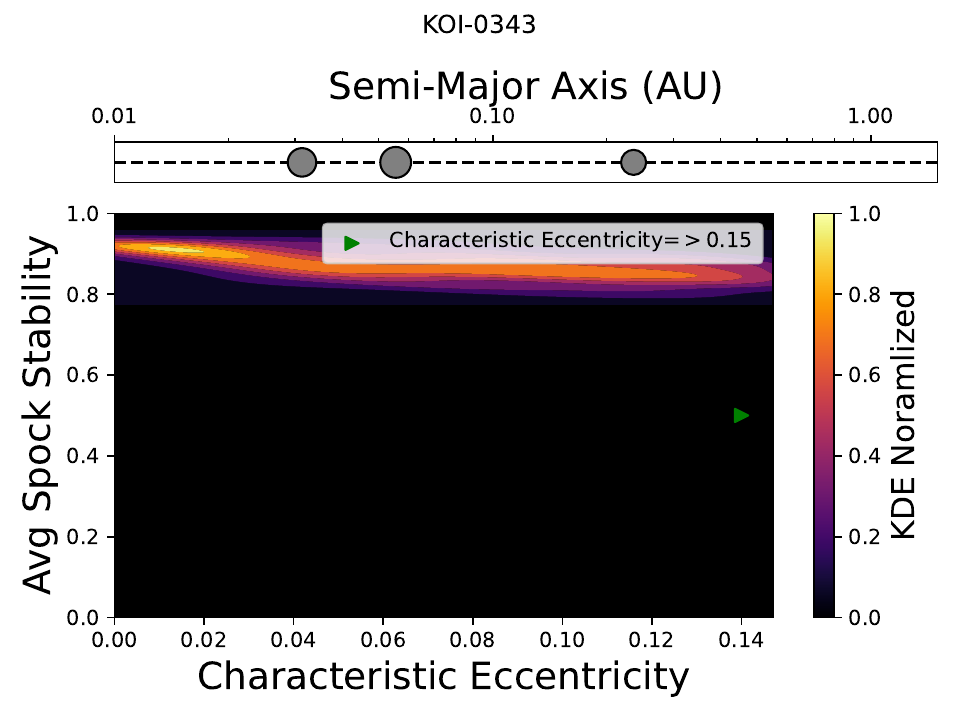}
    \end{subfigure}
\end{figure*}
\begin{figure*}
    \begin{subfigure}{}
        \includegraphics[width=0.48\textwidth]{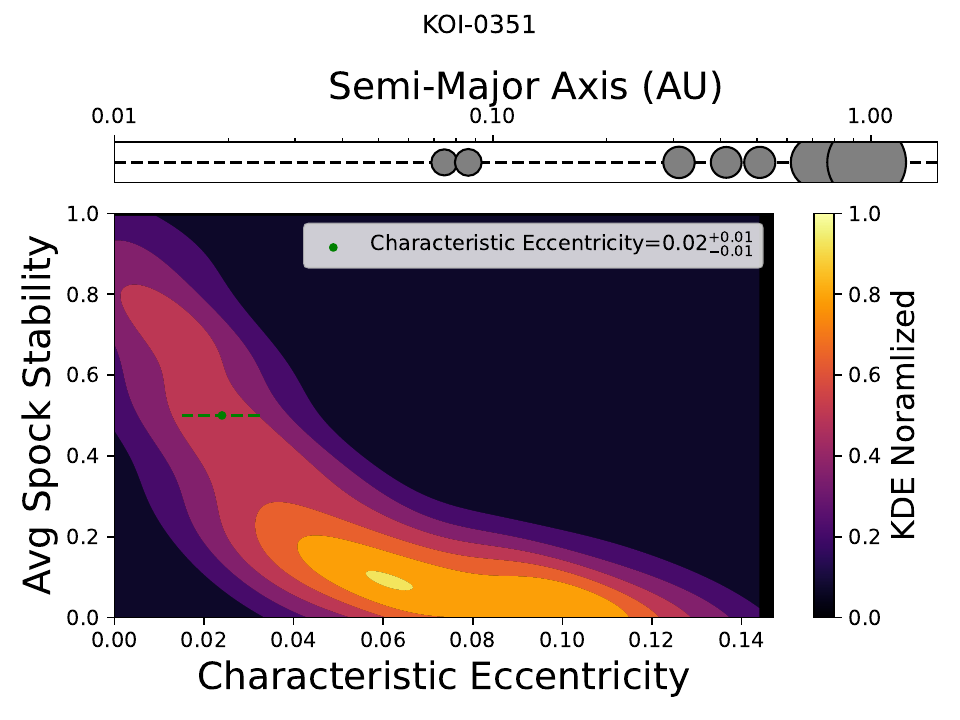}
    \end{subfigure}
    \begin{subfigure}{}
        \includegraphics[width=0.48\textwidth]{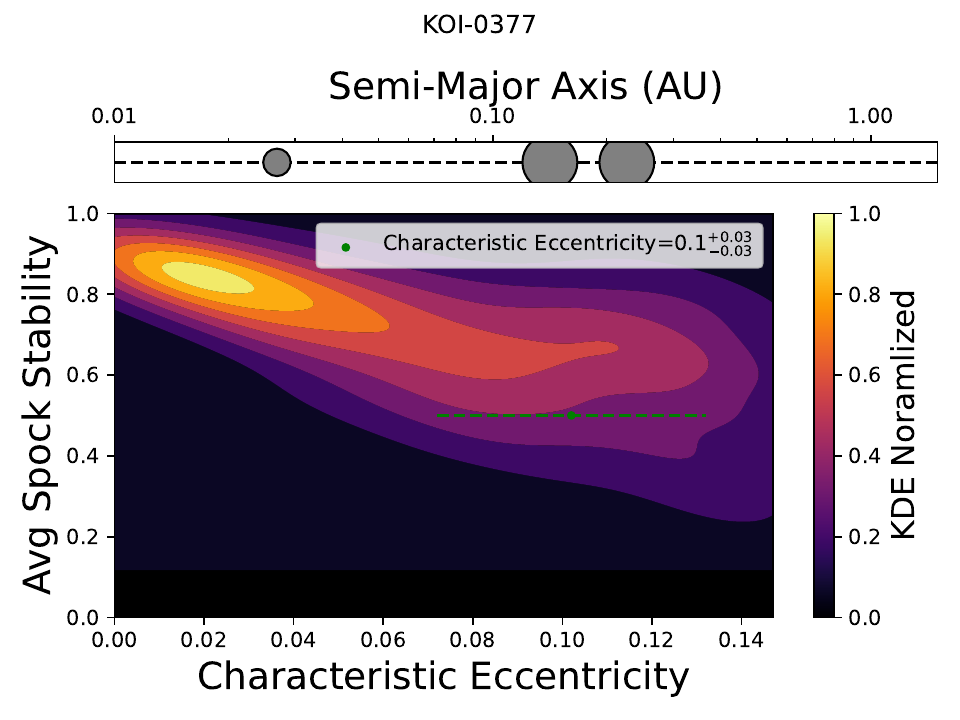}
    \end{subfigure}
    \begin{subfigure}{}
        \includegraphics[width=0.48\textwidth]{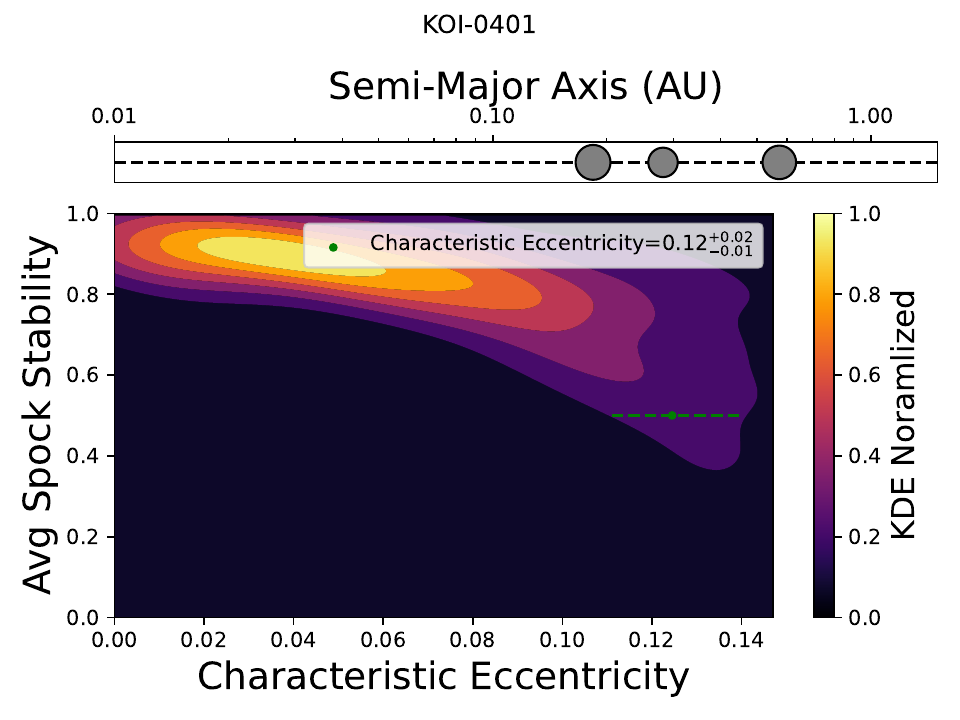}
    \end{subfigure}
    \begin{subfigure}{}
        \includegraphics[width=0.48\textwidth]{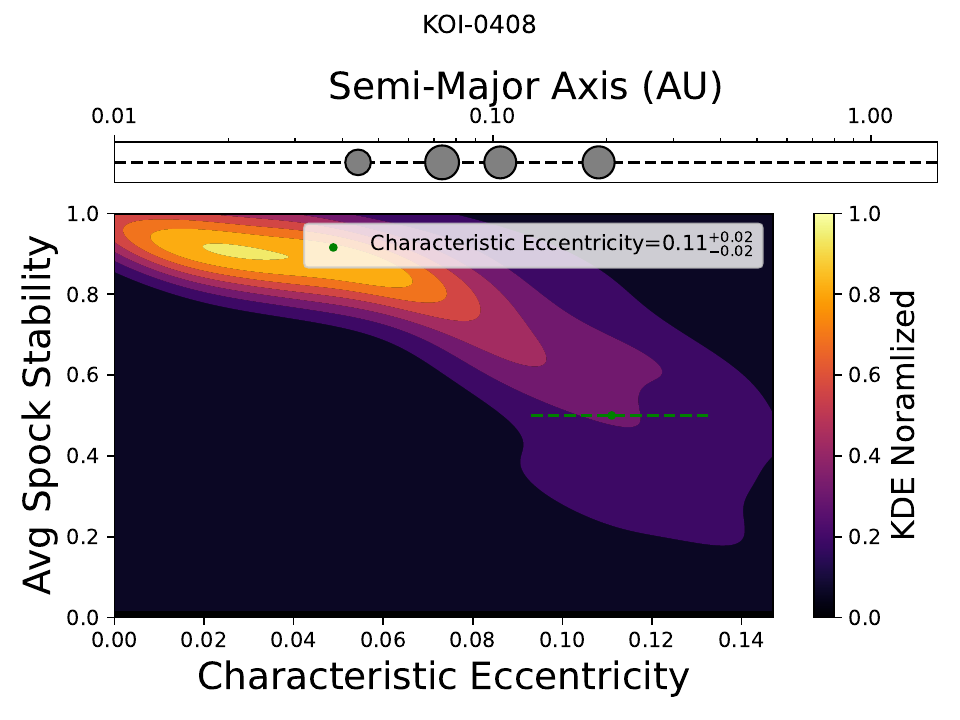}
    \end{subfigure}
    \begin{subfigure}{}
        \includegraphics[width=0.48\textwidth]{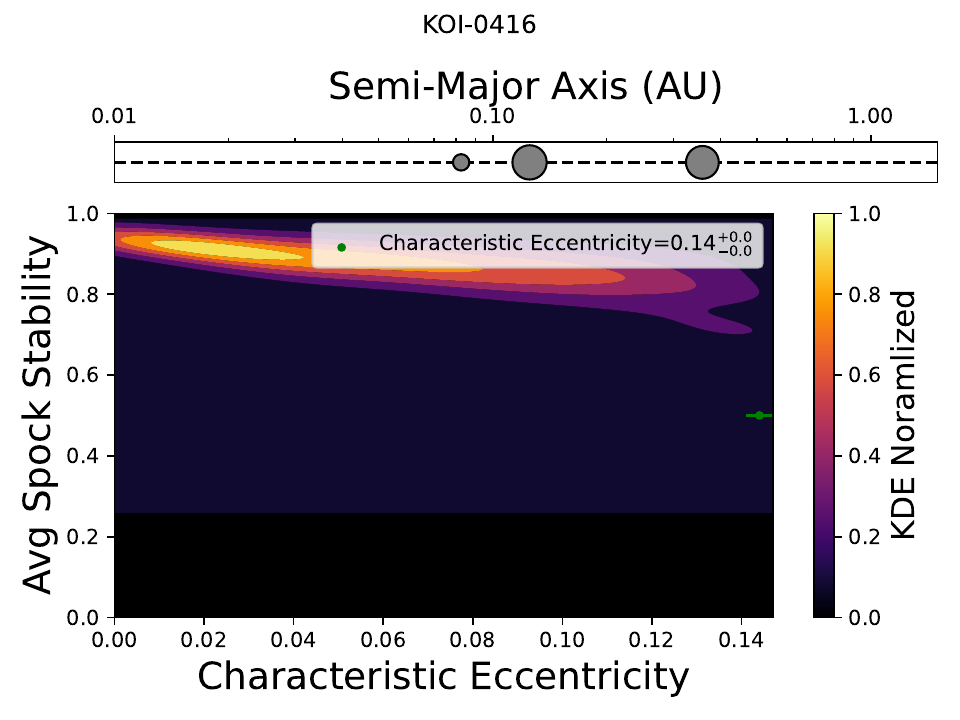}
    \end{subfigure}
    \begin{subfigure}{}
        \includegraphics[width=0.48\textwidth]{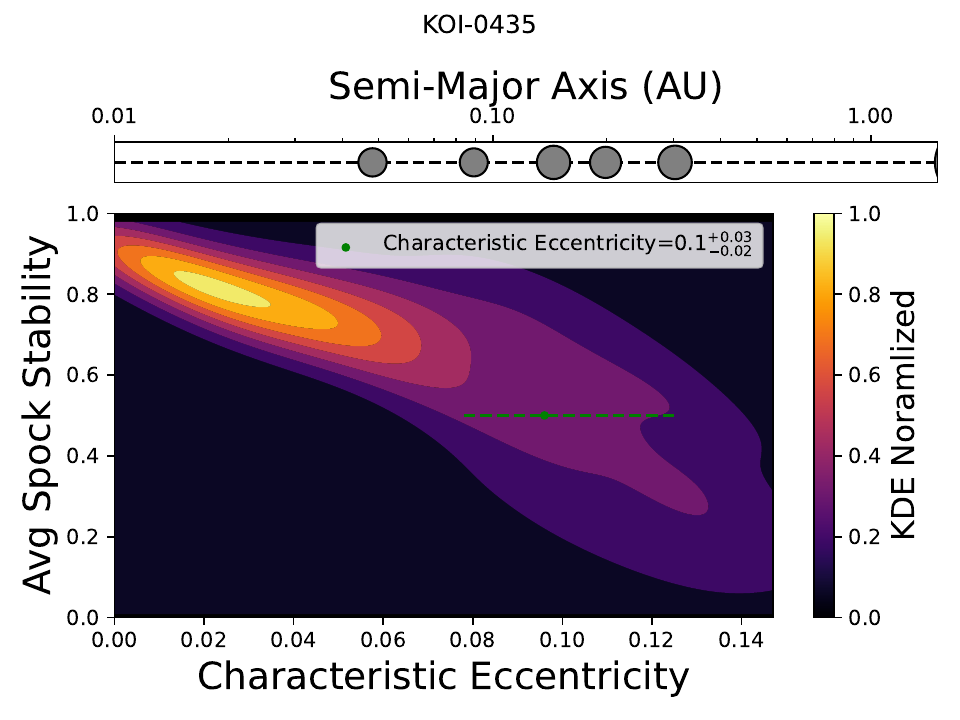}
    \end{subfigure}
\end{figure*}
\begin{figure*}
    \begin{subfigure}{}
        \includegraphics[width=0.48\textwidth]{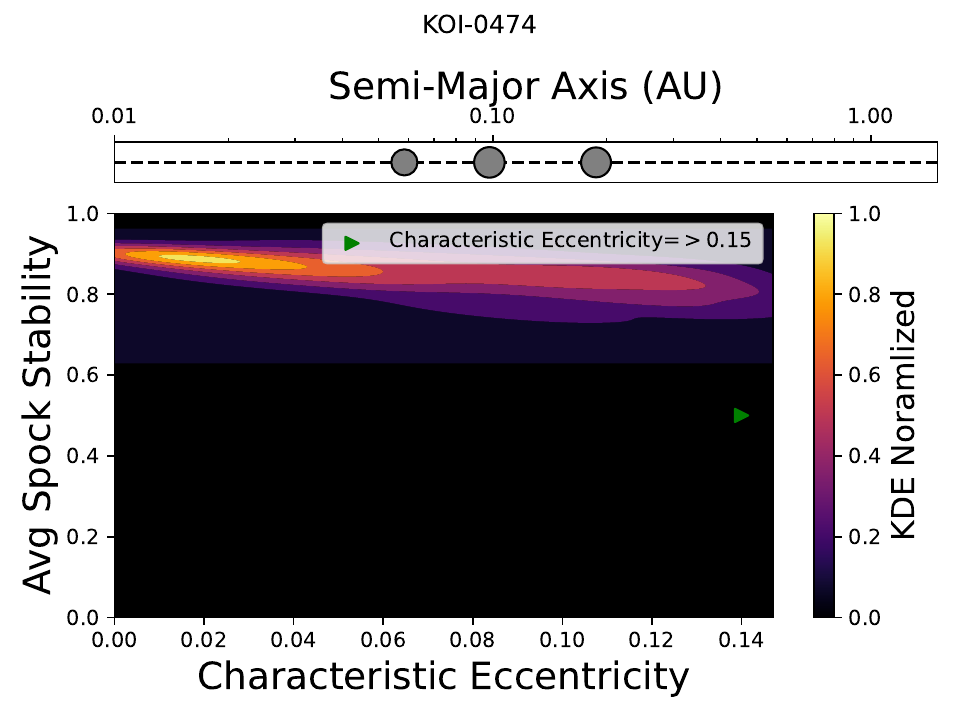}
    \end{subfigure}
    \begin{subfigure}{}
        \includegraphics[width=0.48\textwidth]{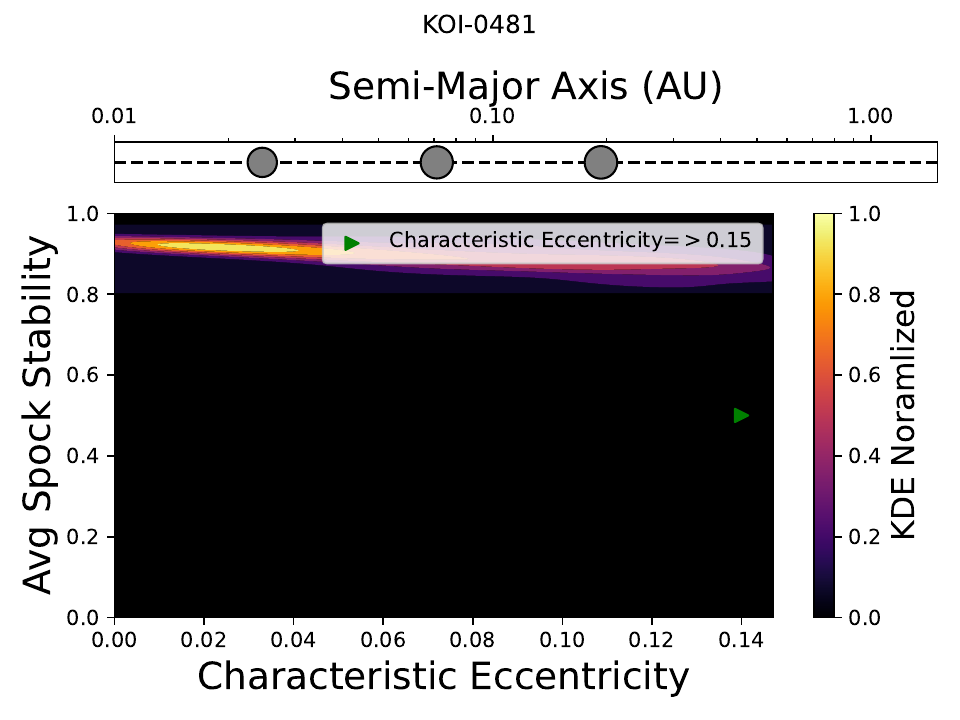}
    \end{subfigure}
    \begin{subfigure}{}
        \includegraphics[width=0.48\textwidth]{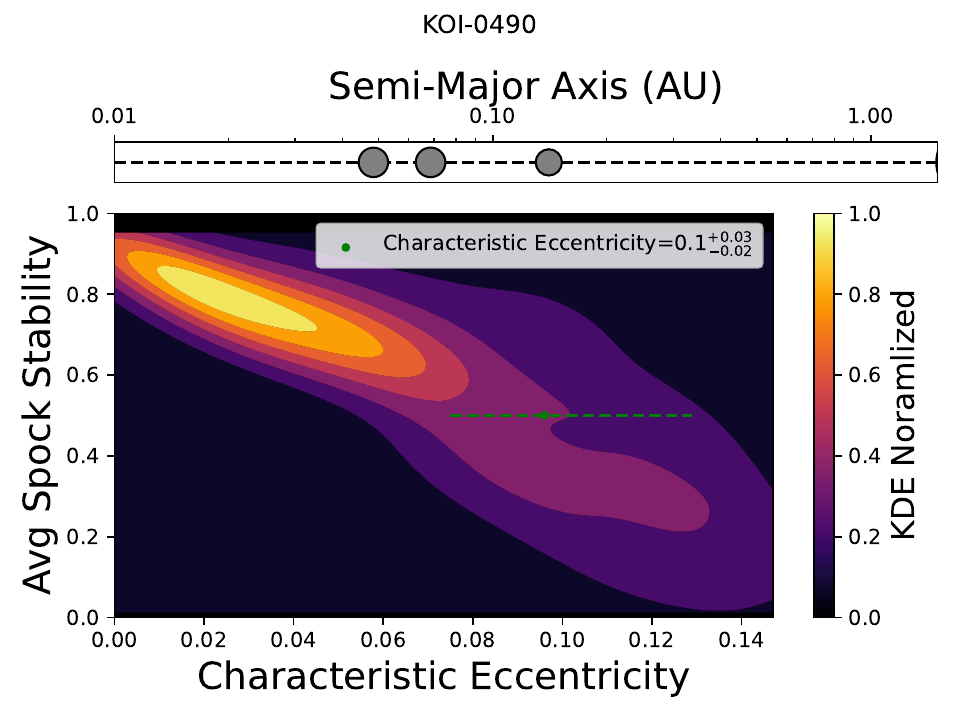}
    \end{subfigure}
    \begin{subfigure}{}
        \includegraphics[width=0.48\textwidth]{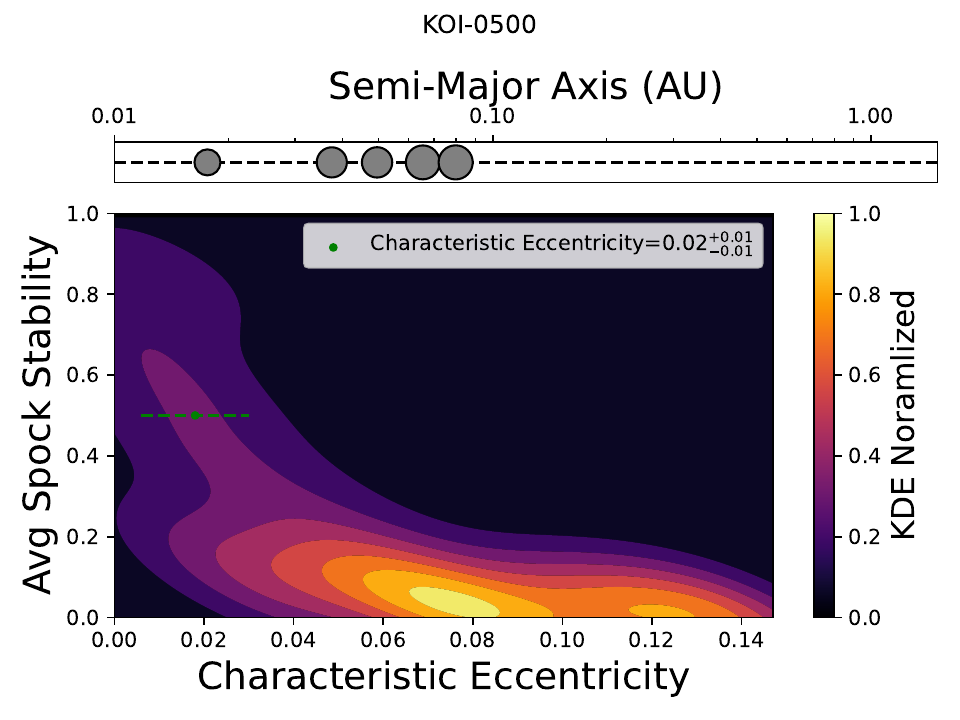}
    \end{subfigure}
    \begin{subfigure}{}
        \includegraphics[width=0.48\textwidth]{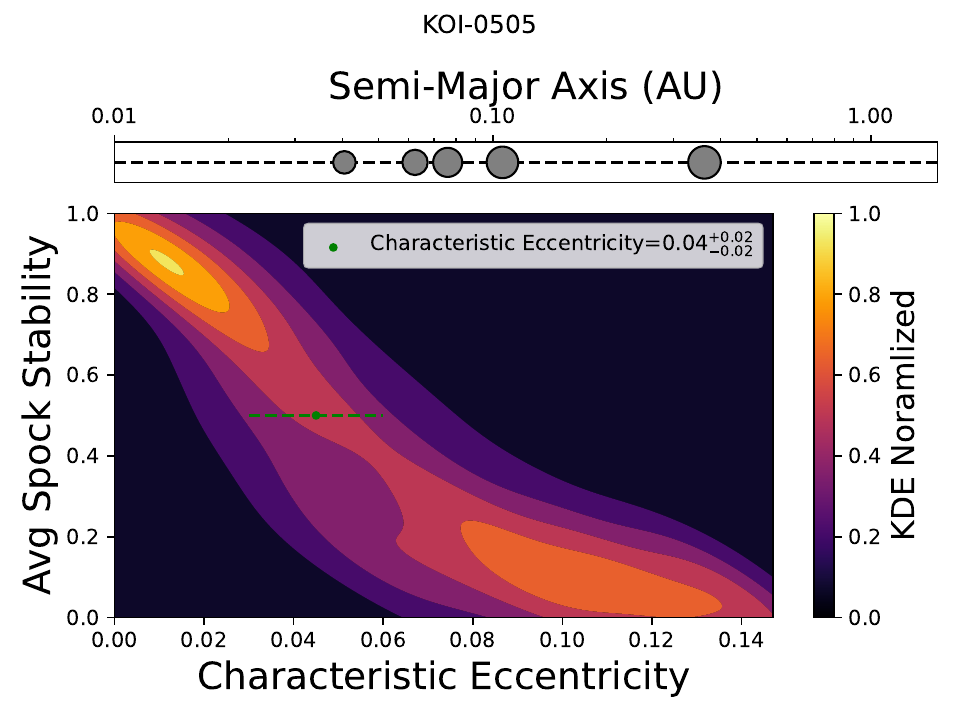}
    \end{subfigure}
    \begin{subfigure}{}
        \includegraphics[width=0.48\textwidth]{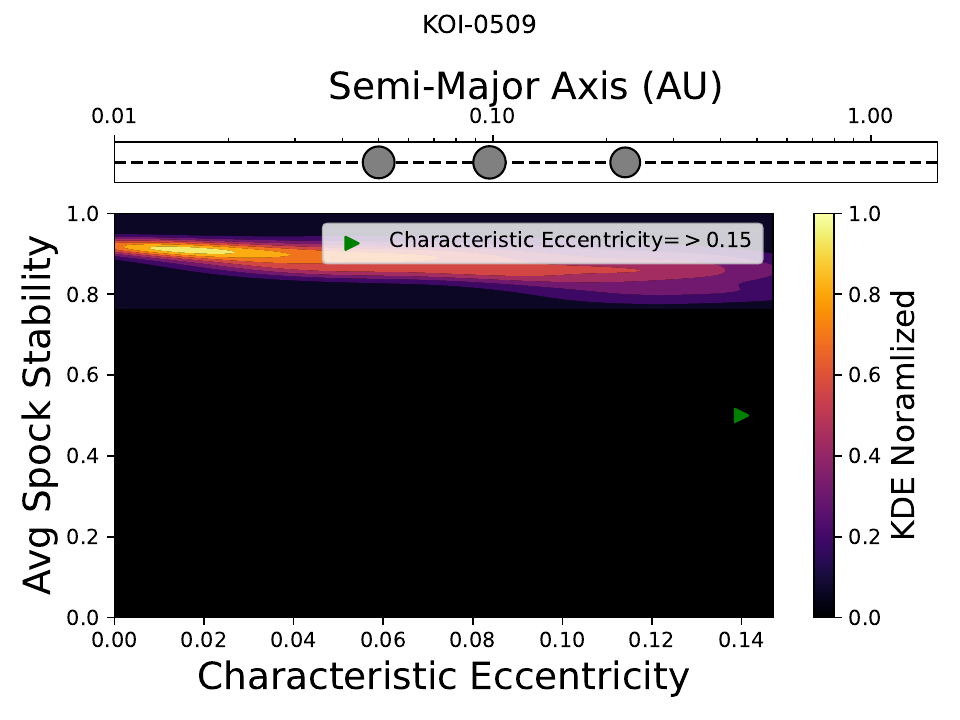}
    \end{subfigure}
\end{figure*}
\begin{figure*}
    \begin{subfigure}{}
        \includegraphics[width=0.48\textwidth]{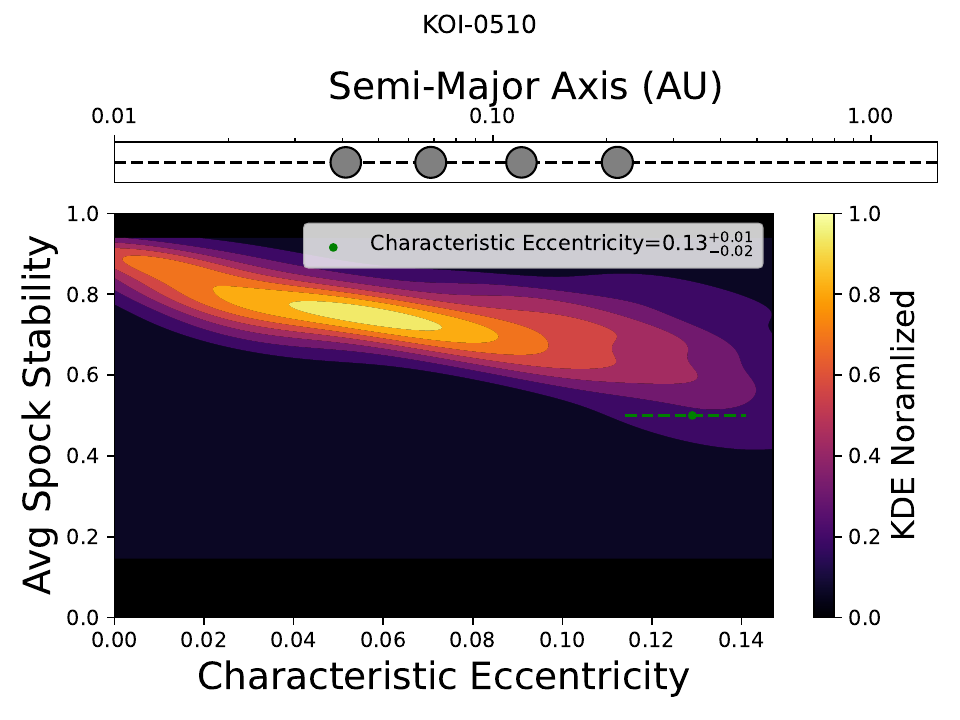}
    \end{subfigure}
    \begin{subfigure}{}
        \includegraphics[width=0.48\textwidth]{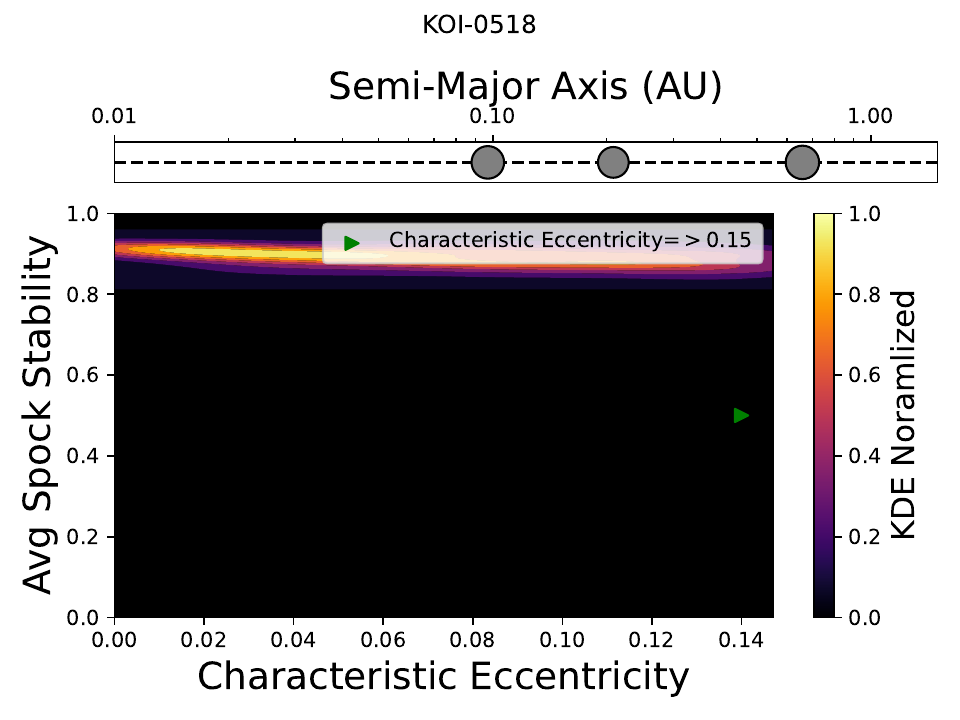}
    \end{subfigure}
    \begin{subfigure}{}
        \includegraphics[width=0.48\textwidth]{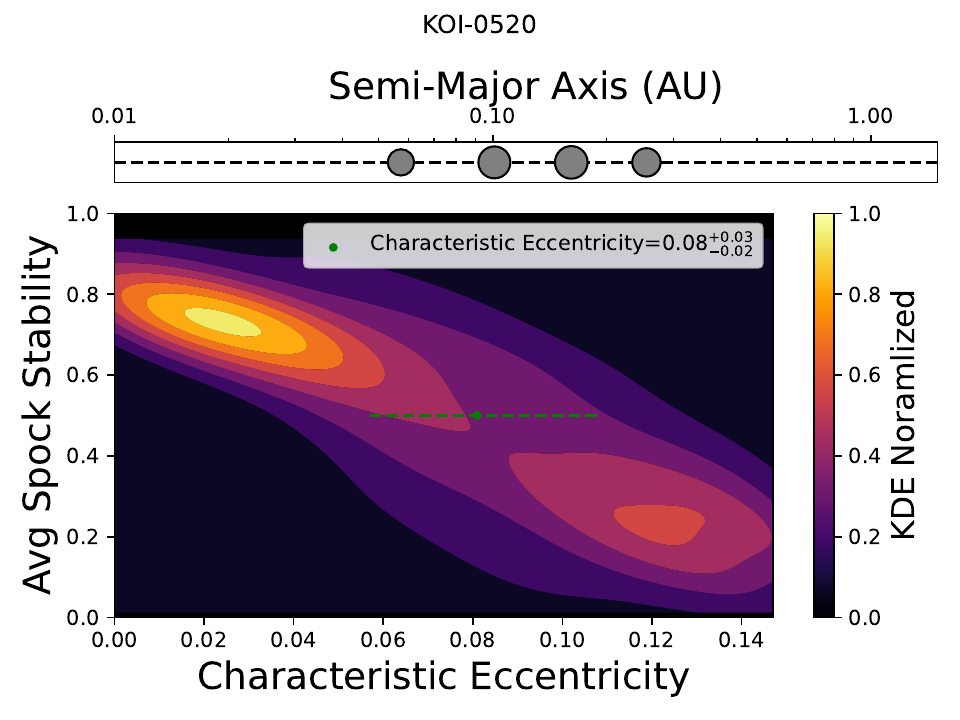}
    \end{subfigure}
    \begin{subfigure}{}
        \includegraphics[width=0.48\textwidth]{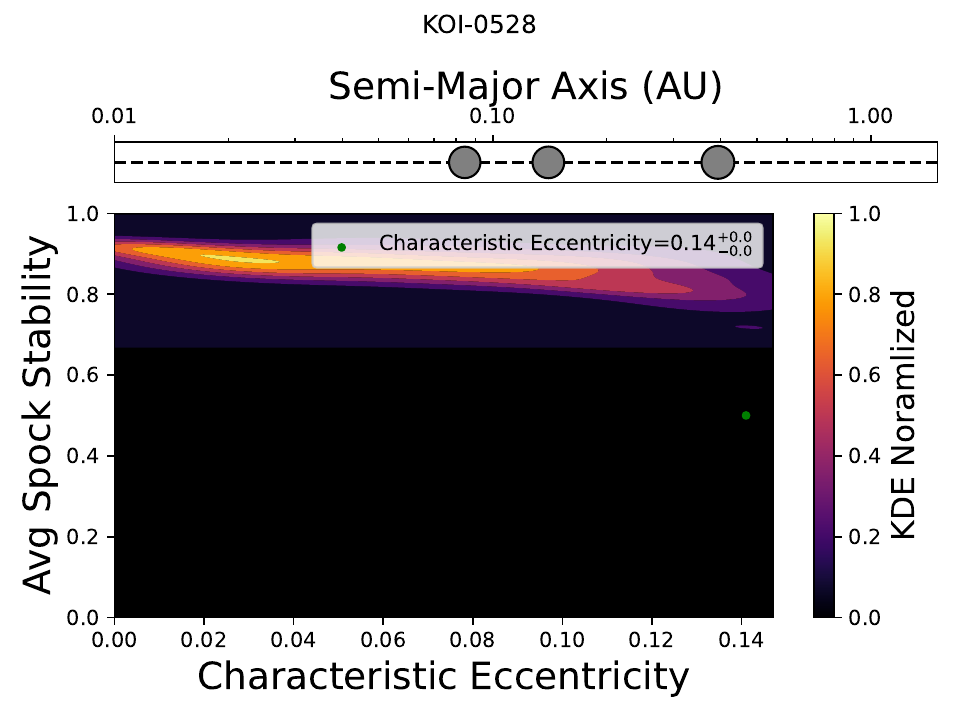}
    \end{subfigure}
    \begin{subfigure}{}
        \includegraphics[width=0.48\textwidth]{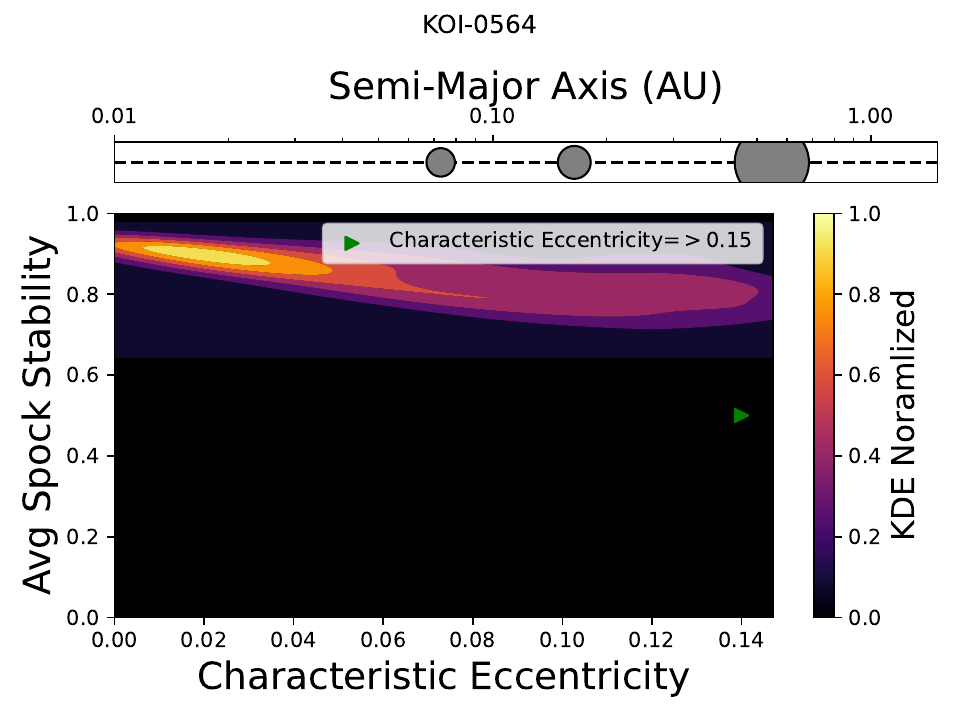}
    \end{subfigure}
    \begin{subfigure}{}
        \includegraphics[width=0.48\textwidth]{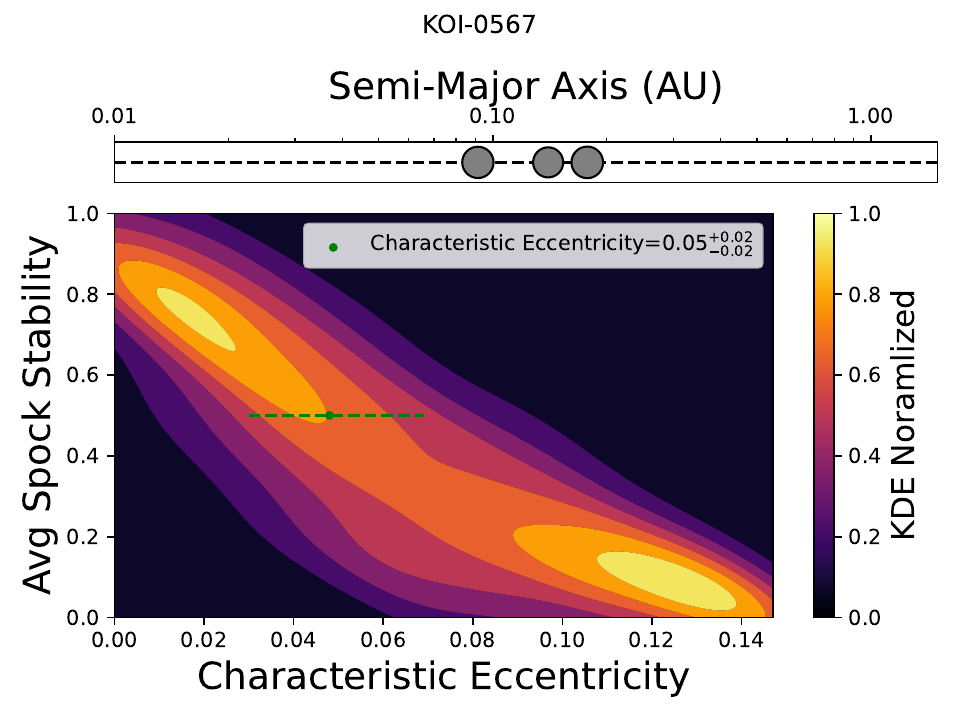}
    \end{subfigure}
\end{figure*}
\begin{figure*}
    \begin{subfigure}{}
        \includegraphics[width=0.48\textwidth]{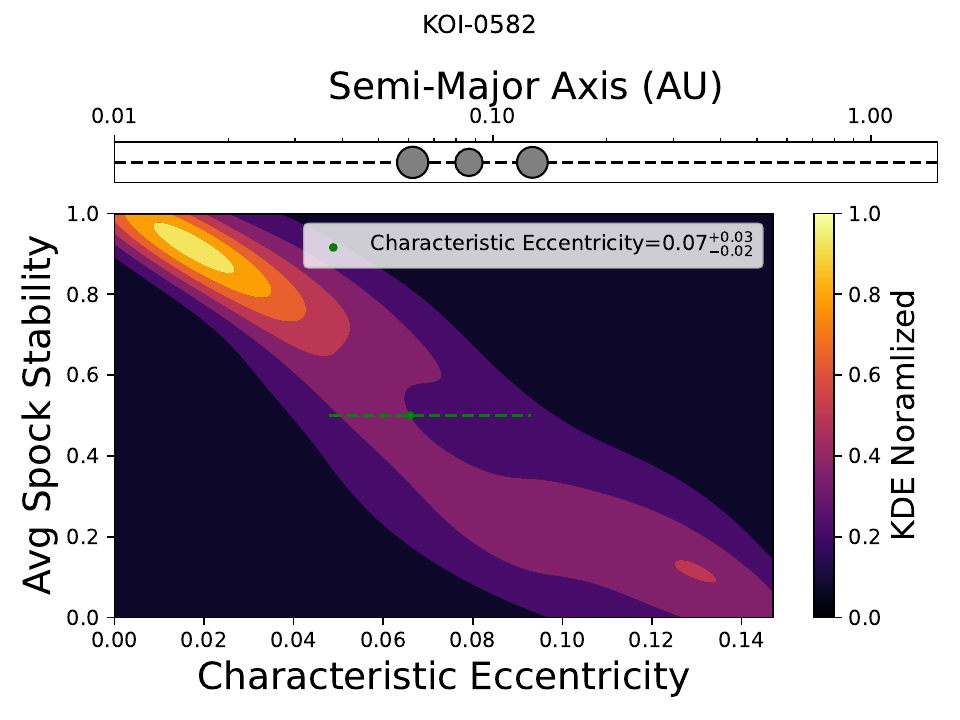}
    \end{subfigure}
    \begin{subfigure}{}
        \includegraphics[width=0.48\textwidth]{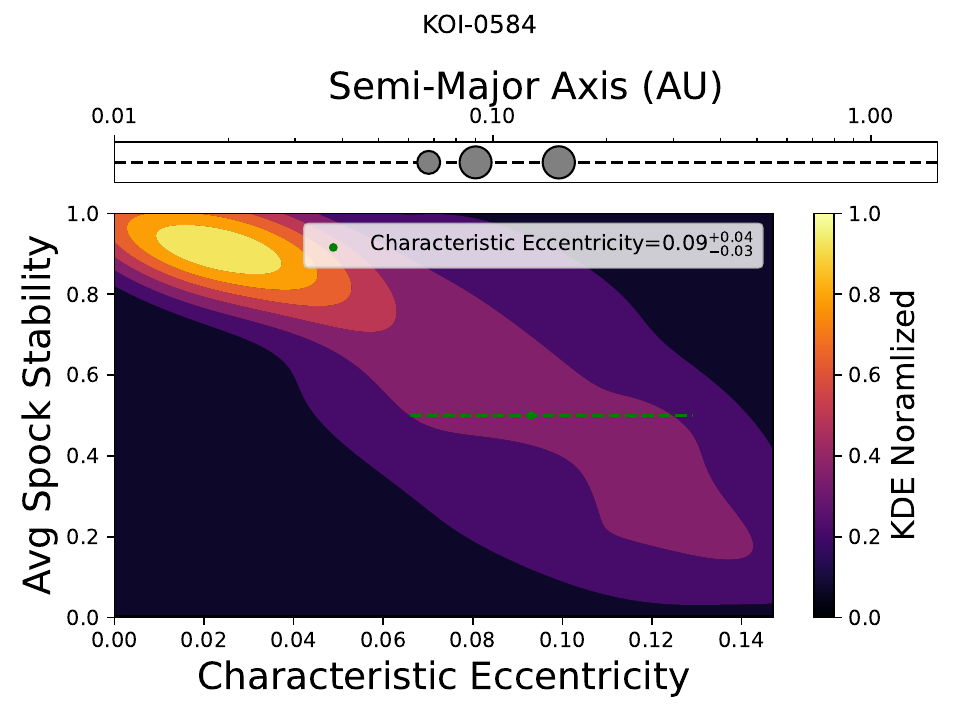}
    \end{subfigure}
    \begin{subfigure}{}
        \includegraphics[width=0.48\textwidth]{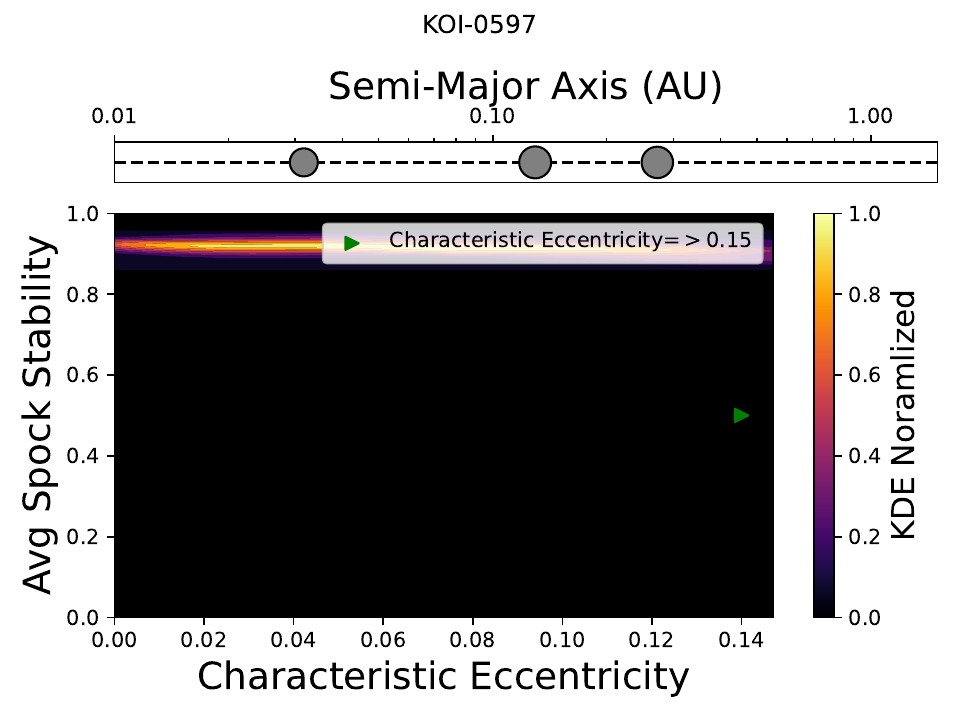}
    \end{subfigure}
    \begin{subfigure}{}
        \includegraphics[width=0.48\textwidth]{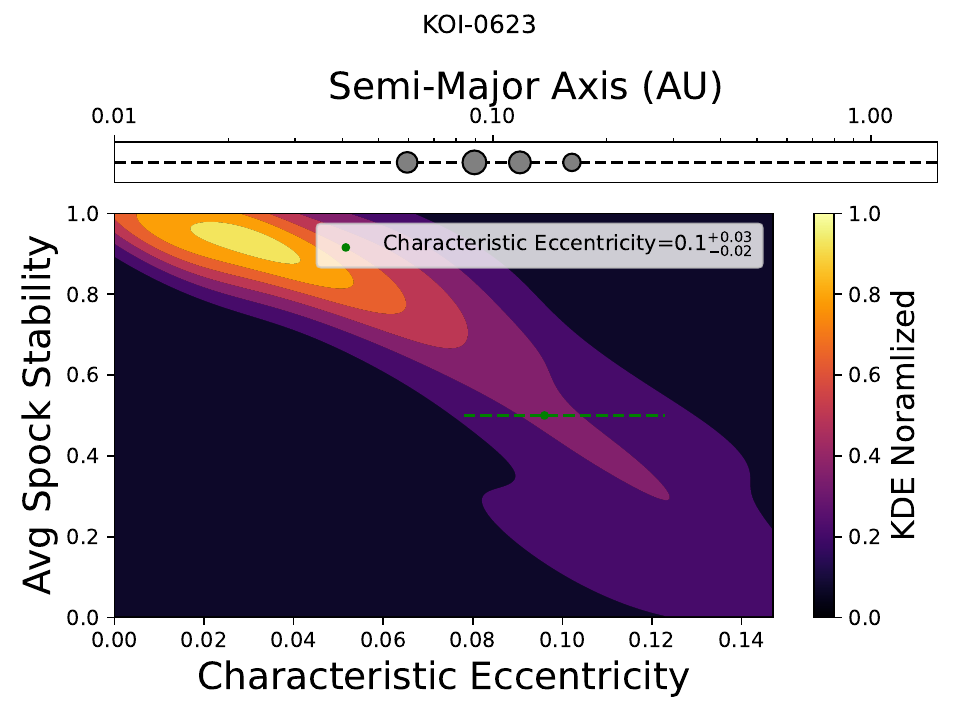}
    \end{subfigure}
    \begin{subfigure}{}
        \includegraphics[width=0.48\textwidth]{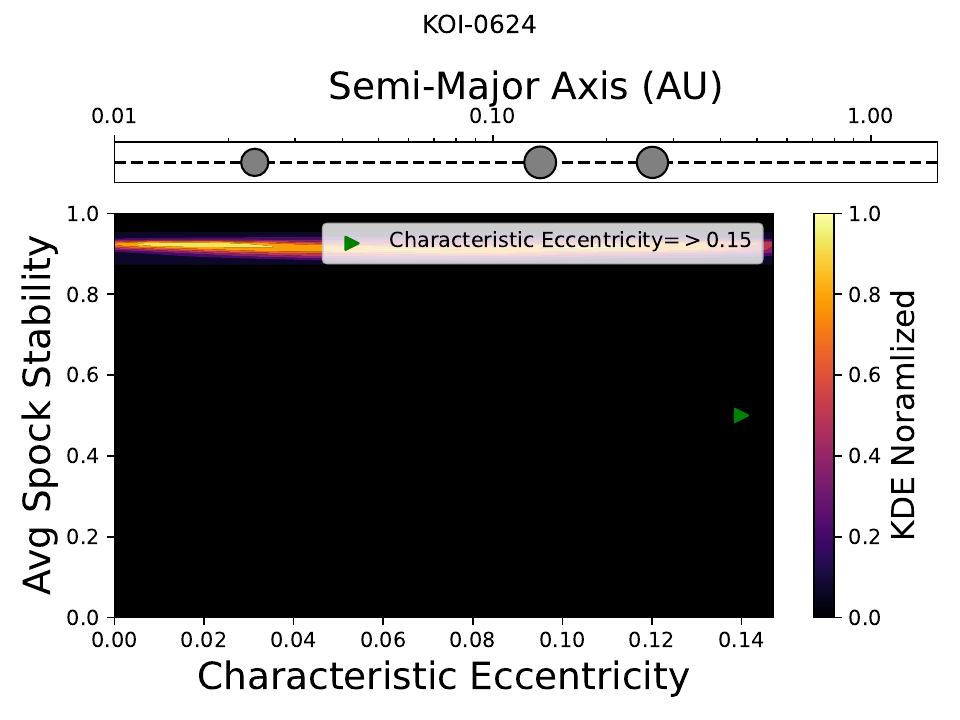}
    \end{subfigure}
    \begin{subfigure}{}
        \includegraphics[width=0.48\textwidth]{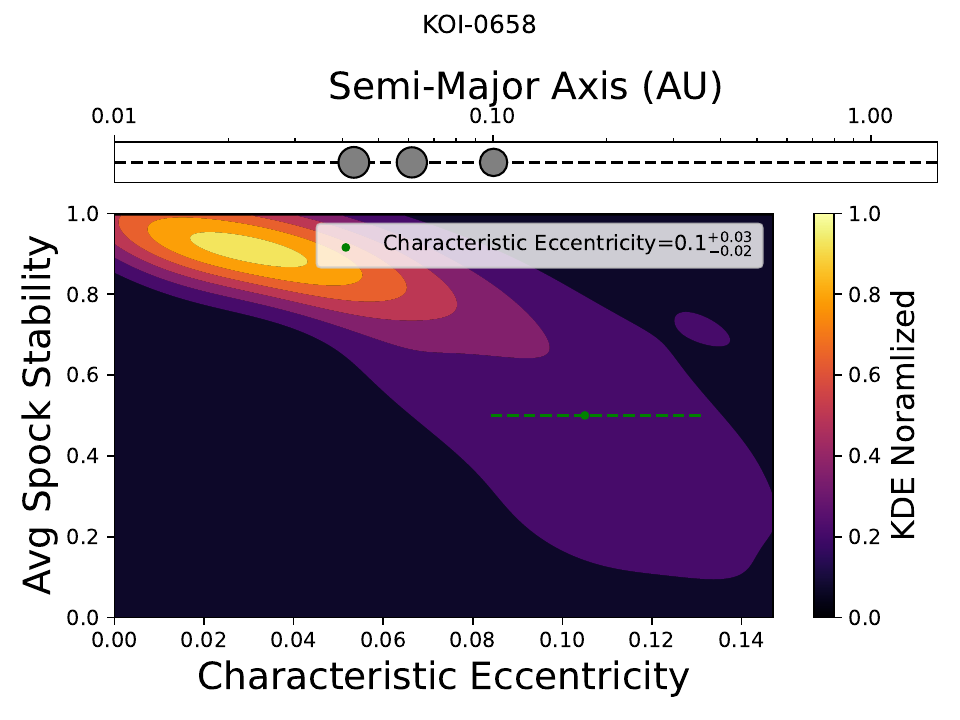}
    \end{subfigure}
\end{figure*}
\begin{figure*}
    \begin{subfigure}{}
        \includegraphics[width=0.48\textwidth]{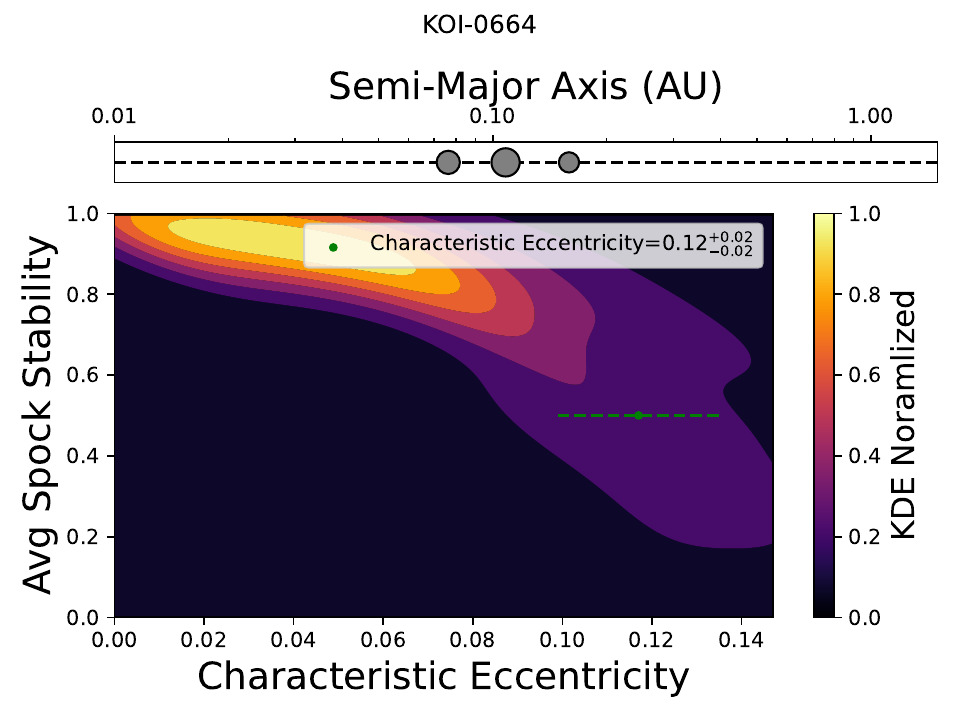}
    \end{subfigure}
    \begin{subfigure}{}
        \includegraphics[width=0.48\textwidth]{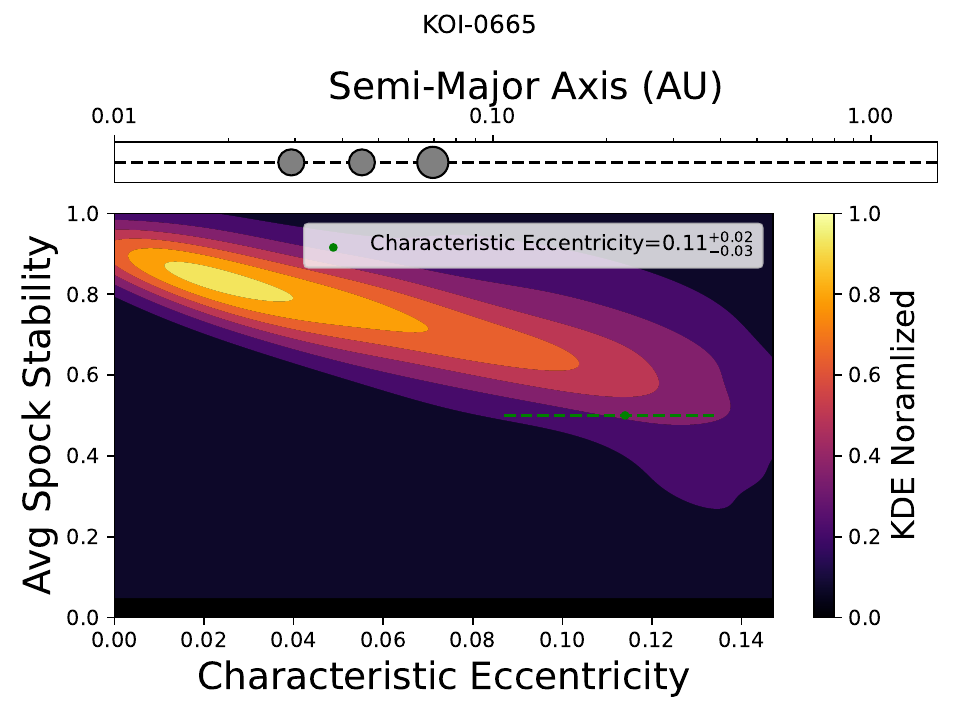}
    \end{subfigure}
    \begin{subfigure}{}
        \includegraphics[width=0.48\textwidth]{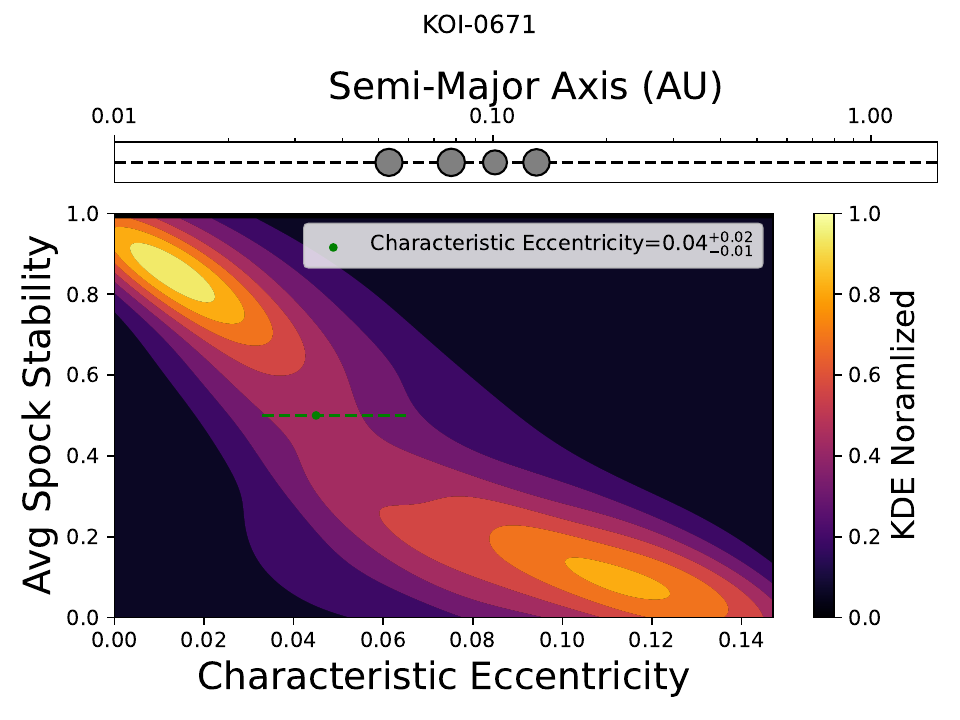}
    \end{subfigure}
    \begin{subfigure}{}
        \includegraphics[width=0.48\textwidth]{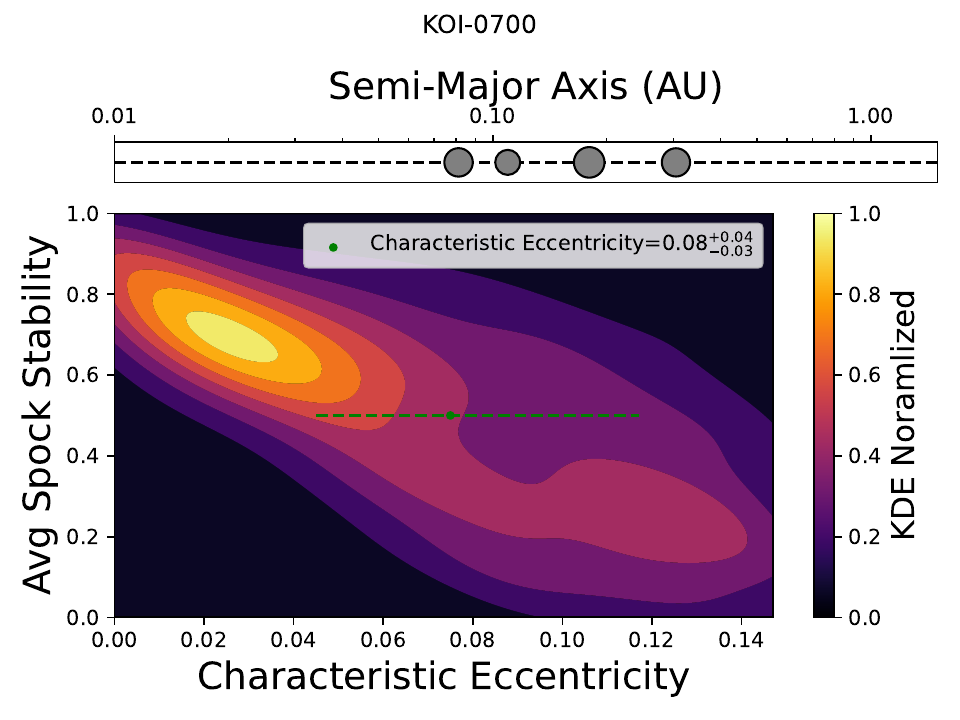}
    \end{subfigure}
    \begin{subfigure}{}
        \includegraphics[width=0.48\textwidth]{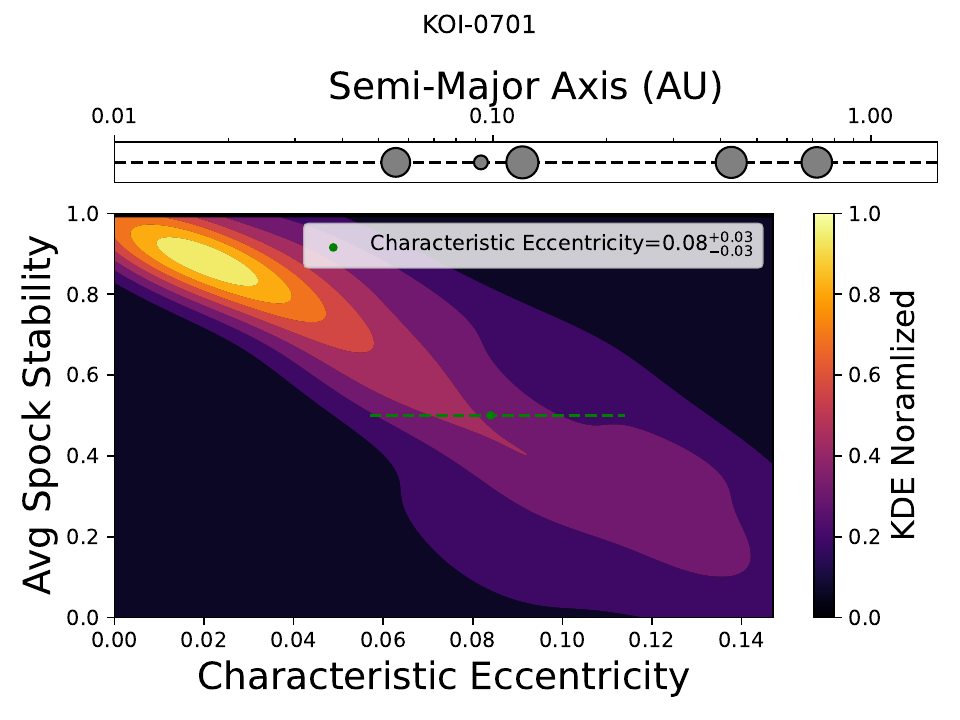}
    \end{subfigure}
    \begin{subfigure}{}
        \includegraphics[width=0.48\textwidth]{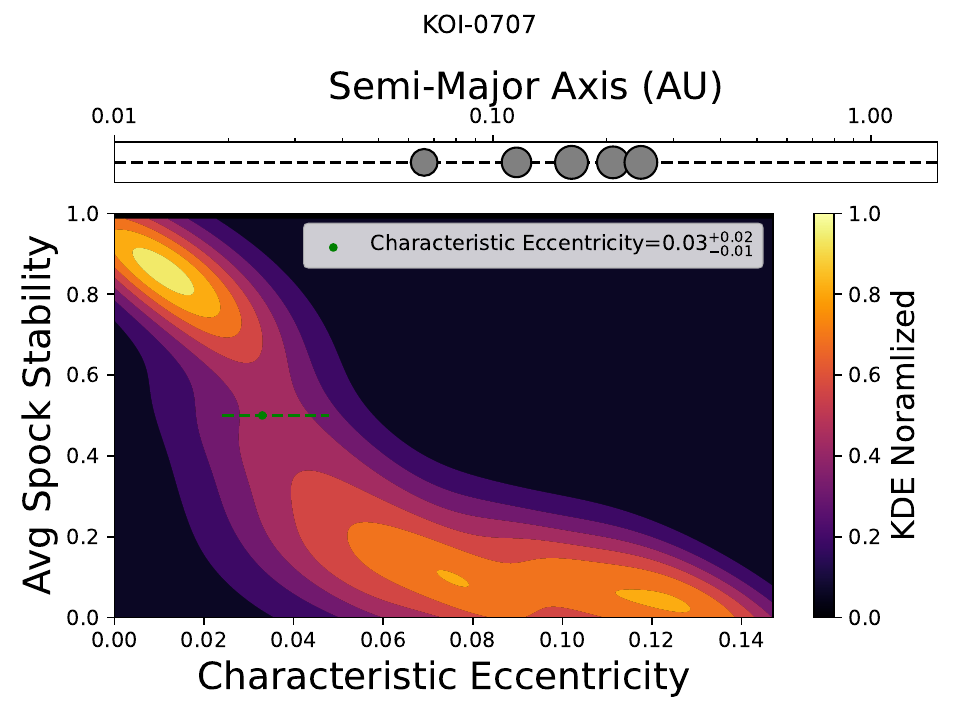}
    \end{subfigure}
\end{figure*}
\begin{figure*}
    \begin{subfigure}{}
        \includegraphics[width=0.48\textwidth]{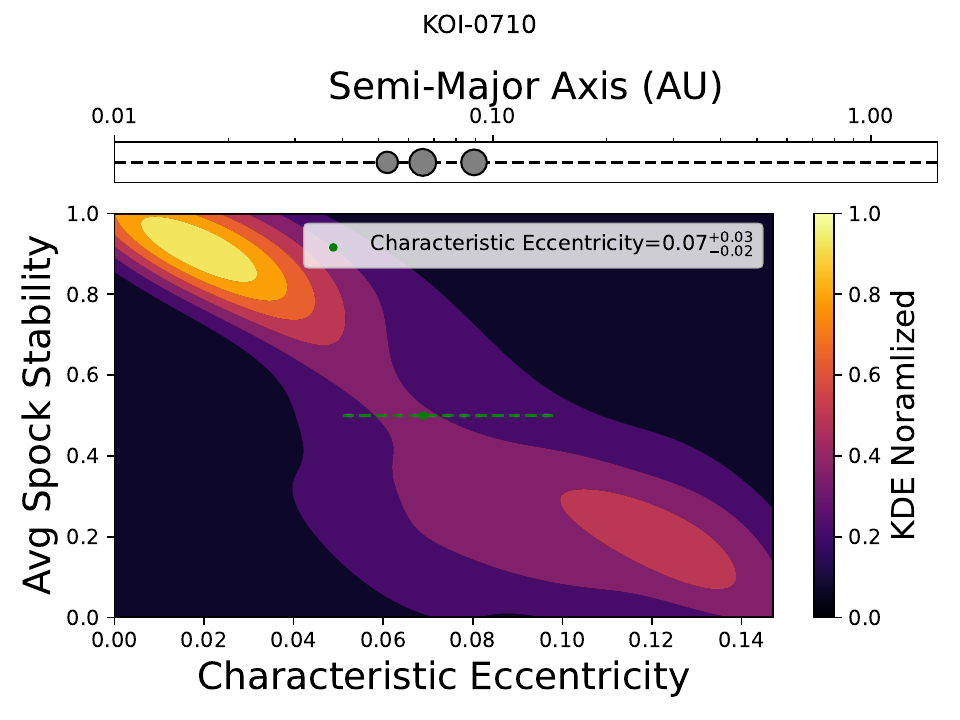}
    \end{subfigure}
    \begin{subfigure}{}
        \includegraphics[width=0.48\textwidth]{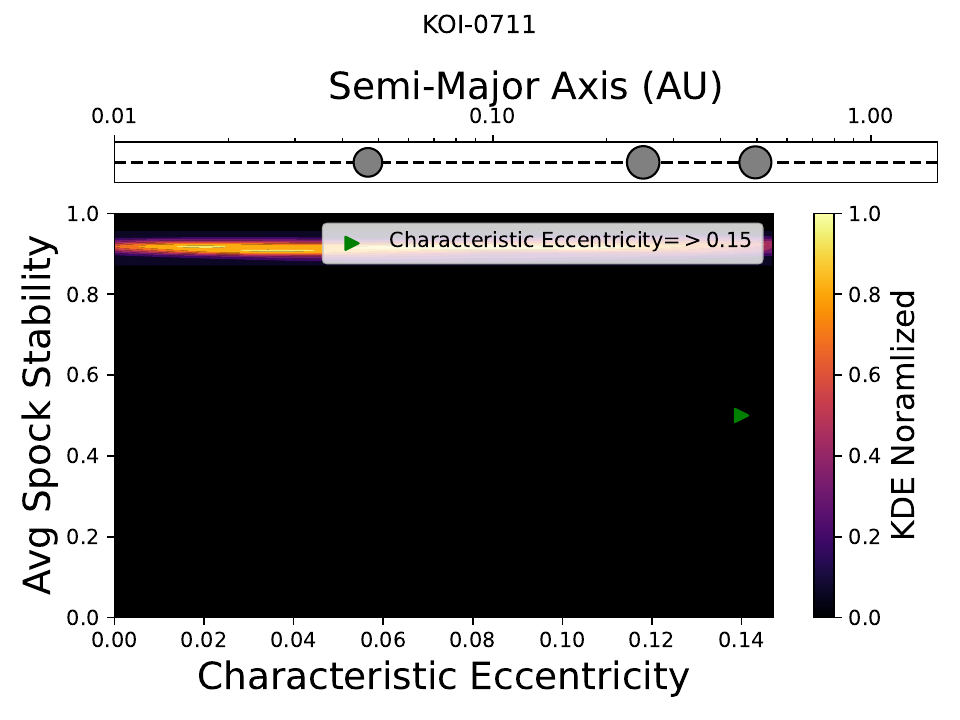}
    \end{subfigure}
    \begin{subfigure}{}
        \includegraphics[width=0.48\textwidth]{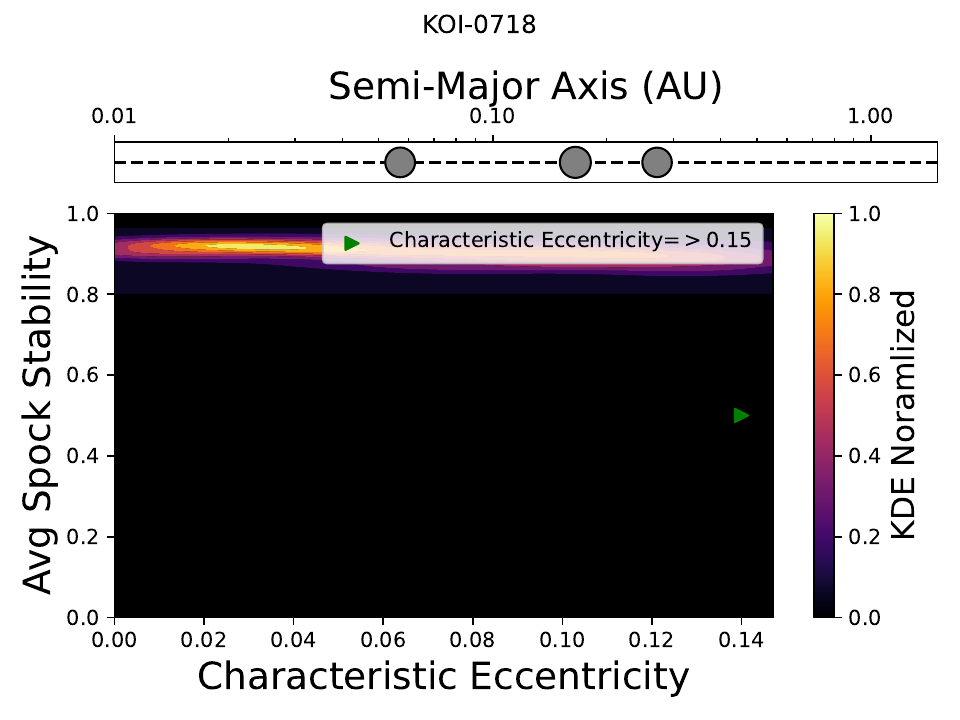}
    \end{subfigure}
    \begin{subfigure}{}
        \includegraphics[width=0.48\textwidth]{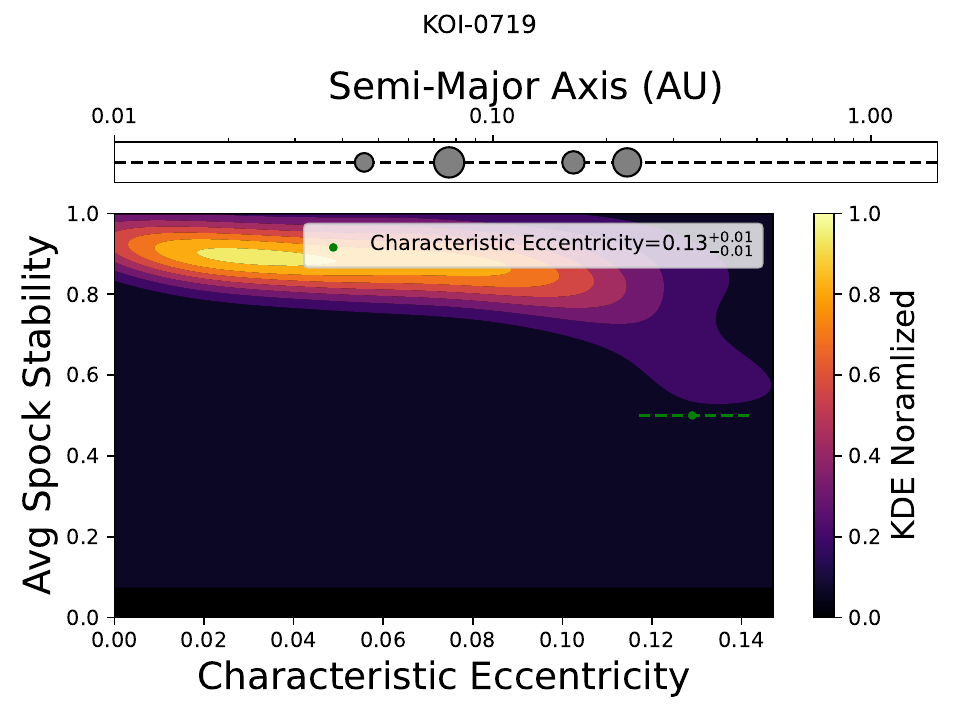}
    \end{subfigure}
    \begin{subfigure}{}
        \includegraphics[width=0.48\textwidth]{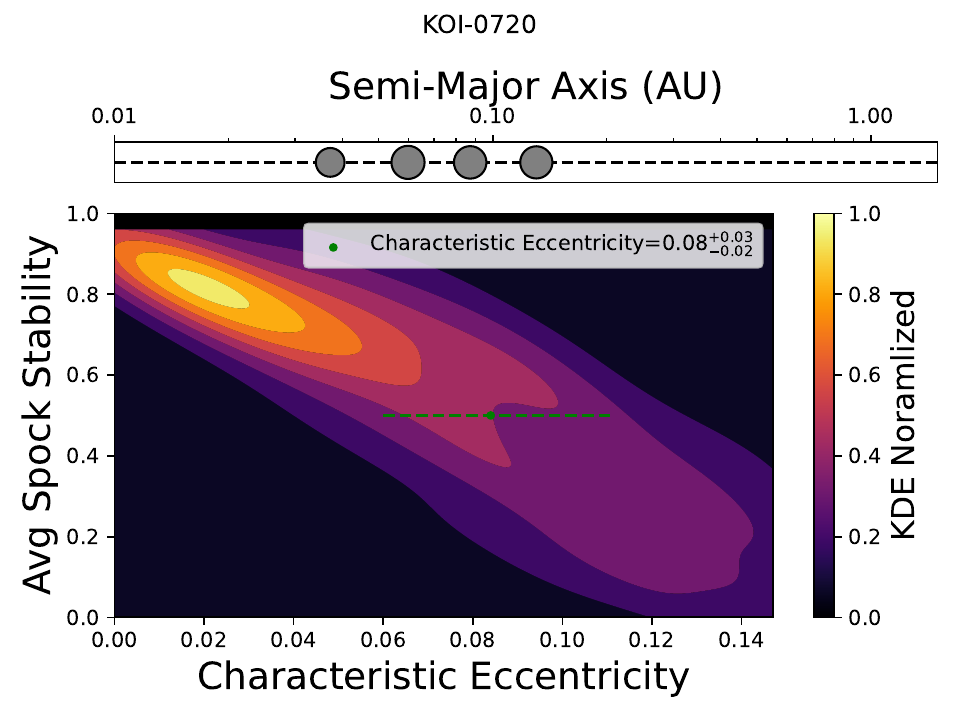}
    \end{subfigure}
    \begin{subfigure}{}
        \includegraphics[width=0.48\textwidth]{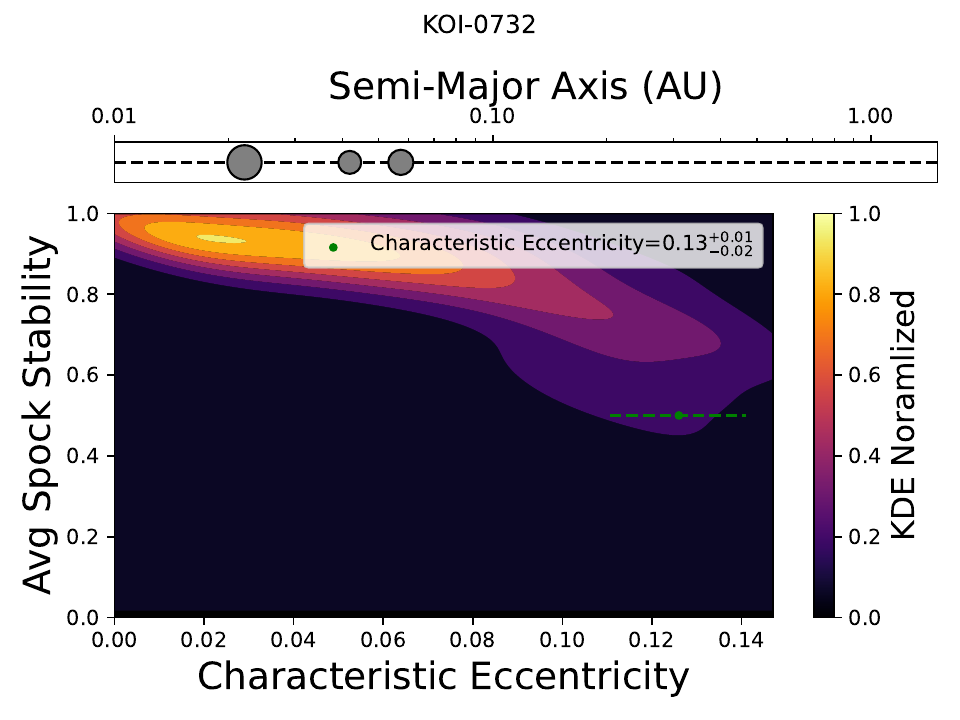}
    \end{subfigure}
\end{figure*}
\begin{figure*}
    \begin{subfigure}{}
        \includegraphics[width=0.48\textwidth]{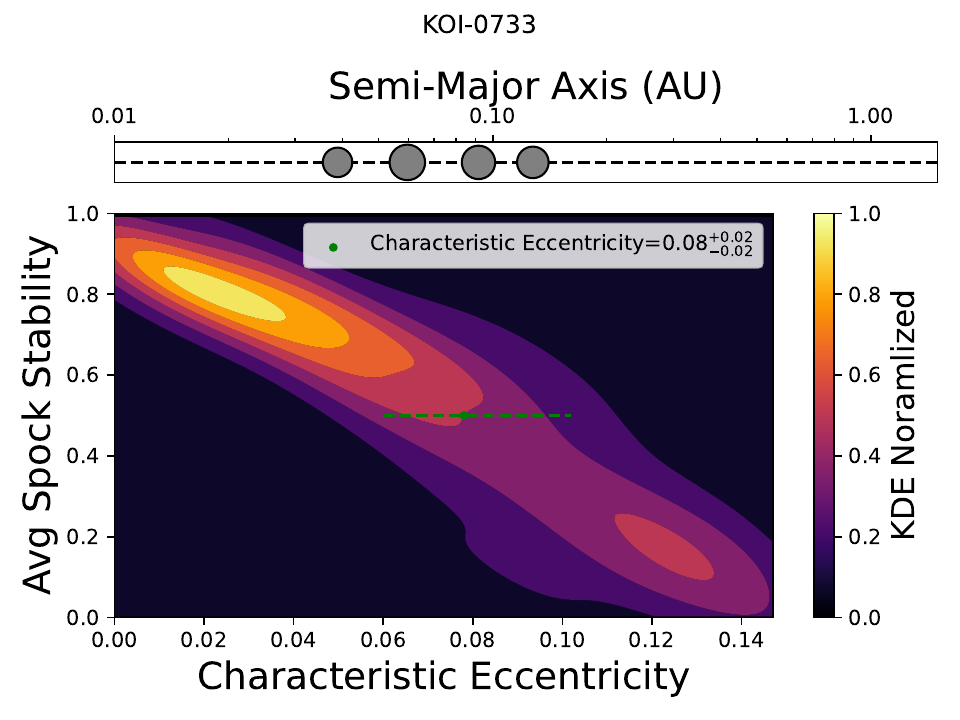}
    \end{subfigure}
    \begin{subfigure}{}
        \includegraphics[width=0.48\textwidth]{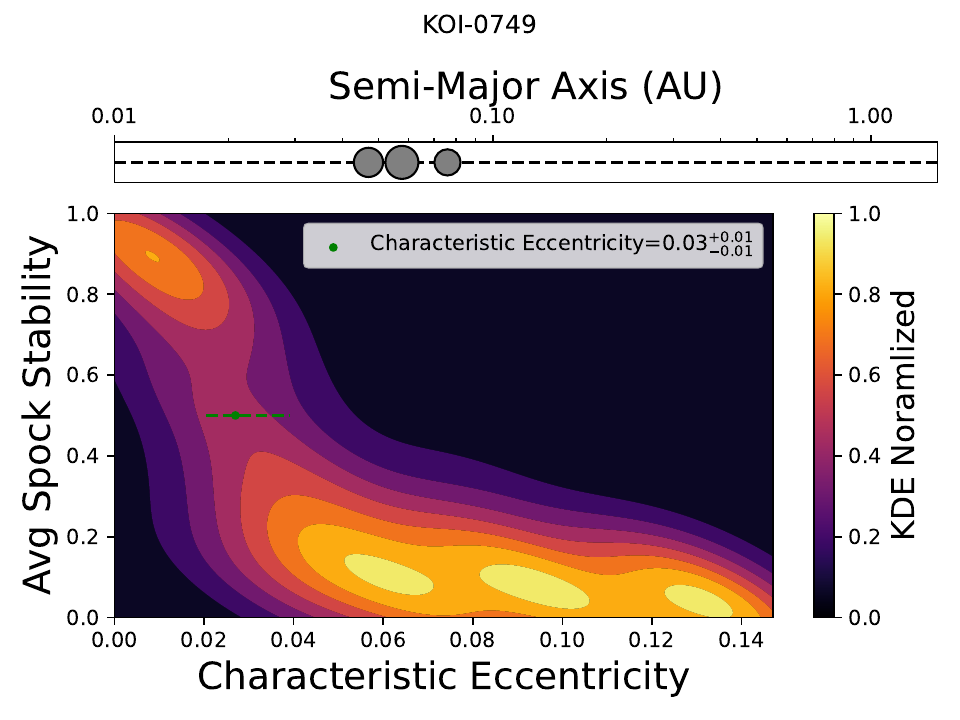}
    \end{subfigure}
    \begin{subfigure}{}
        \includegraphics[width=0.48\textwidth]{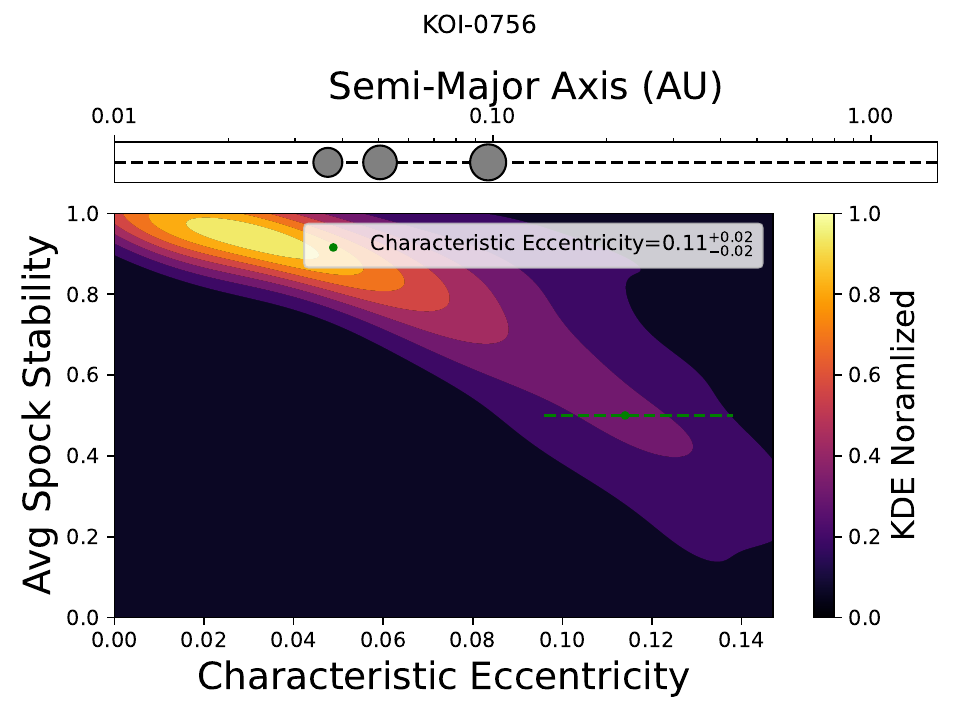}
    \end{subfigure}
    \begin{subfigure}{}
        \includegraphics[width=0.48\textwidth]{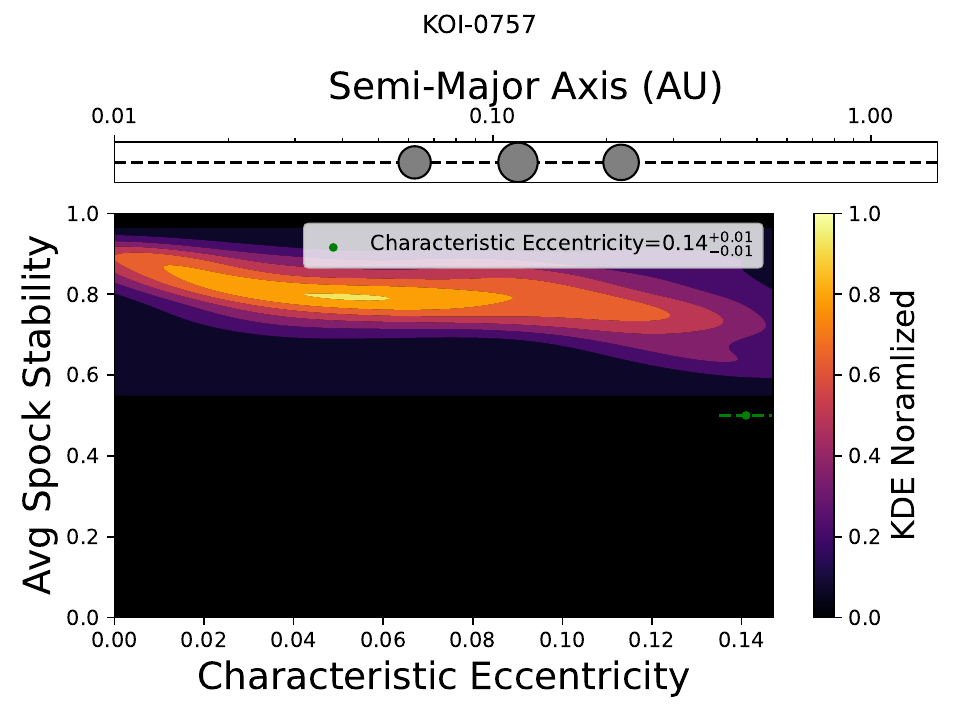}
    \end{subfigure}
    \begin{subfigure}{}
        \includegraphics[width=0.48\textwidth]{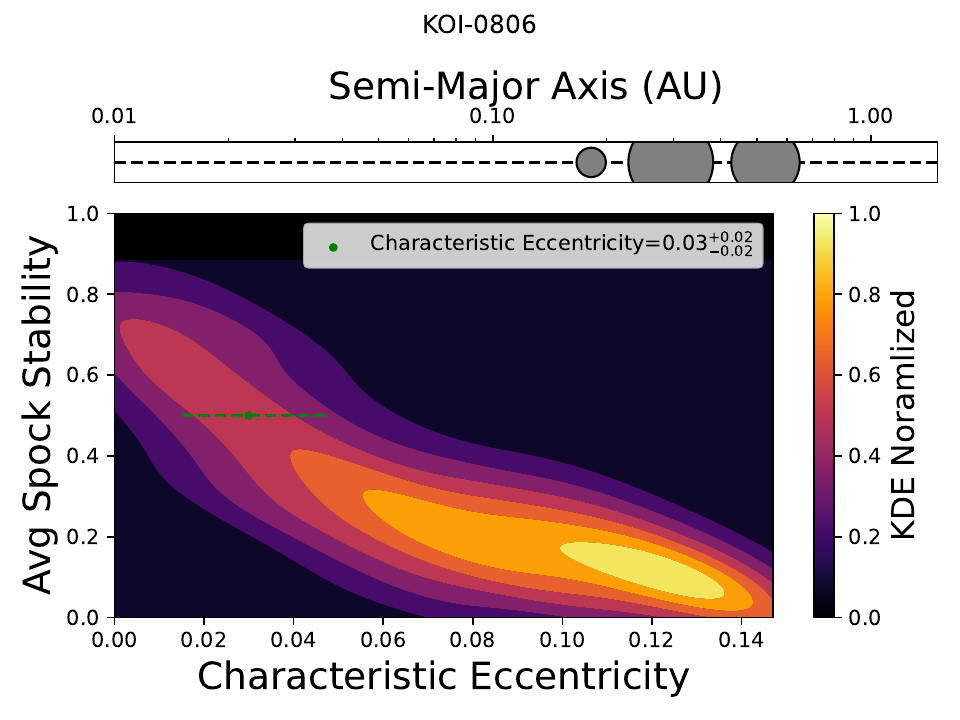}
    \end{subfigure}
    \begin{subfigure}{}
        \includegraphics[width=0.48\textwidth]{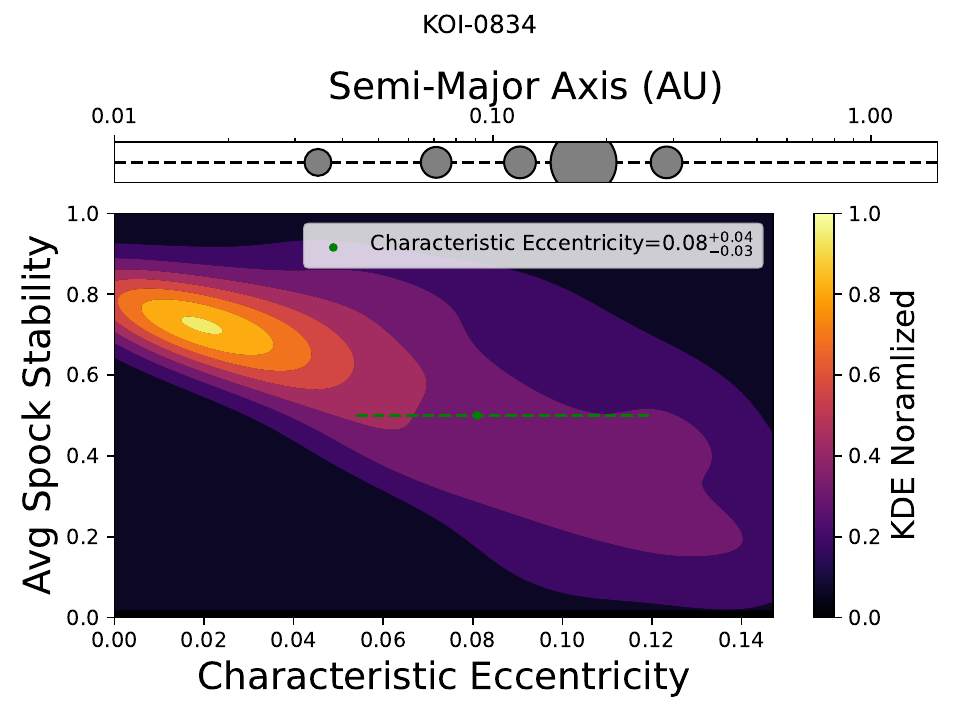}
    \end{subfigure}
\end{figure*}
\begin{figure*}
    \begin{subfigure}{}
        \includegraphics[width=0.48\textwidth]{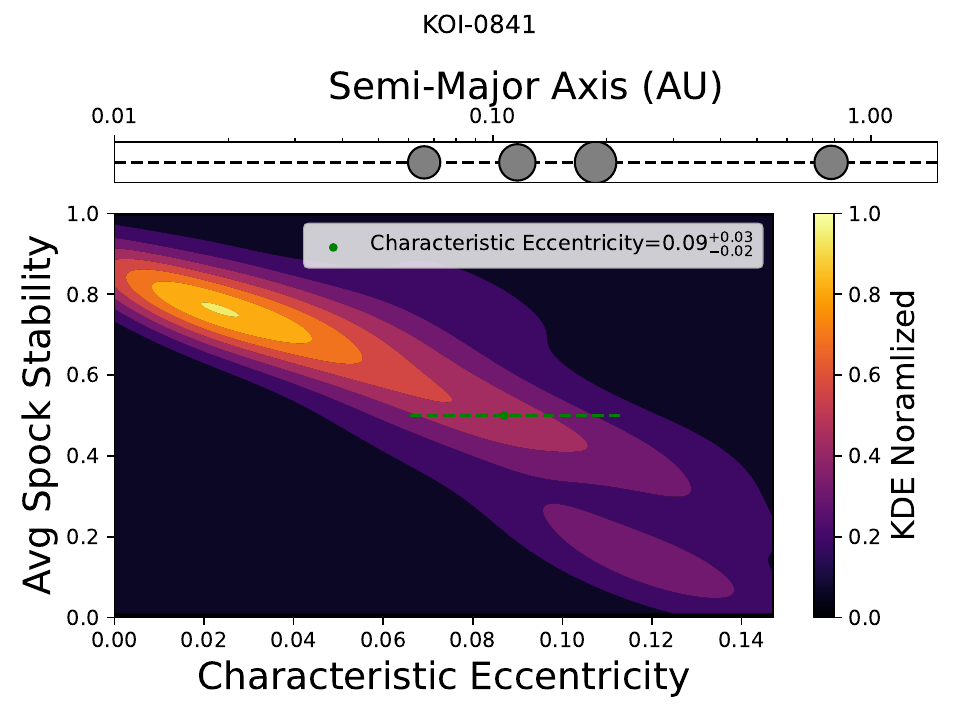}
    \end{subfigure}
    \begin{subfigure}{}
        \includegraphics[width=0.48\textwidth]{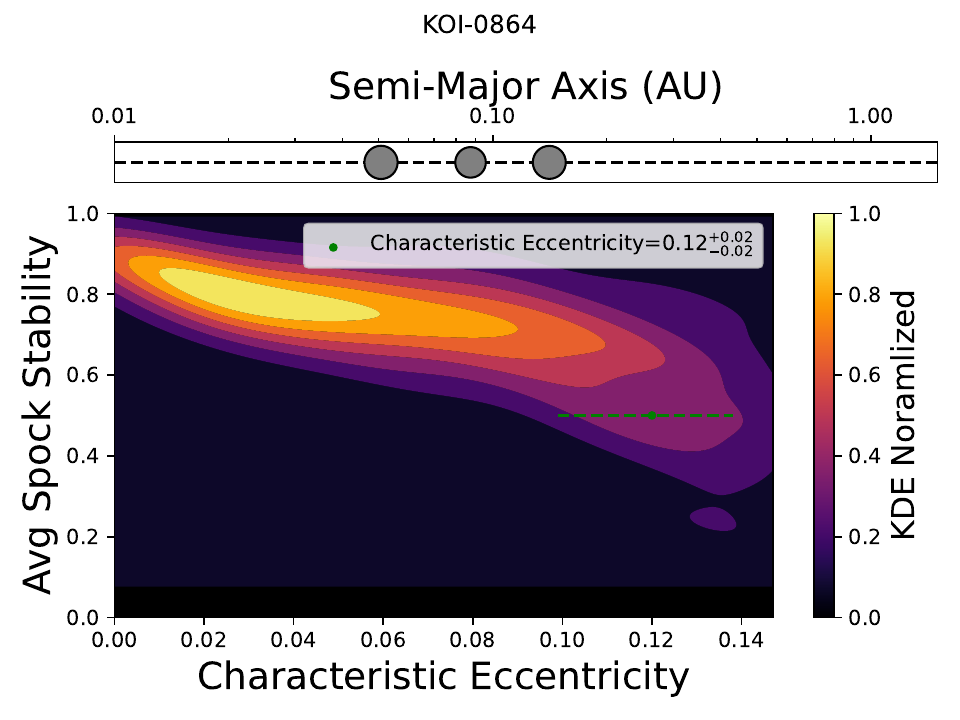}
    \end{subfigure}
    \begin{subfigure}{}
        \includegraphics[width=0.48\textwidth]{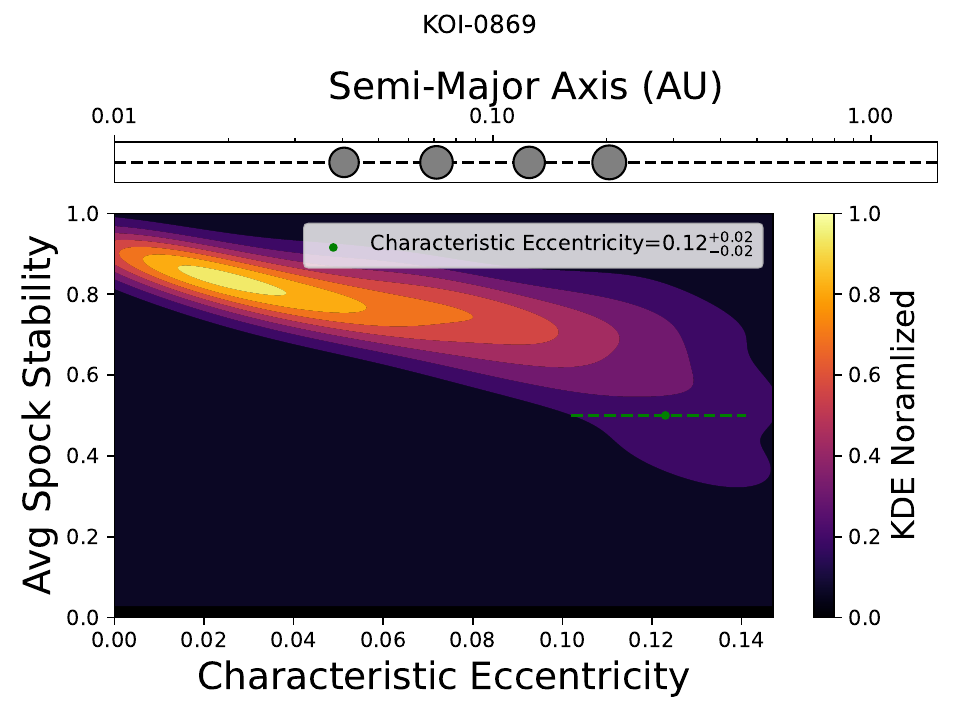}
    \end{subfigure}
    \begin{subfigure}{}
        \includegraphics[width=0.48\textwidth]{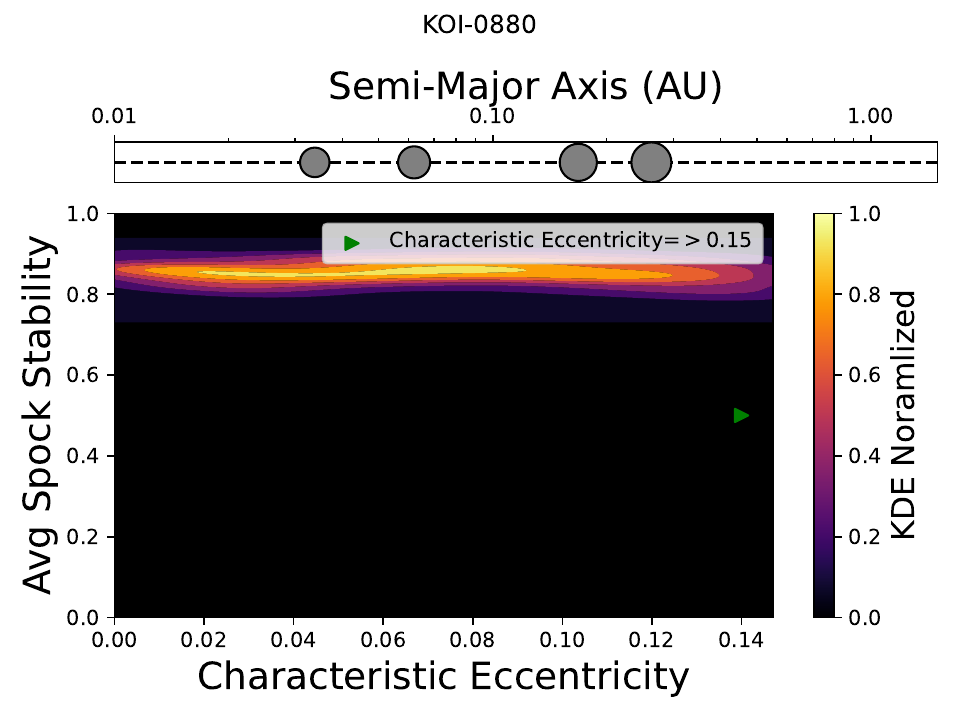}
    \end{subfigure}
    \begin{subfigure}{}
        \includegraphics[width=0.48\textwidth]{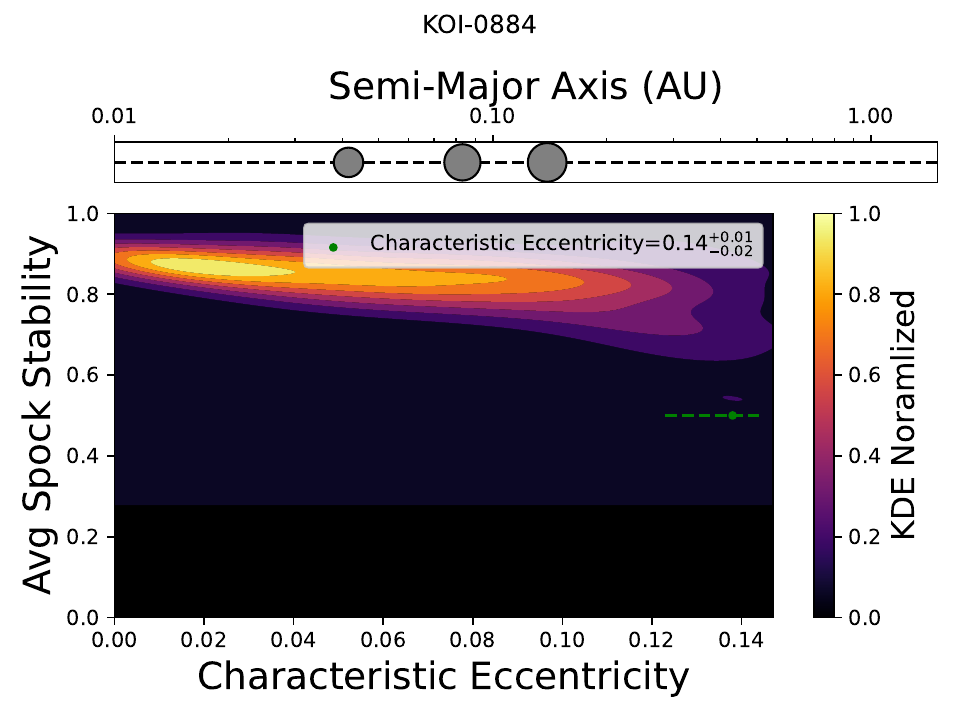}
    \end{subfigure}
    \begin{subfigure}{}
        \includegraphics[width=0.48\textwidth]{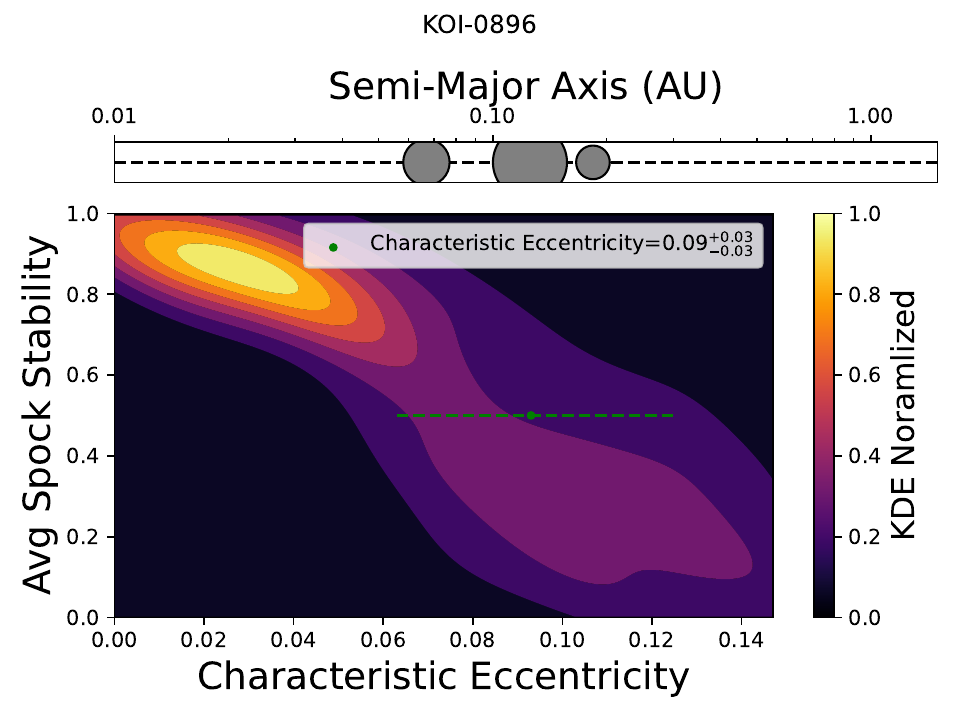}
    \end{subfigure}
\end{figure*}
\begin{figure*}
    \begin{subfigure}{}
        \includegraphics[width=0.48\textwidth]{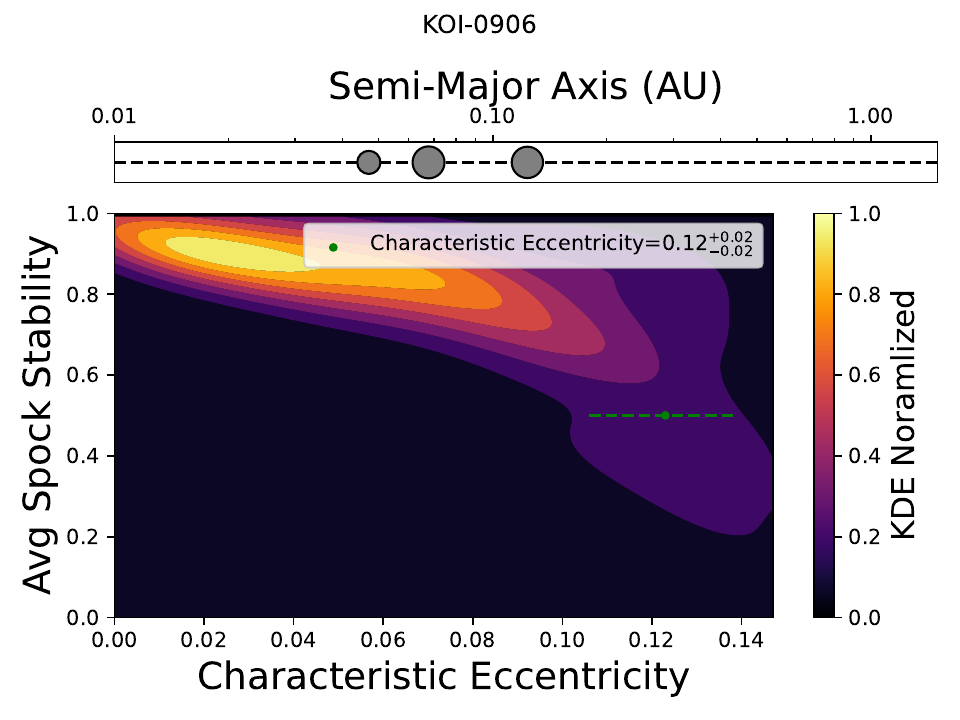}
    \end{subfigure}
    \begin{subfigure}{}
        \includegraphics[width=0.48\textwidth]{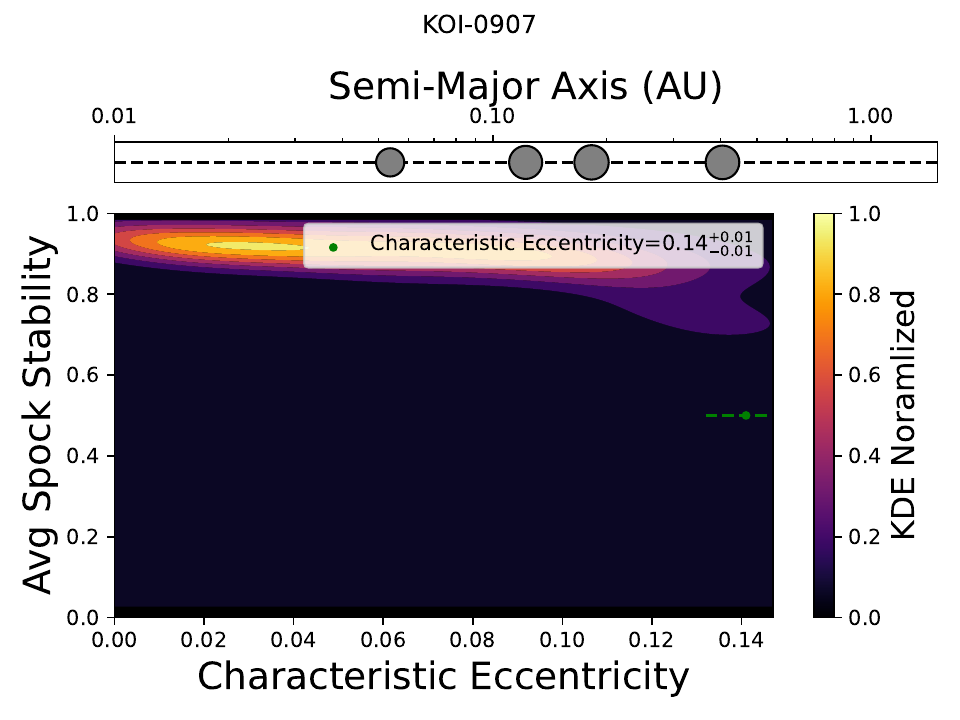}
    \end{subfigure}
    \begin{subfigure}{}
        \includegraphics[width=0.48\textwidth]{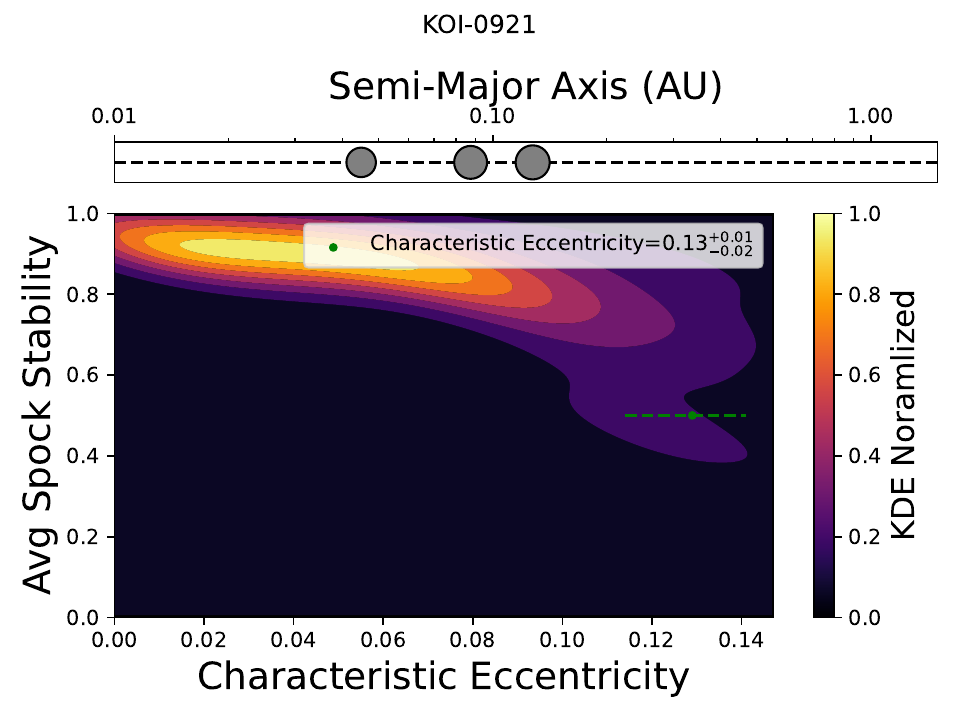}
    \end{subfigure}
    \begin{subfigure}{}
        \includegraphics[width=0.48\textwidth]{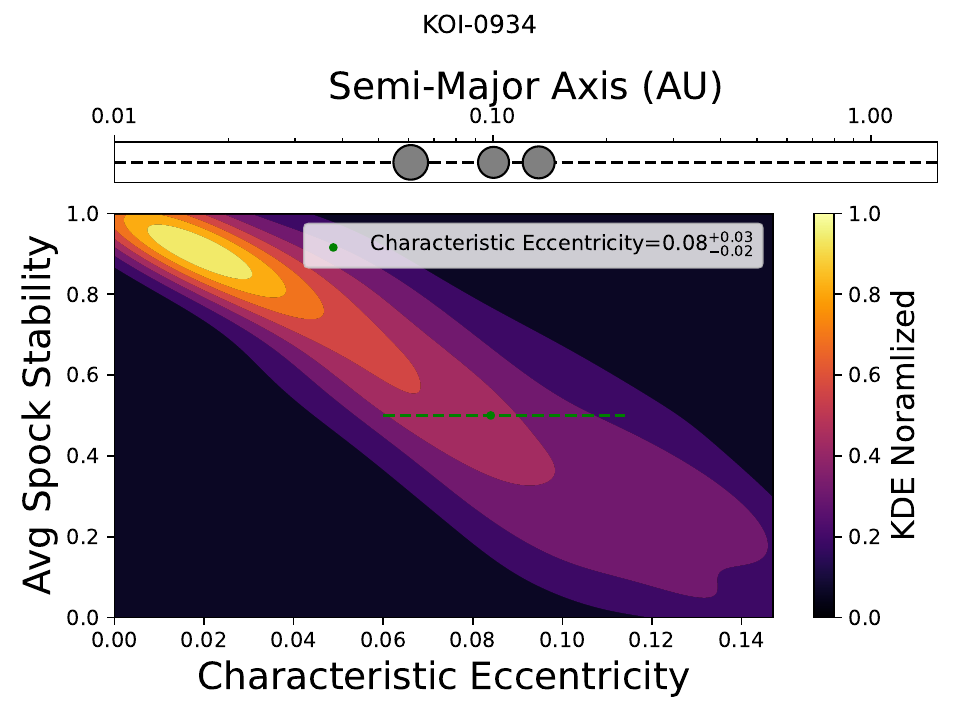}
    \end{subfigure}
    \begin{subfigure}{}
        \includegraphics[width=0.48\textwidth]{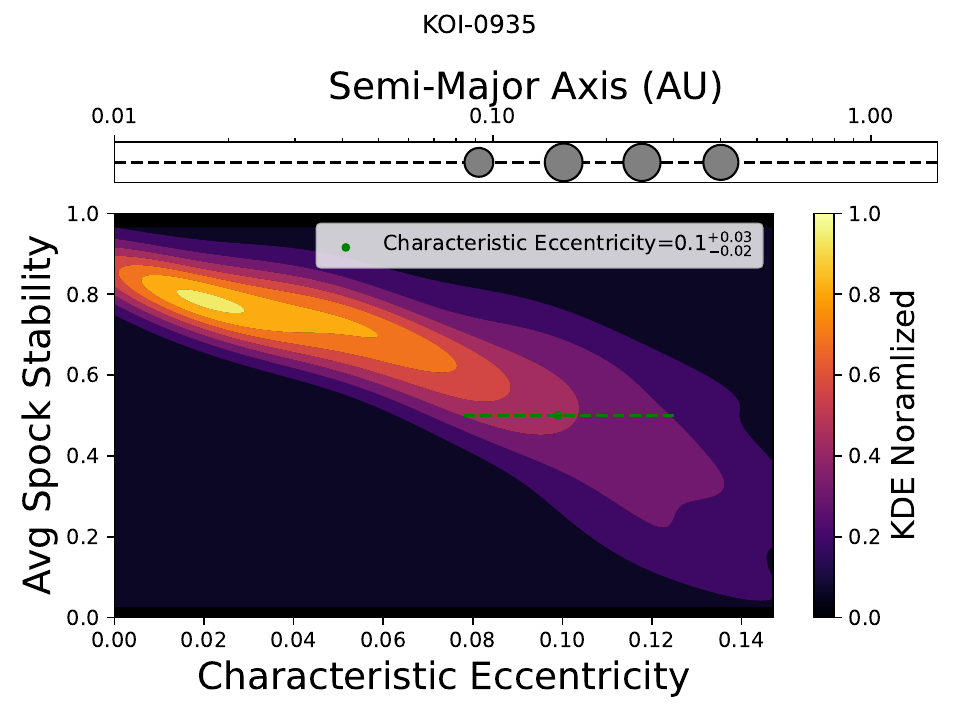}
    \end{subfigure}
    \begin{subfigure}{}
        \includegraphics[width=0.48\textwidth]{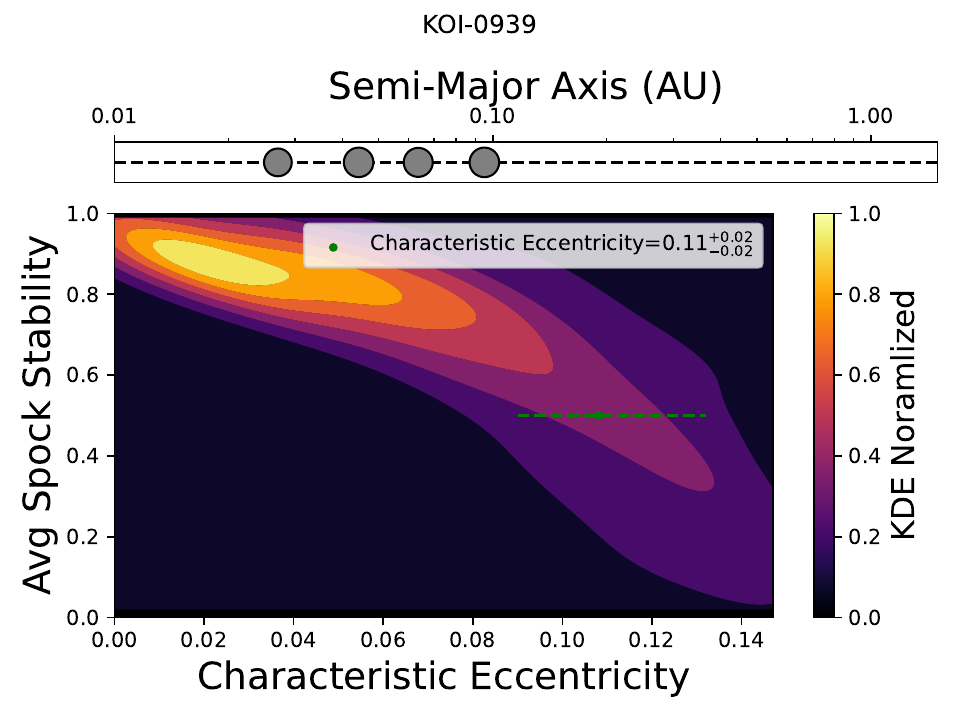}
    \end{subfigure}
\end{figure*}
\begin{figure*}
    \begin{subfigure}{}
        \includegraphics[width=0.48\textwidth]{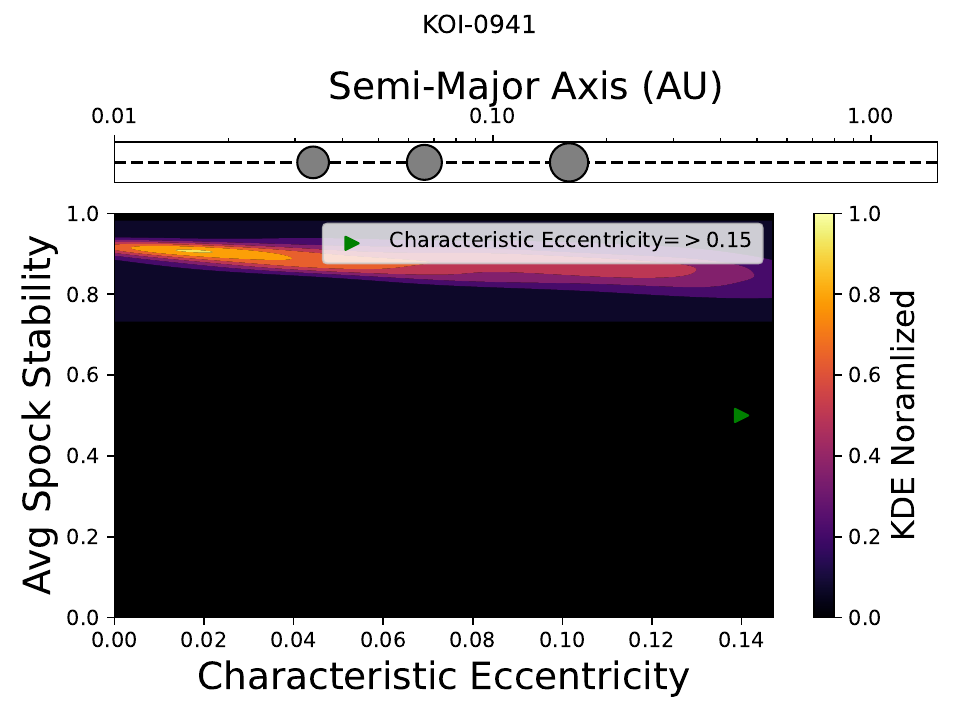}
    \end{subfigure}
    \begin{subfigure}{}
        \includegraphics[width=0.48\textwidth]{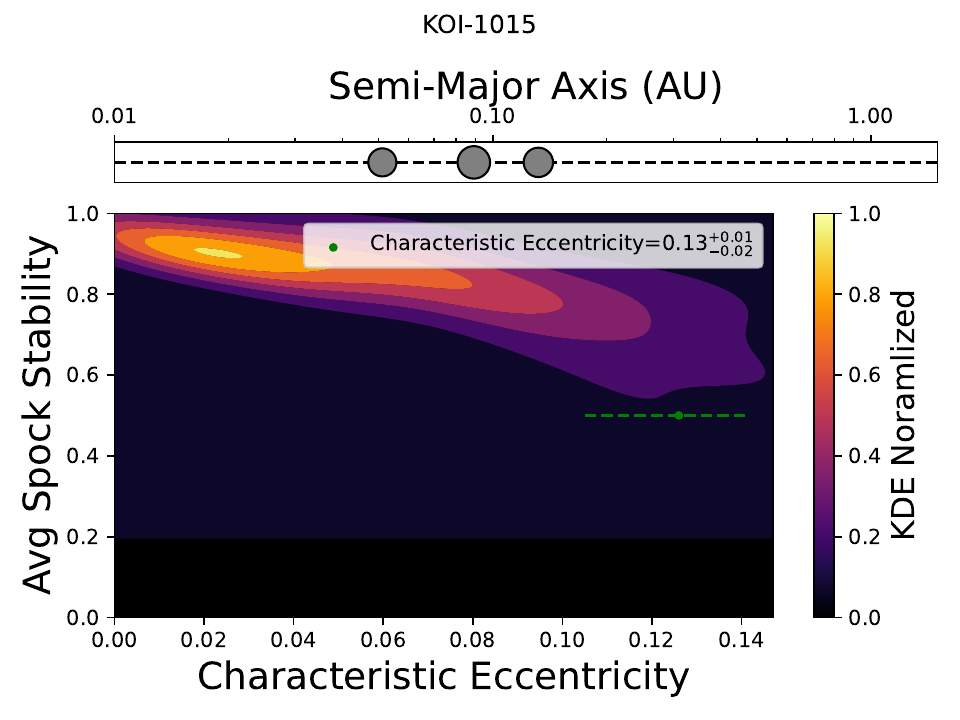}
    \end{subfigure}
    \begin{subfigure}{}
        \includegraphics[width=0.48\textwidth]{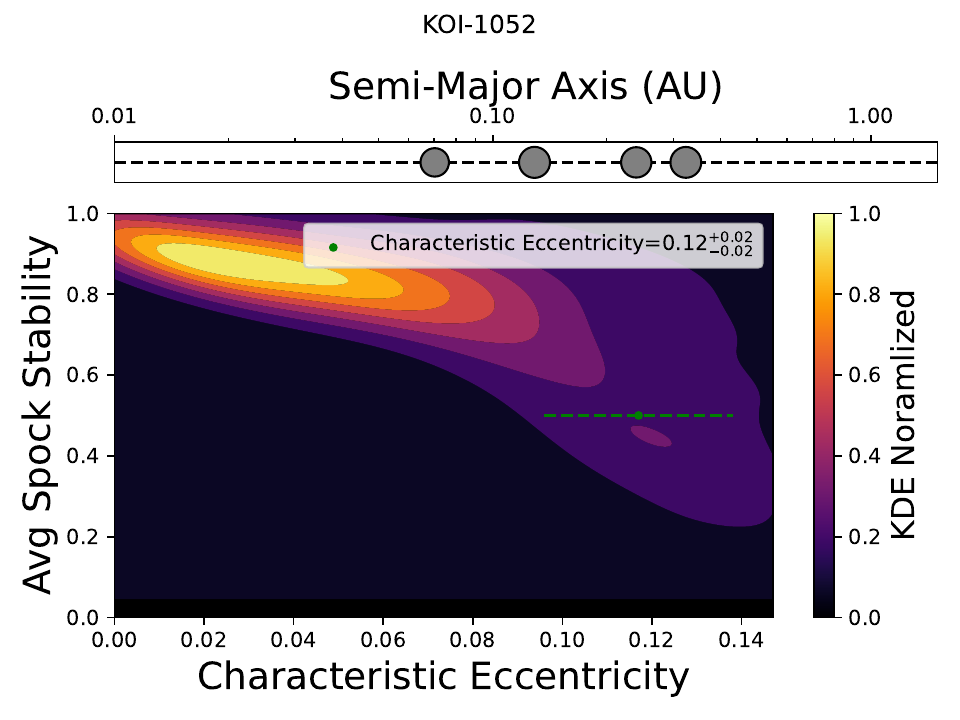}
    \end{subfigure}
    \begin{subfigure}{}
        \includegraphics[width=0.48\textwidth]{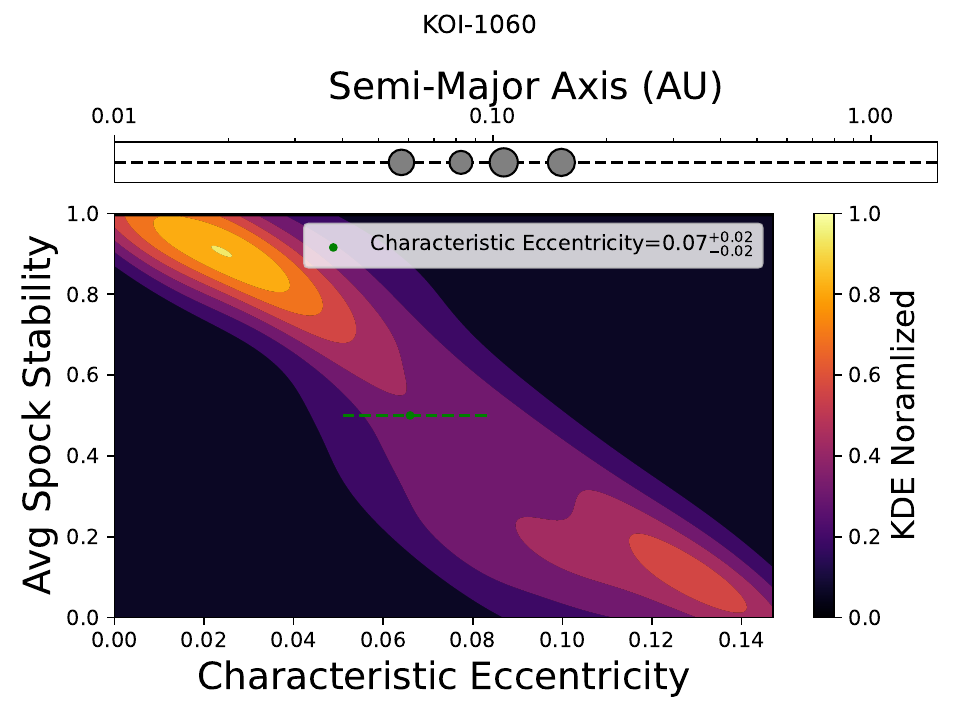}
    \end{subfigure}
    \begin{subfigure}{}
        \includegraphics[width=0.48\textwidth]{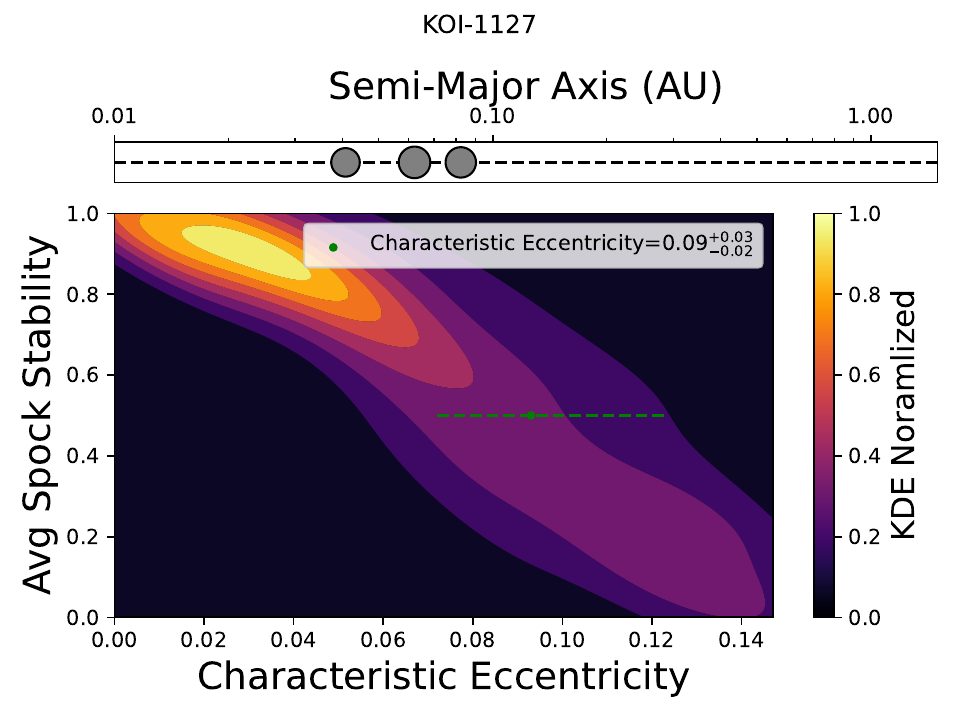}
    \end{subfigure}
    \begin{subfigure}{}
        \includegraphics[width=0.48\textwidth]{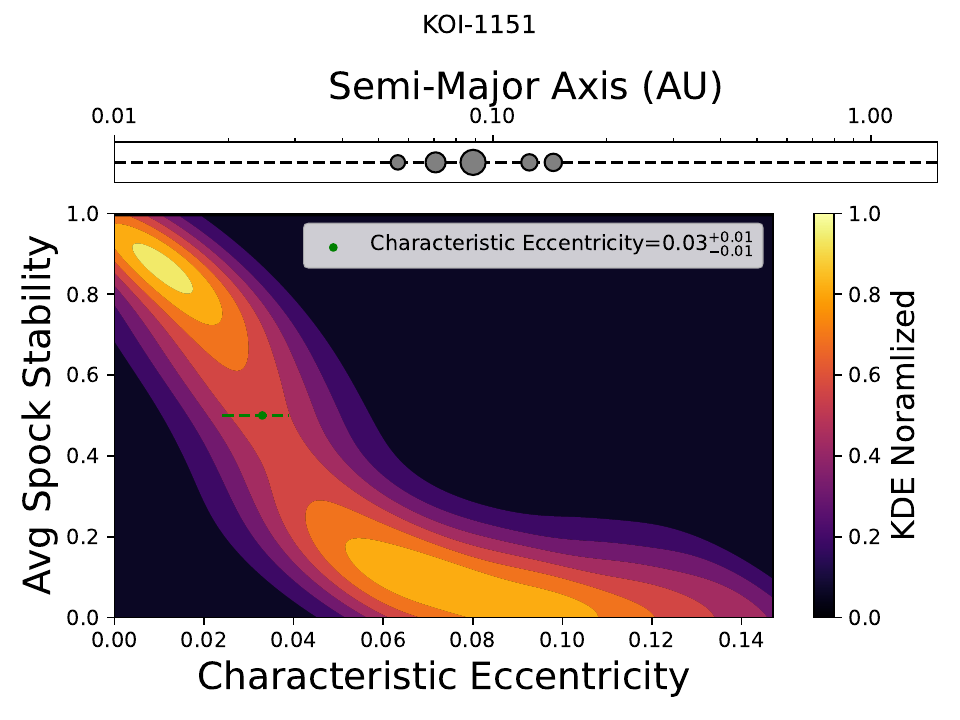}
    \end{subfigure}
\end{figure*}
\begin{figure*}
    \begin{subfigure}{}
        \includegraphics[width=0.48\textwidth]{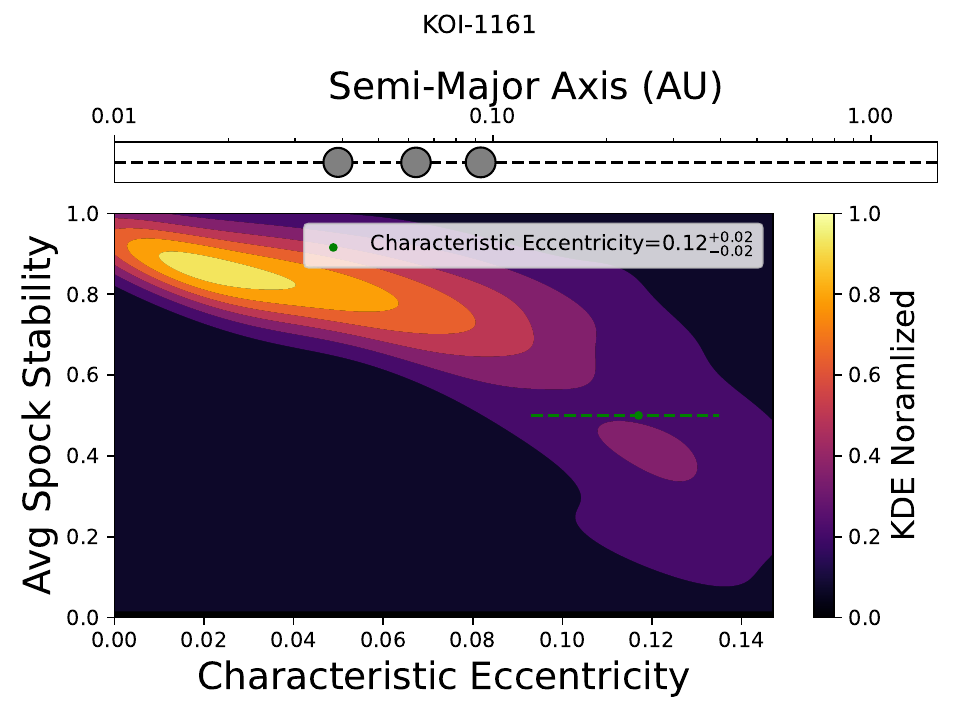}
    \end{subfigure}
    \begin{subfigure}{}
        \includegraphics[width=0.48\textwidth]{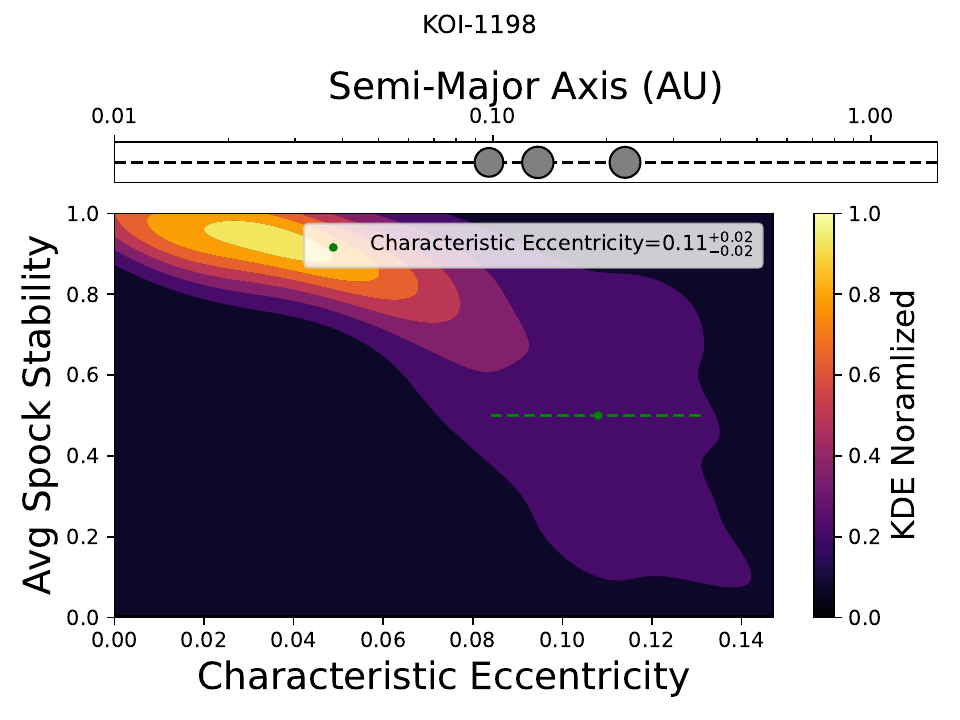}
    \end{subfigure}
    \begin{subfigure}{}
        \includegraphics[width=0.48\textwidth]{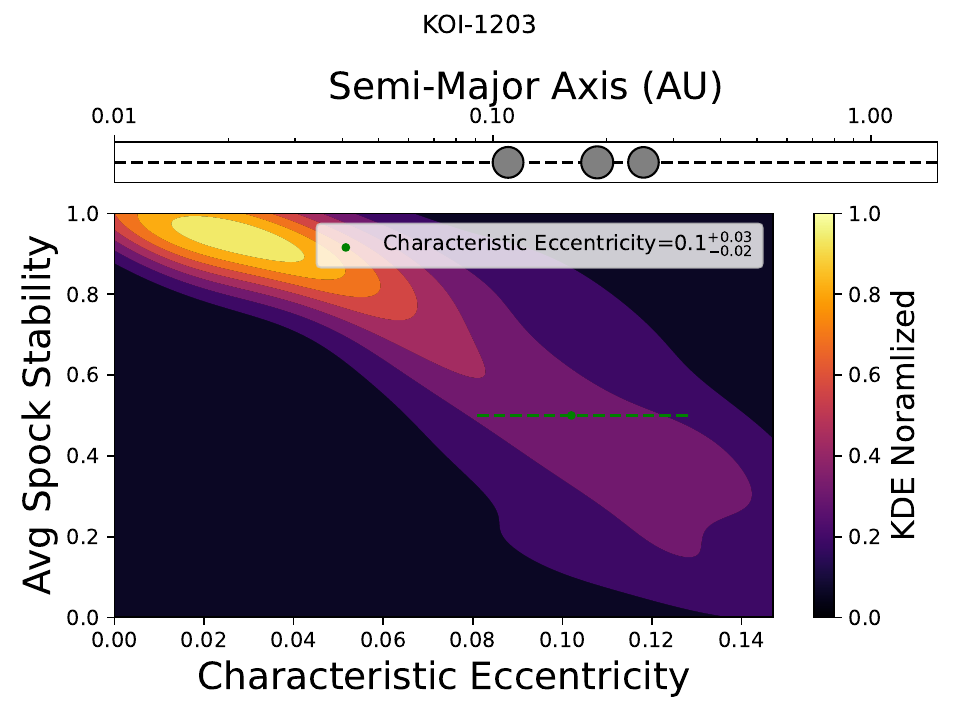}
    \end{subfigure}
    \begin{subfigure}{}
        \includegraphics[width=0.48\textwidth]{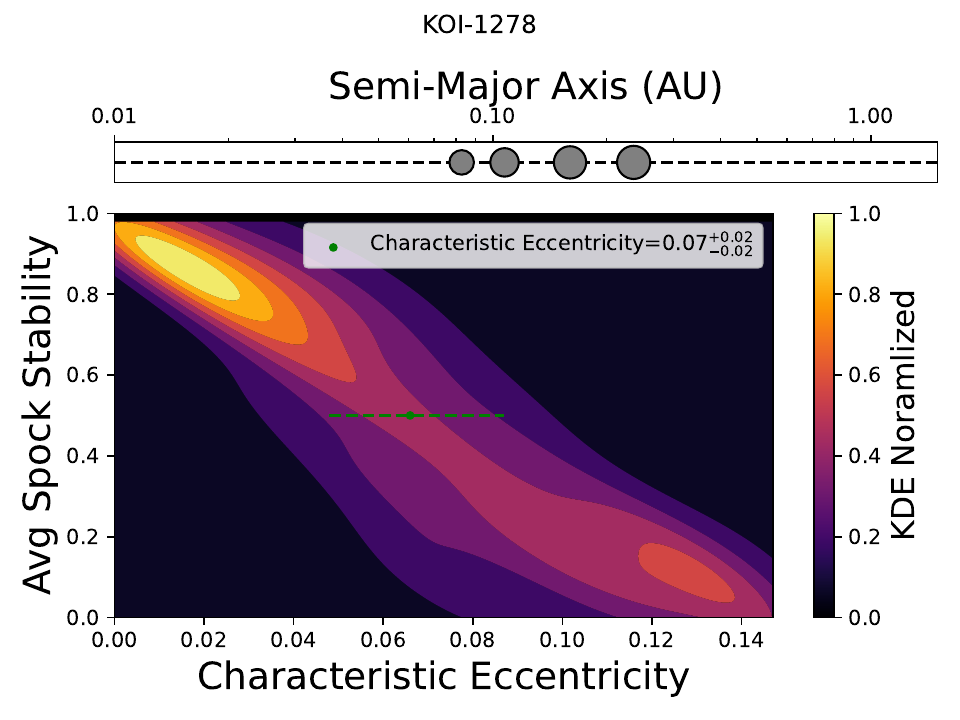}
    \end{subfigure}
    \begin{subfigure}{}
        \includegraphics[width=0.48\textwidth]{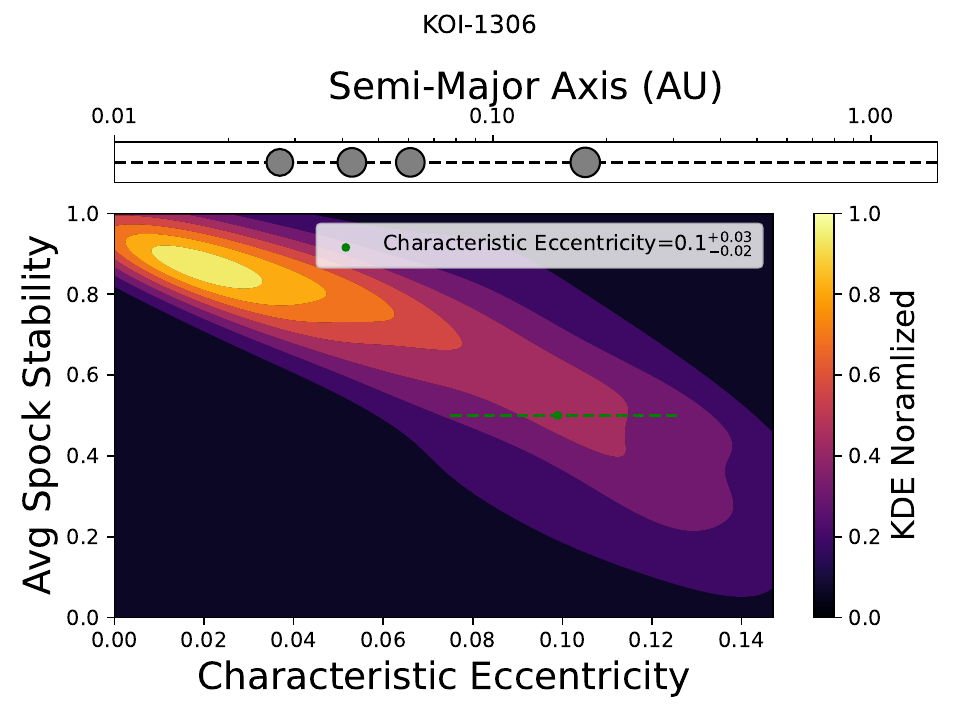}
    \end{subfigure}
    \begin{subfigure}{}
        \includegraphics[width=0.48\textwidth]{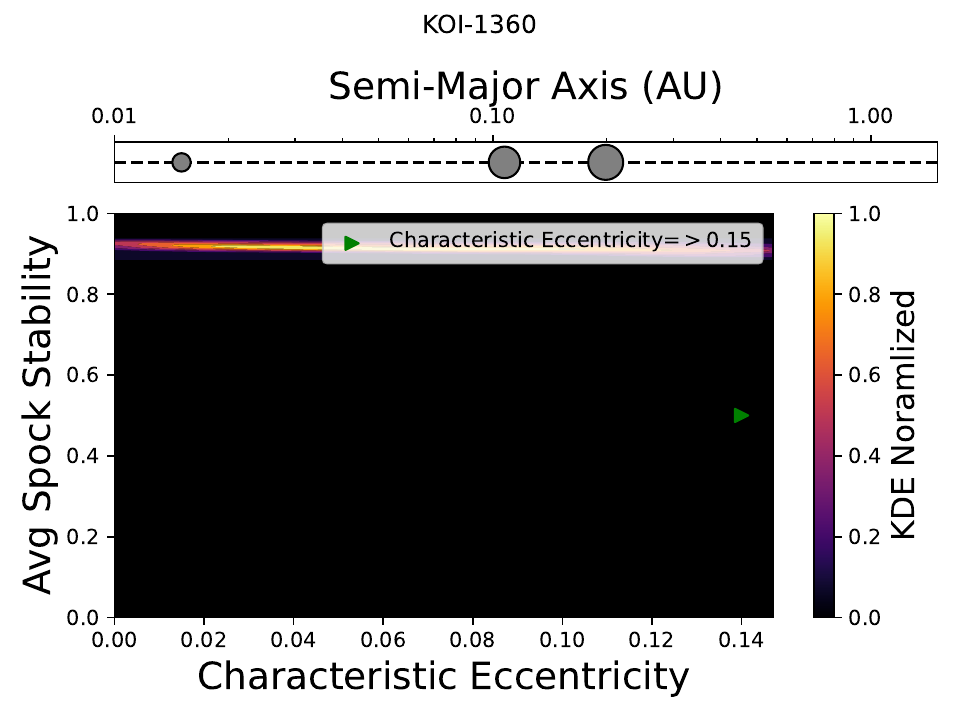}
    \end{subfigure}
\end{figure*}
\begin{figure*}
    \begin{subfigure}{}
        \includegraphics[width=0.48\textwidth]{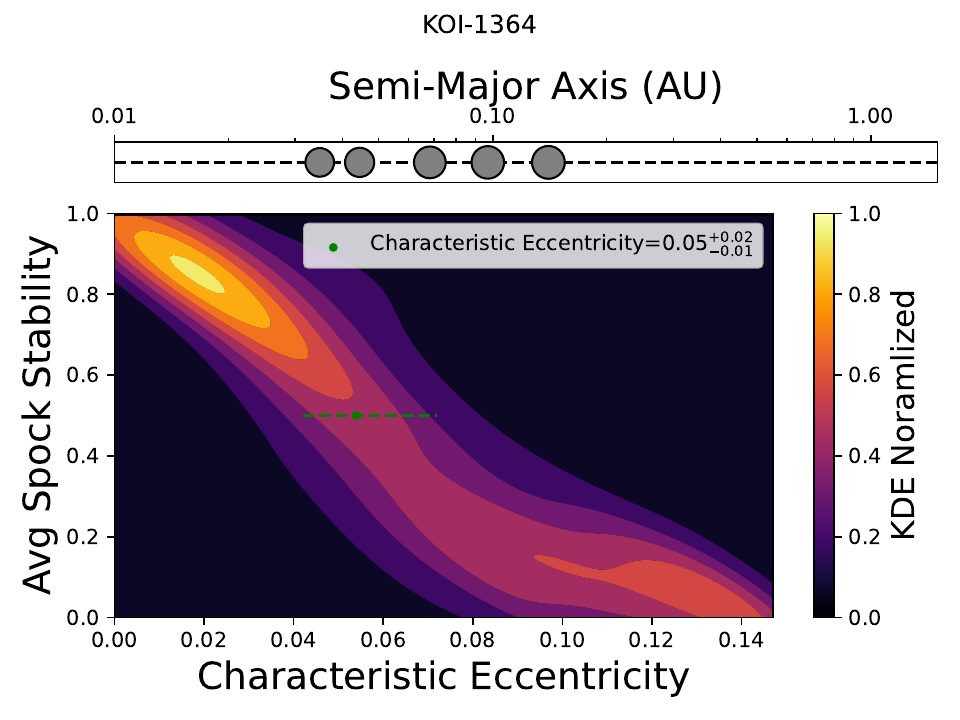}
    \end{subfigure}
    \begin{subfigure}{}
        \includegraphics[width=0.48\textwidth]{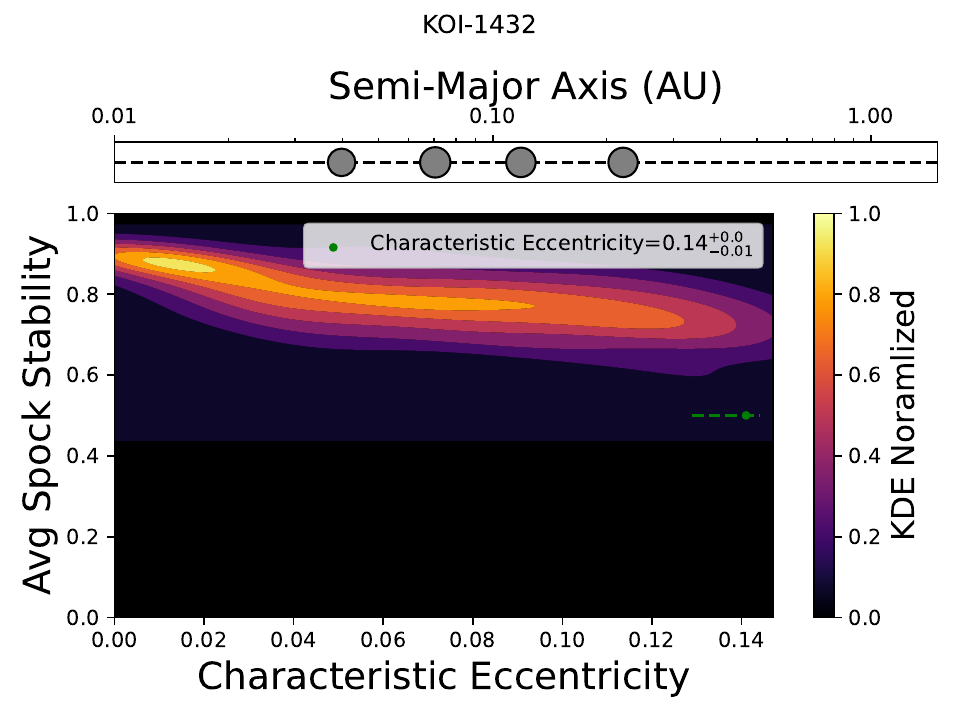}
    \end{subfigure}
    \begin{subfigure}{}
        \includegraphics[width=0.48\textwidth]{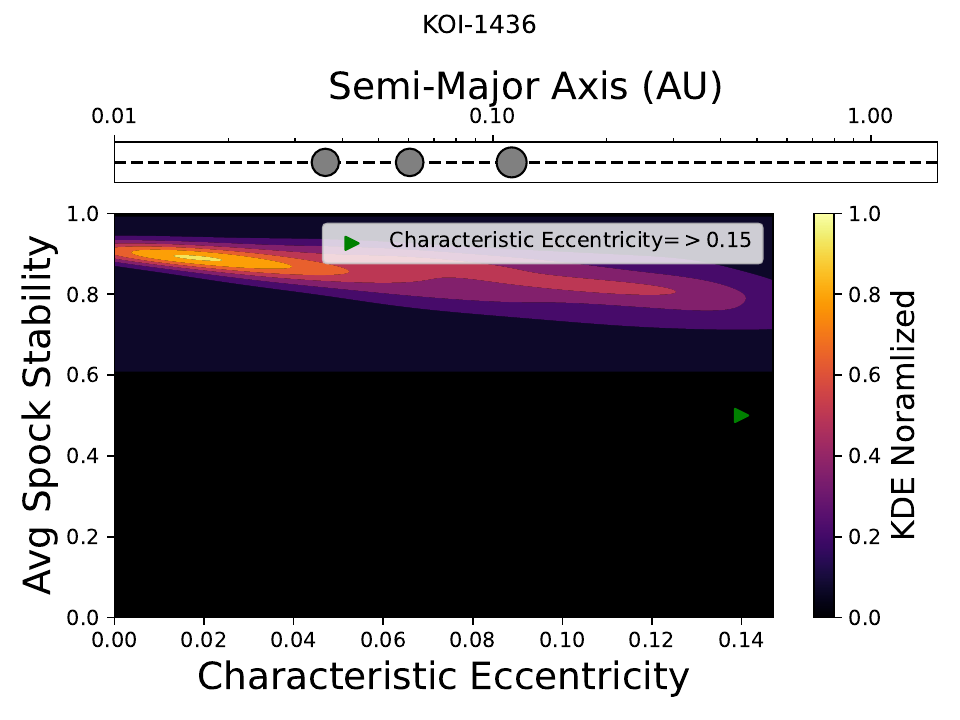}
    \end{subfigure}
    \begin{subfigure}{}
        \includegraphics[width=0.48\textwidth]{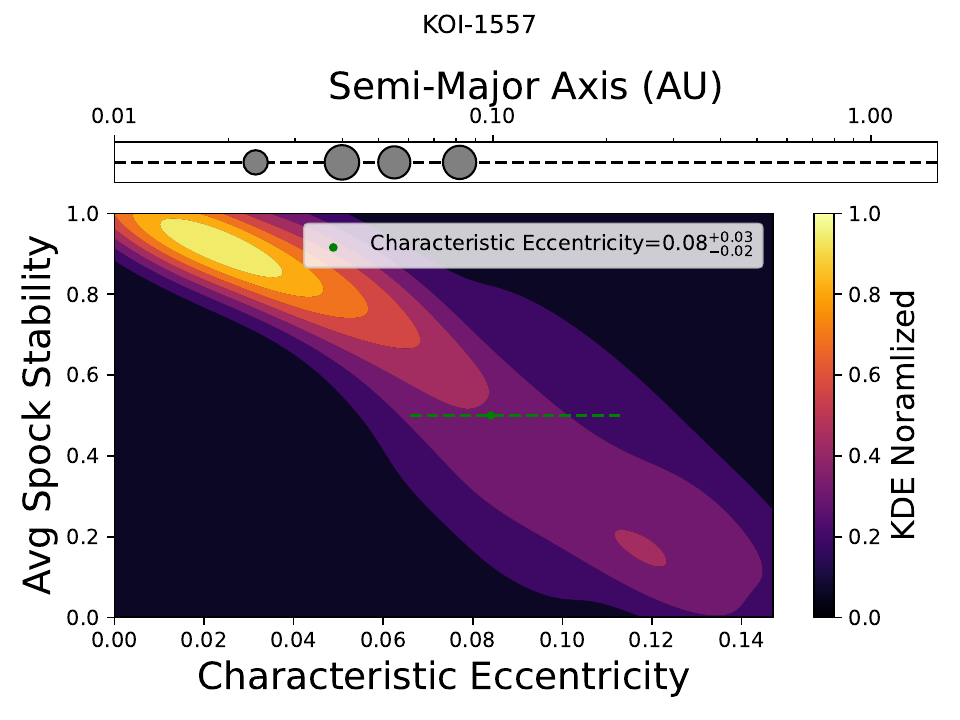}
    \end{subfigure}
    \begin{subfigure}{}
        \includegraphics[width=0.48\textwidth]{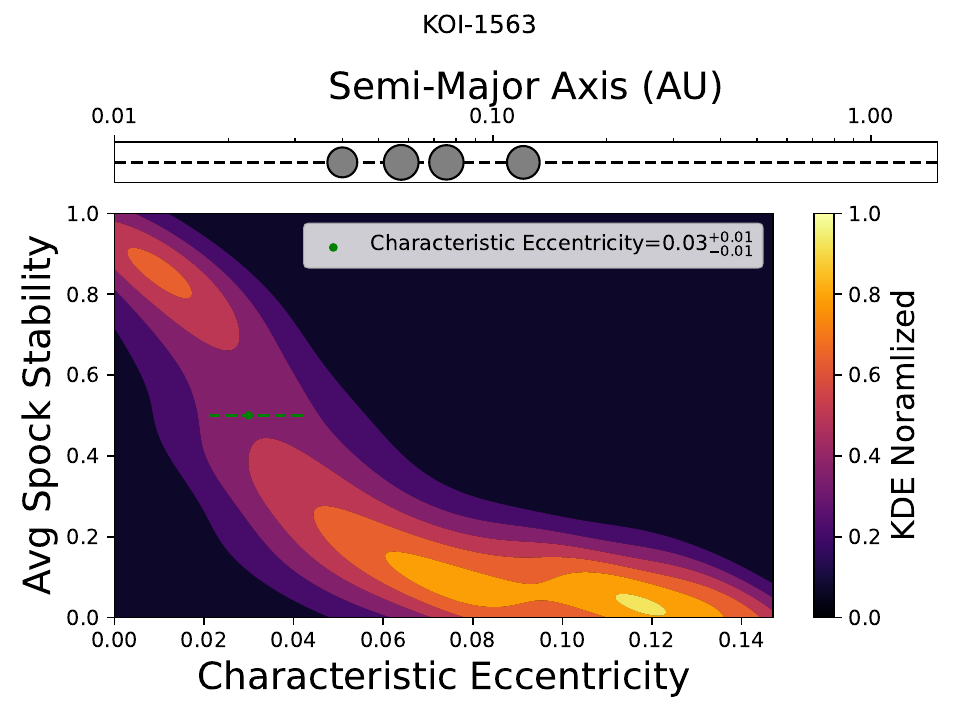}
    \end{subfigure}
    \begin{subfigure}{}
        \includegraphics[width=0.48\textwidth]{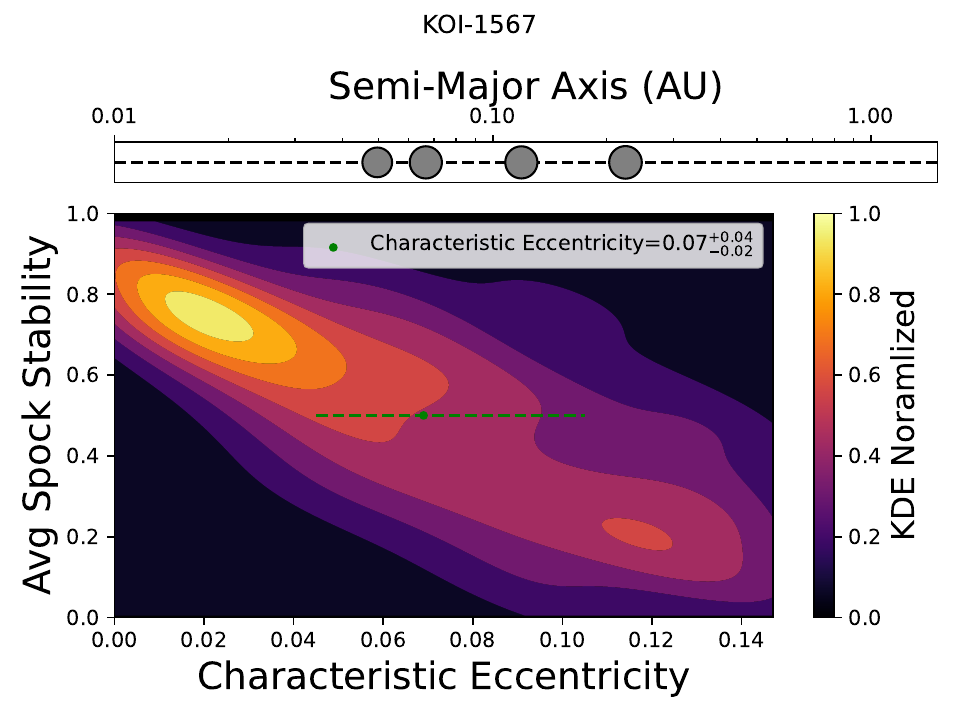}
    \end{subfigure}
\end{figure*}
\begin{figure*}
    \begin{subfigure}{}
        \includegraphics[width=0.48\textwidth]{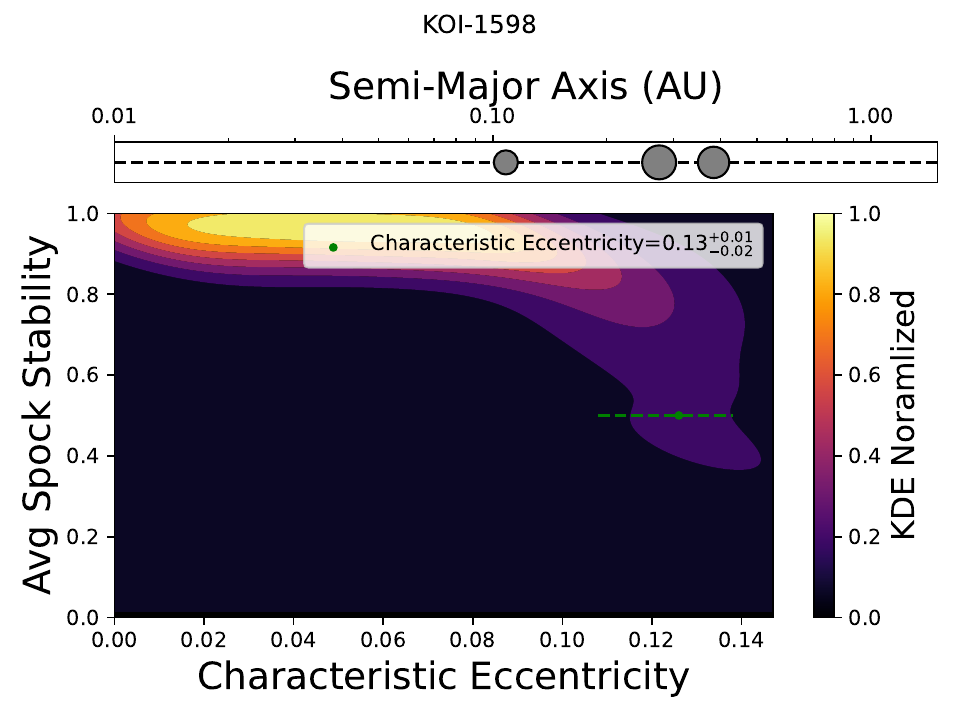}
    \end{subfigure}
    \begin{subfigure}{}
        \includegraphics[width=0.48\textwidth]{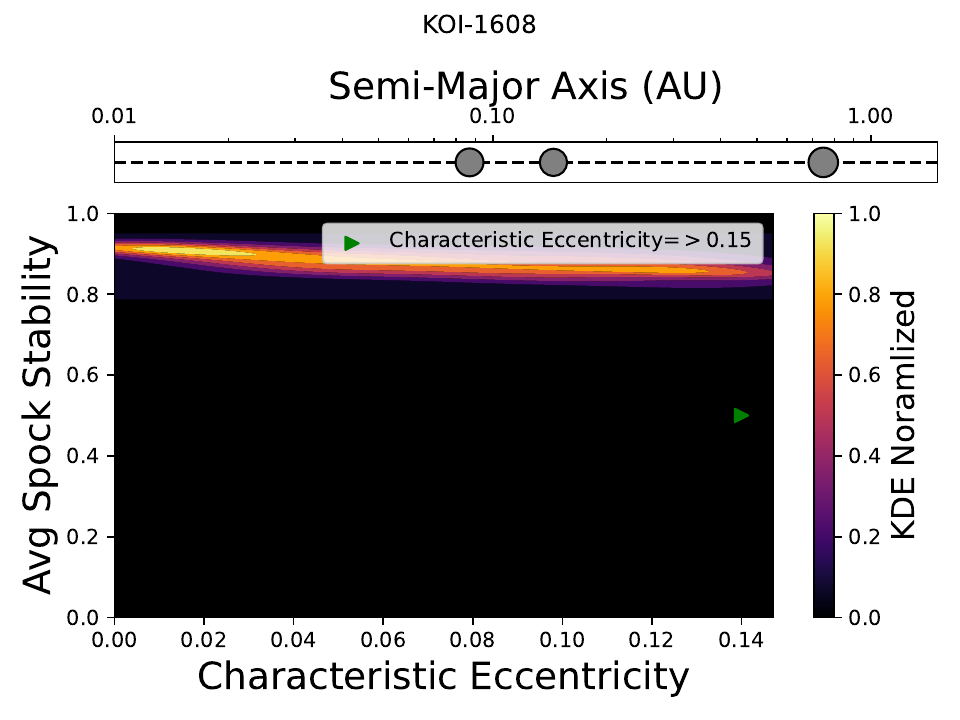}
    \end{subfigure}
   \begin{subfigure}{}
        \includegraphics[width=0.48\textwidth]{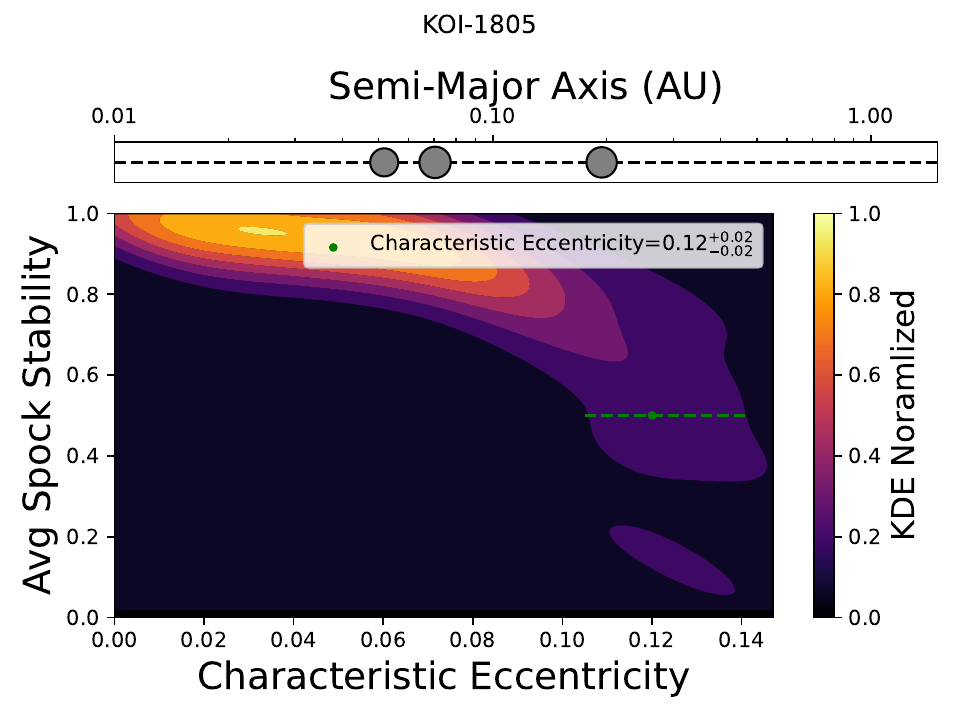}
    \end{subfigure}
    \begin{subfigure}{}
        \includegraphics[width=0.48\textwidth]{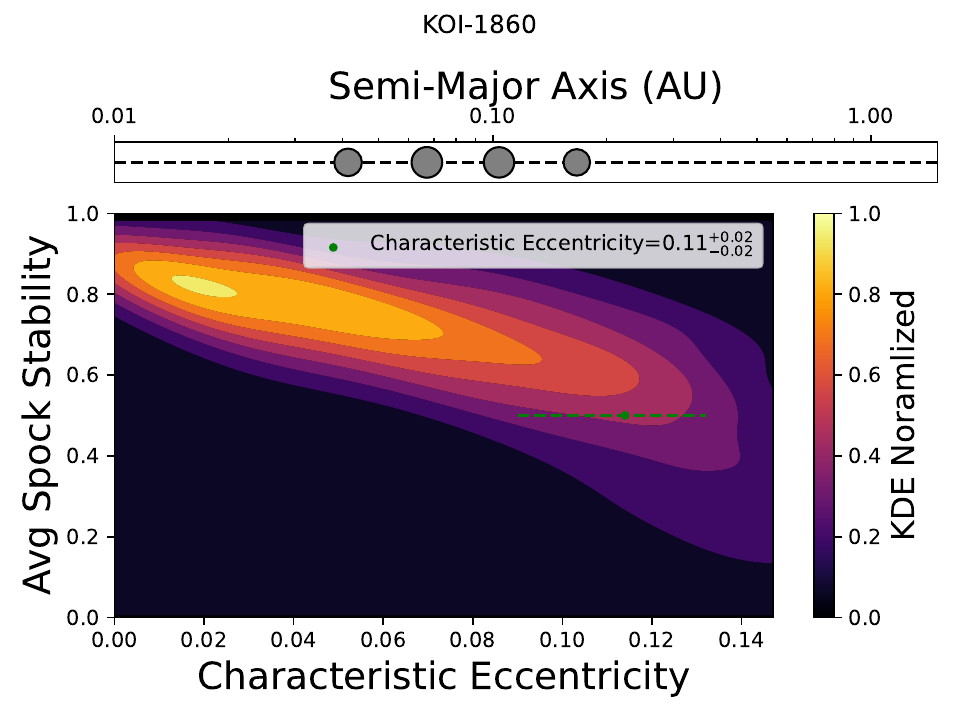}
    \end{subfigure}
    \begin{subfigure}{}
        \includegraphics[width=0.48\textwidth]{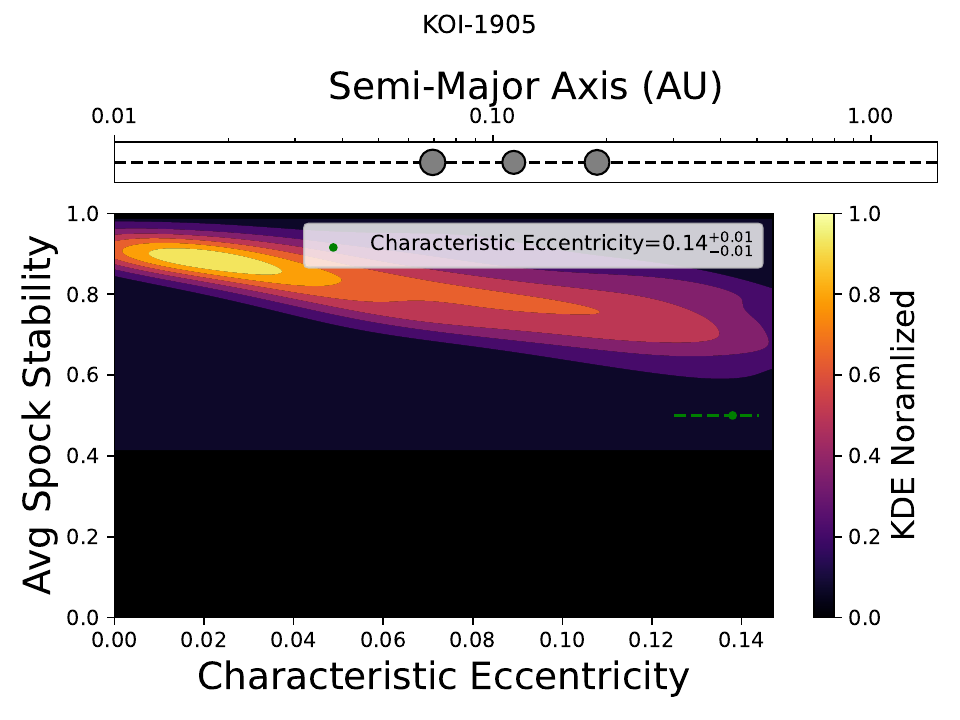}
    \end{subfigure}
    \begin{subfigure}{}
        \includegraphics[width=0.48\textwidth]{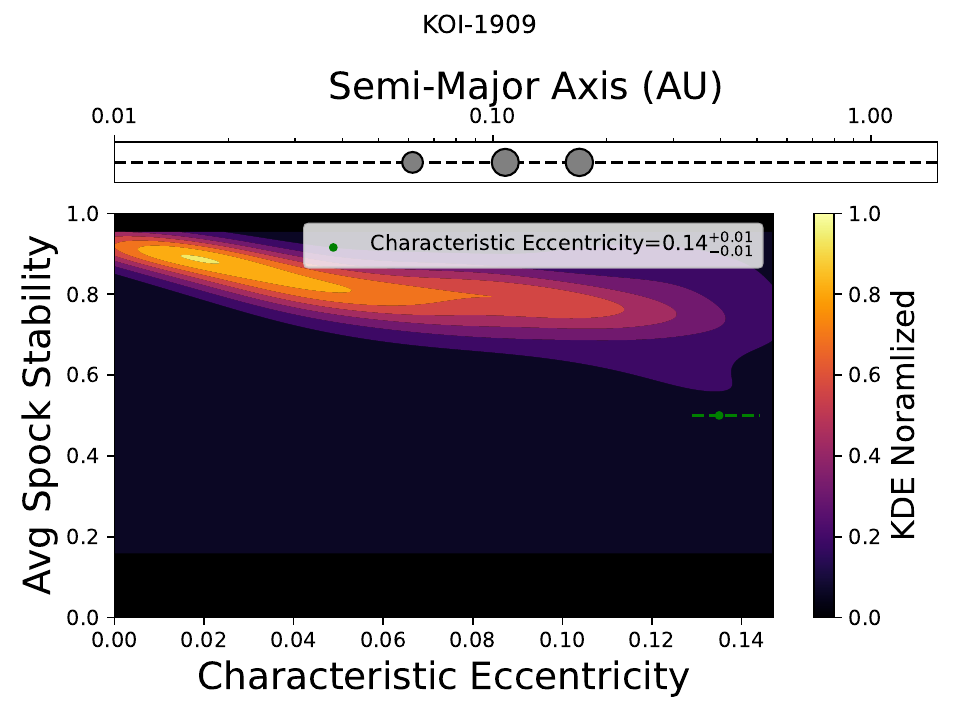}
    \end{subfigure}
\end{figure*}
\begin{figure*}
    \begin{subfigure}{}
        \includegraphics[width=0.48\textwidth]{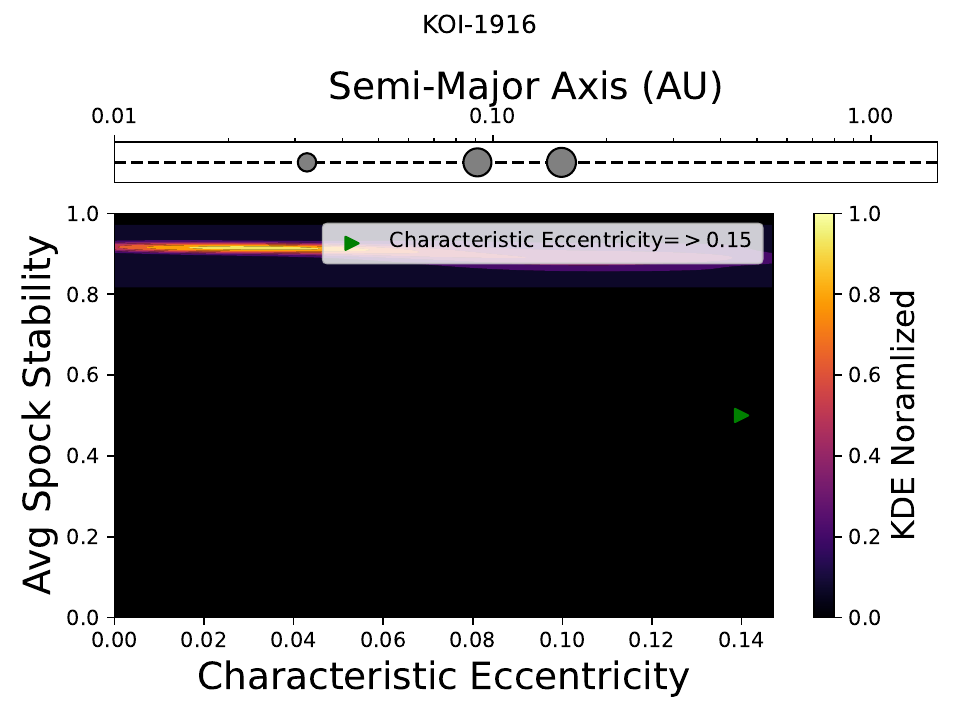}
    \end{subfigure}
    \begin{subfigure}{}
        \includegraphics[width=0.48\textwidth]{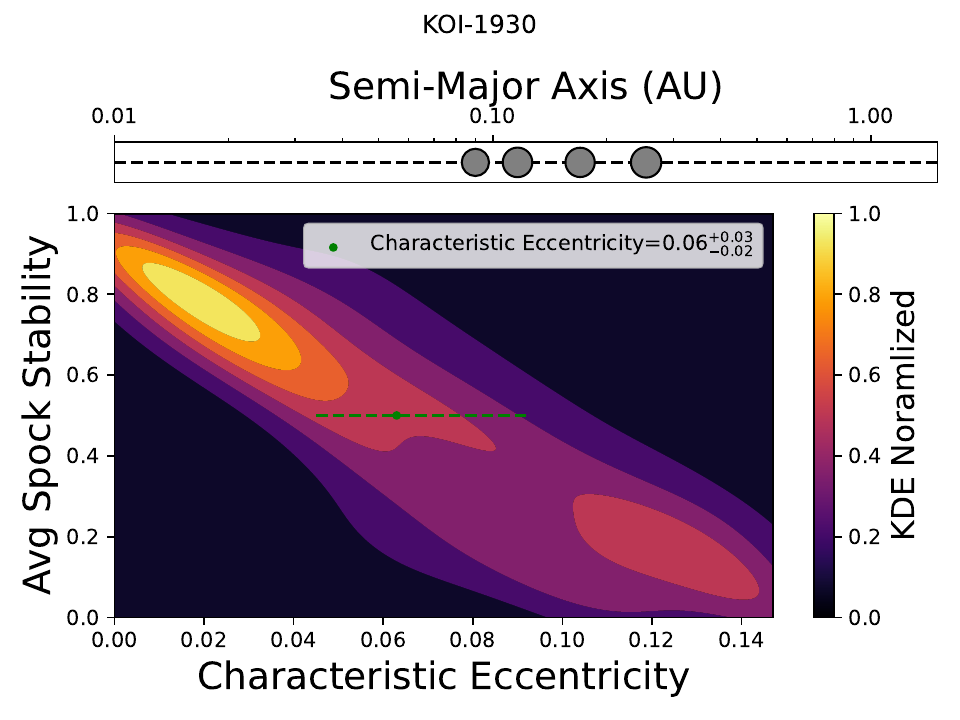}
    \end{subfigure}
    \begin{subfigure}{}
        \includegraphics[width=0.48\textwidth]{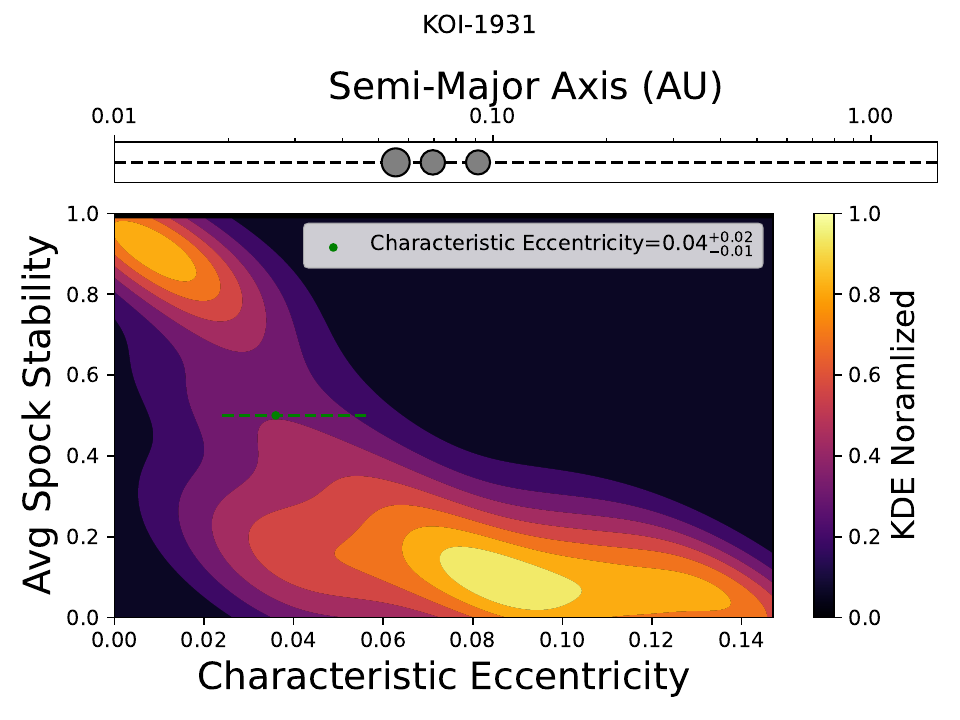}
    \end{subfigure}
    \begin{subfigure}{}
        \includegraphics[width=0.48\textwidth]{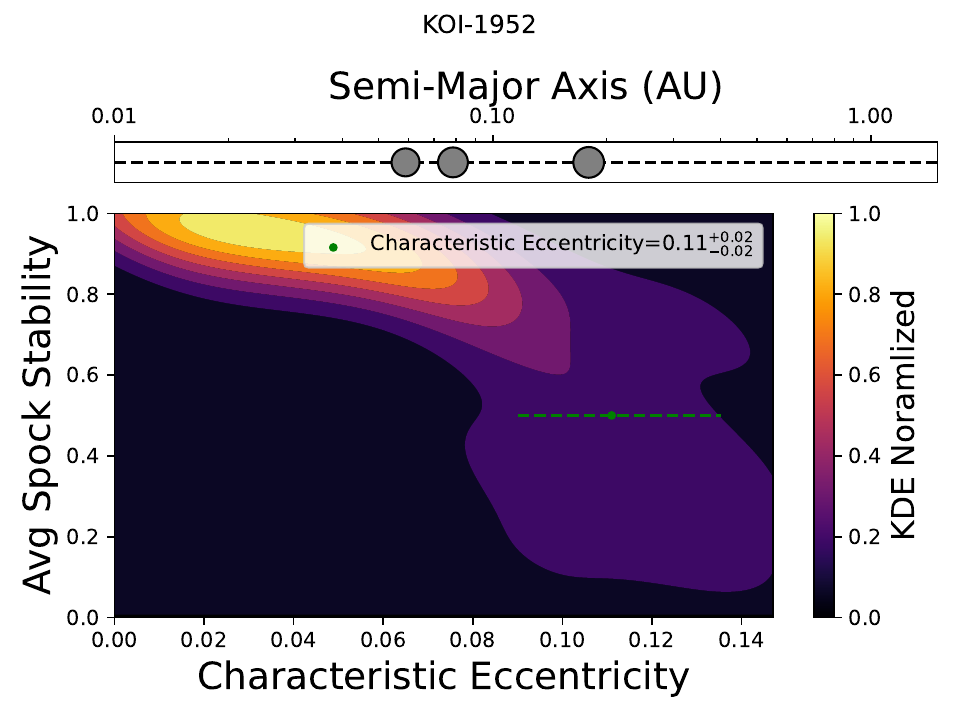}
    \end{subfigure}
    \begin{subfigure}{}
        \includegraphics[width=0.48\textwidth]{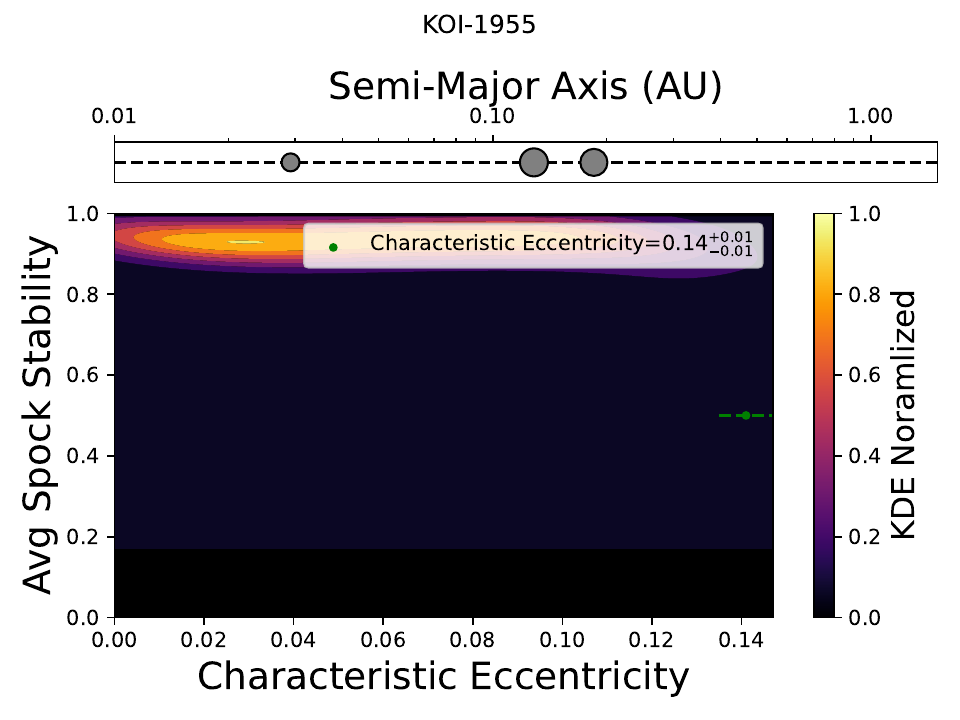}
    \end{subfigure}
    \begin{subfigure}{}
        \includegraphics[width=0.48\textwidth]{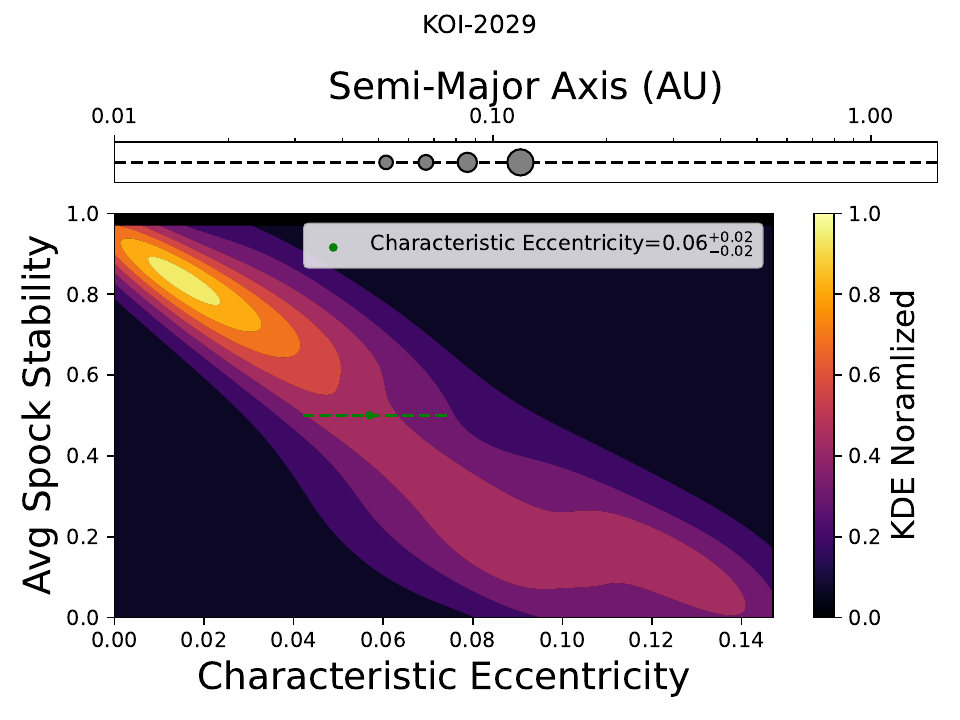}
    \end{subfigure}
\end{figure*}
\begin{figure*}
    \begin{subfigure}{}
        \includegraphics[width=0.48\textwidth]{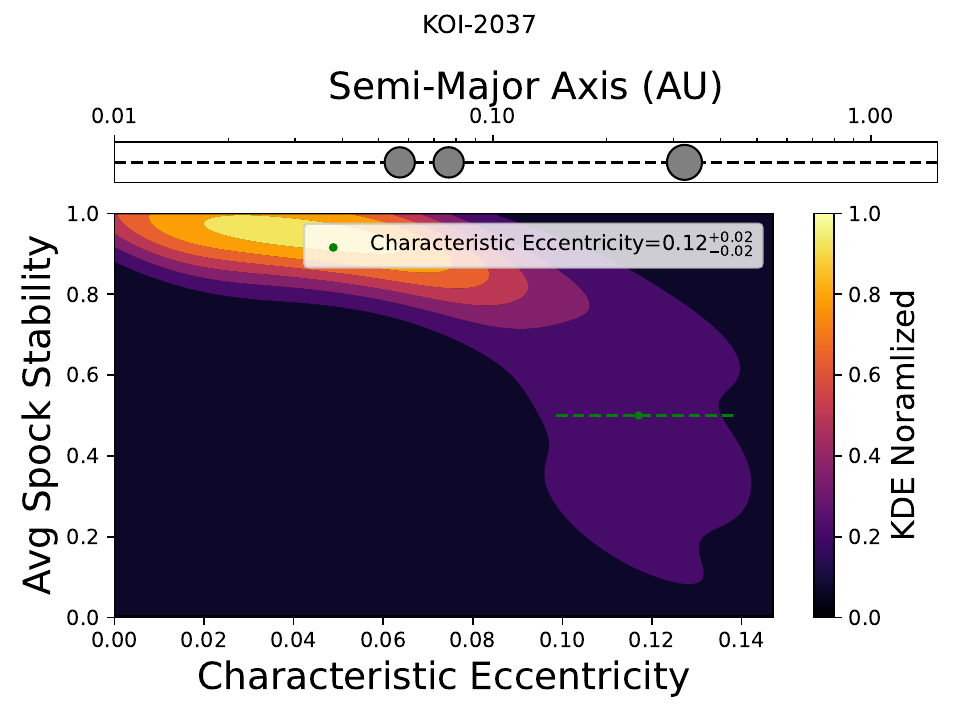}
    \end{subfigure}
    \begin{subfigure}{}
        \includegraphics[width=0.48\textwidth]{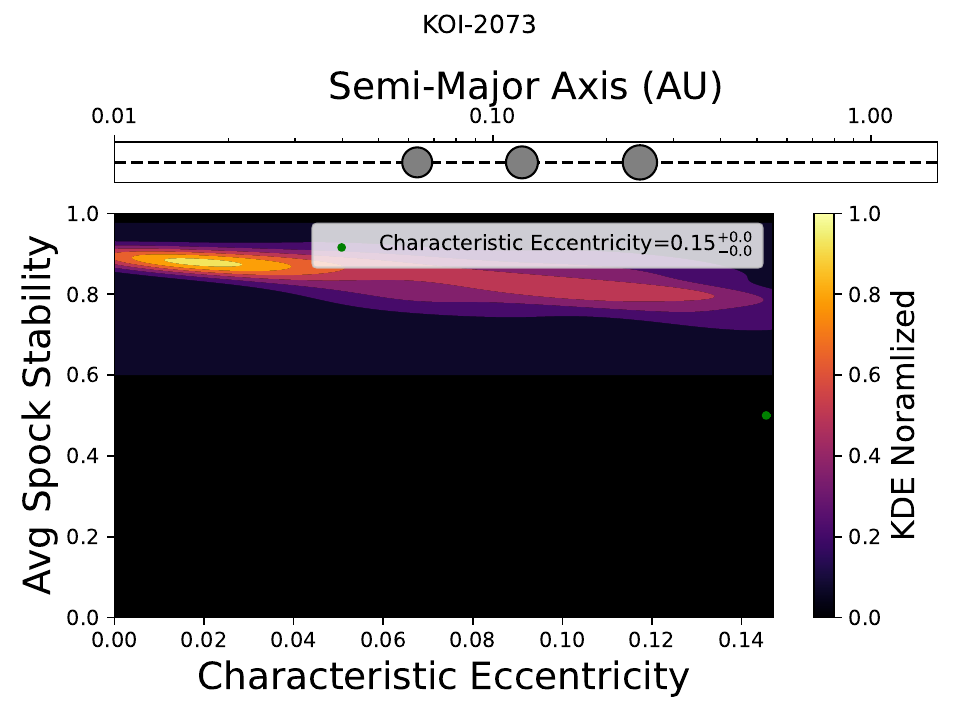}
    \end{subfigure}
    \begin{subfigure}{}
        \includegraphics[width=0.48\textwidth]{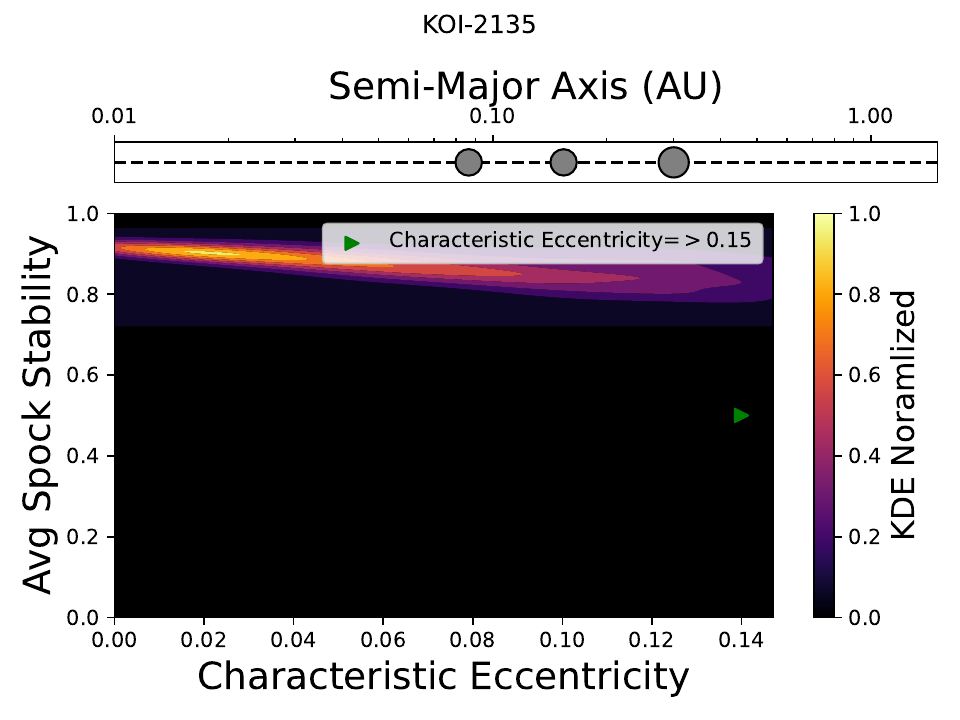}
    \end{subfigure}
    \begin{subfigure}{}
        \includegraphics[width=0.48\textwidth]{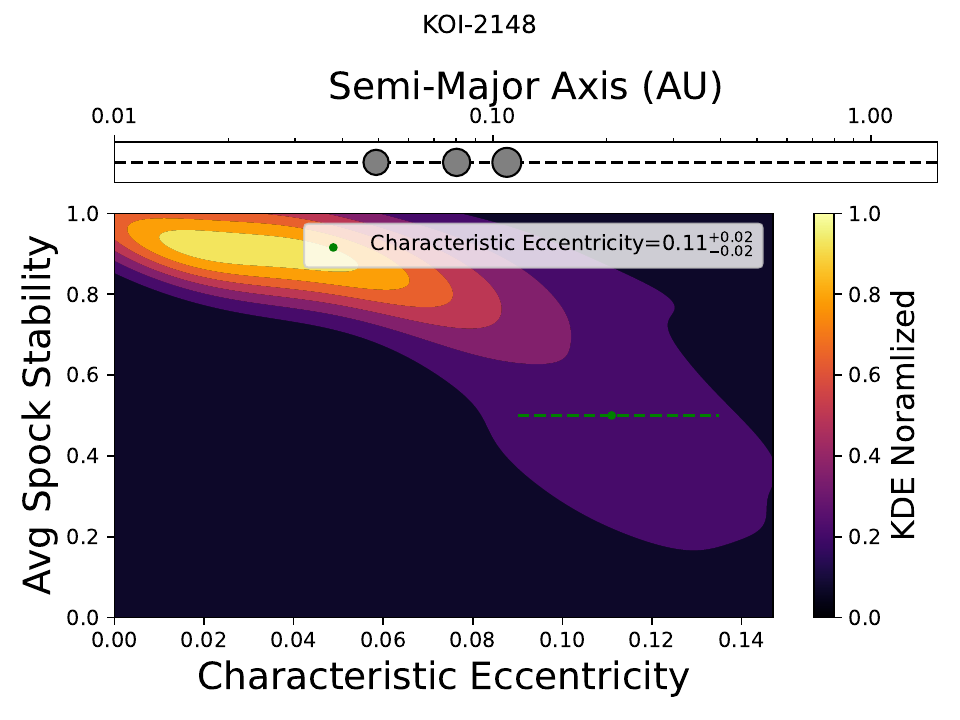}
    \end{subfigure}
    \begin{subfigure}{}
        \includegraphics[width=0.48\textwidth]{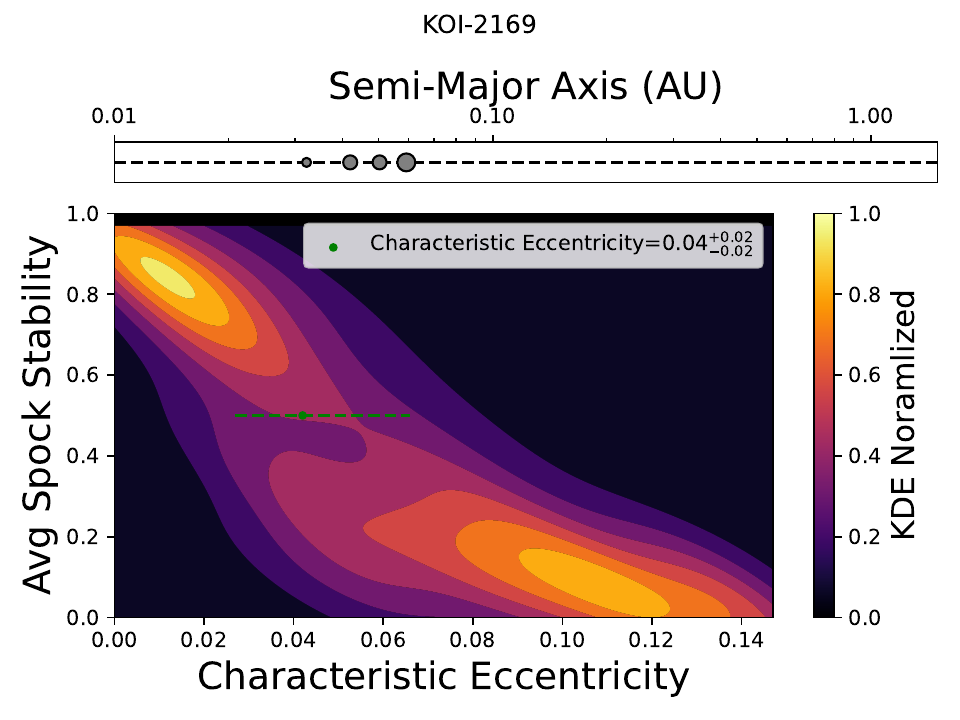}
    \end{subfigure}
    \begin{subfigure}{}
        \includegraphics[width=0.48\textwidth]{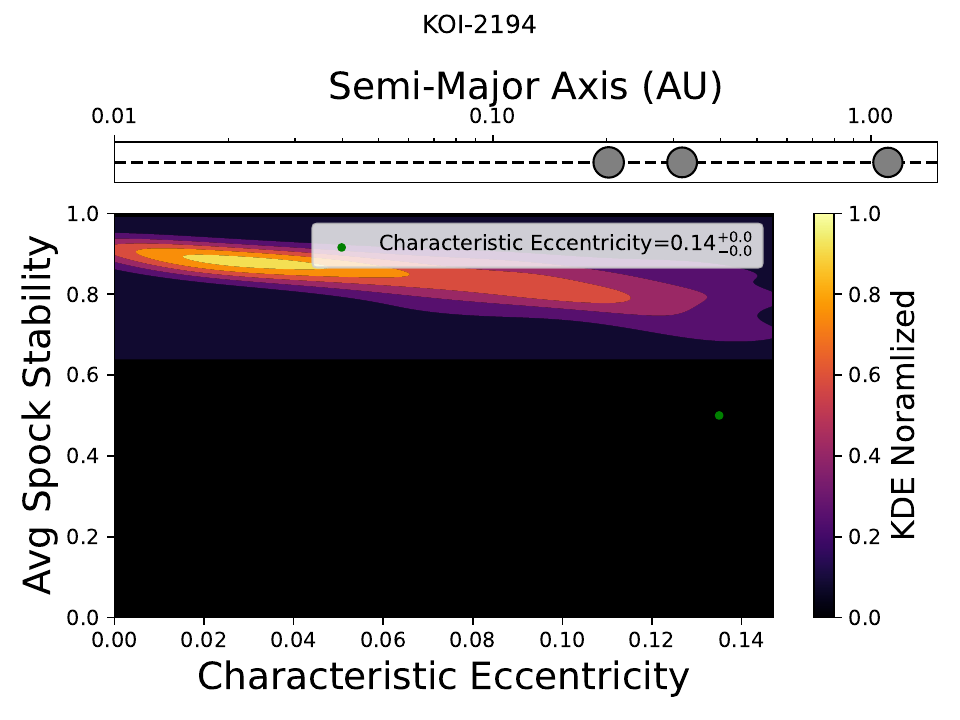}
    \end{subfigure}
\end{figure*}
\begin{figure*}
    \begin{subfigure}{}
        \includegraphics[width=0.48\textwidth]{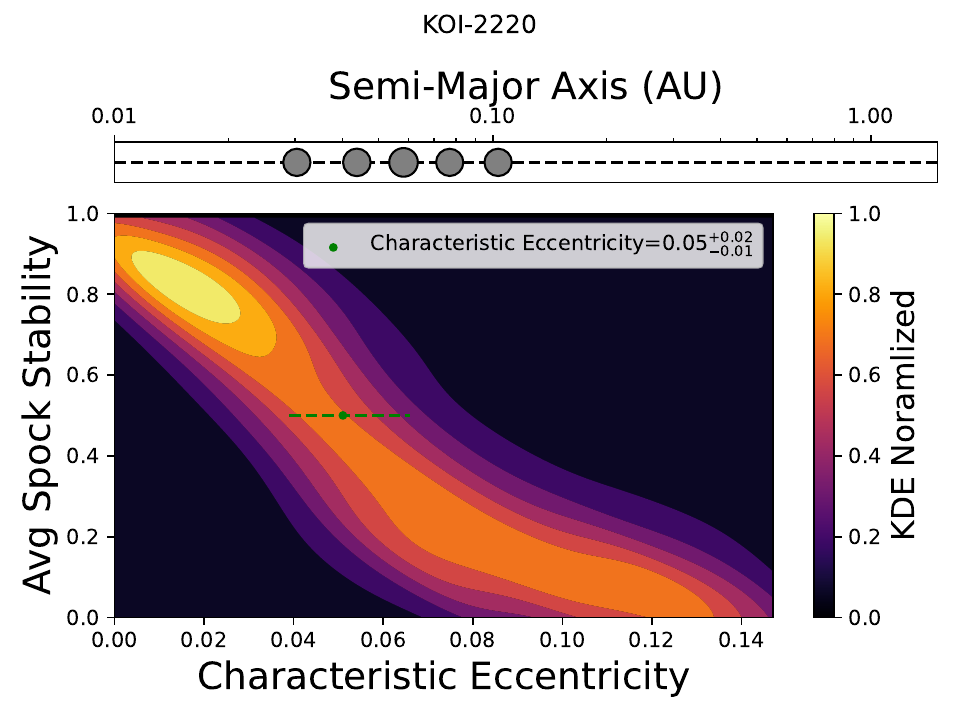}
    \end{subfigure}
    \begin{subfigure}{}
        \includegraphics[width=0.48\textwidth]{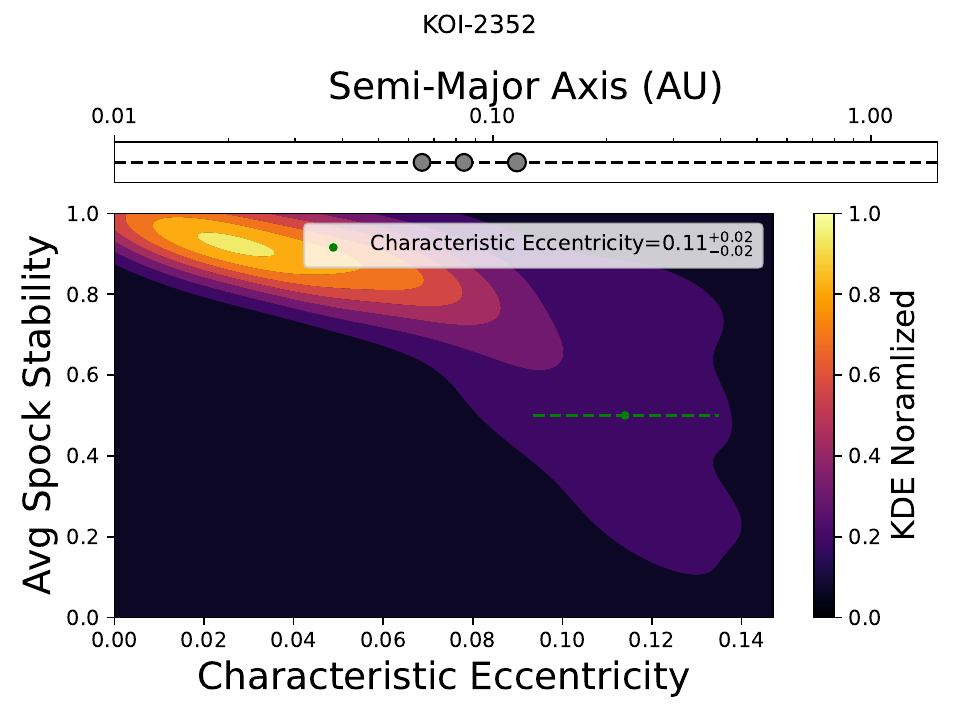}
    \end{subfigure}
    \begin{subfigure}{}
        \includegraphics[width=0.48\textwidth]{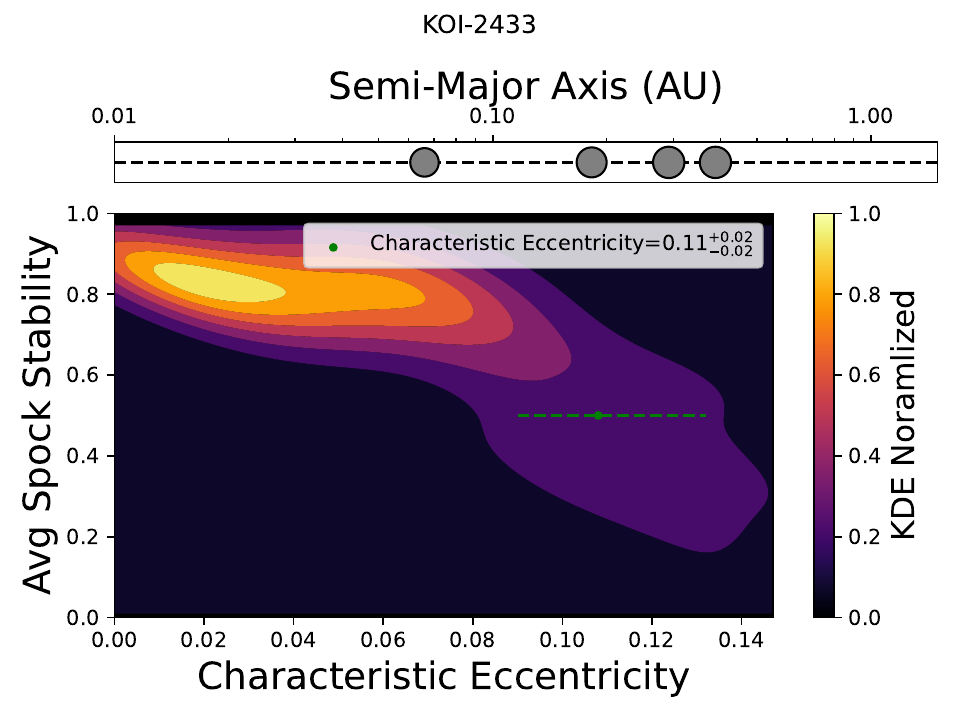}
    \end{subfigure}
    \begin{subfigure}{}
        \includegraphics[width=0.48\textwidth]{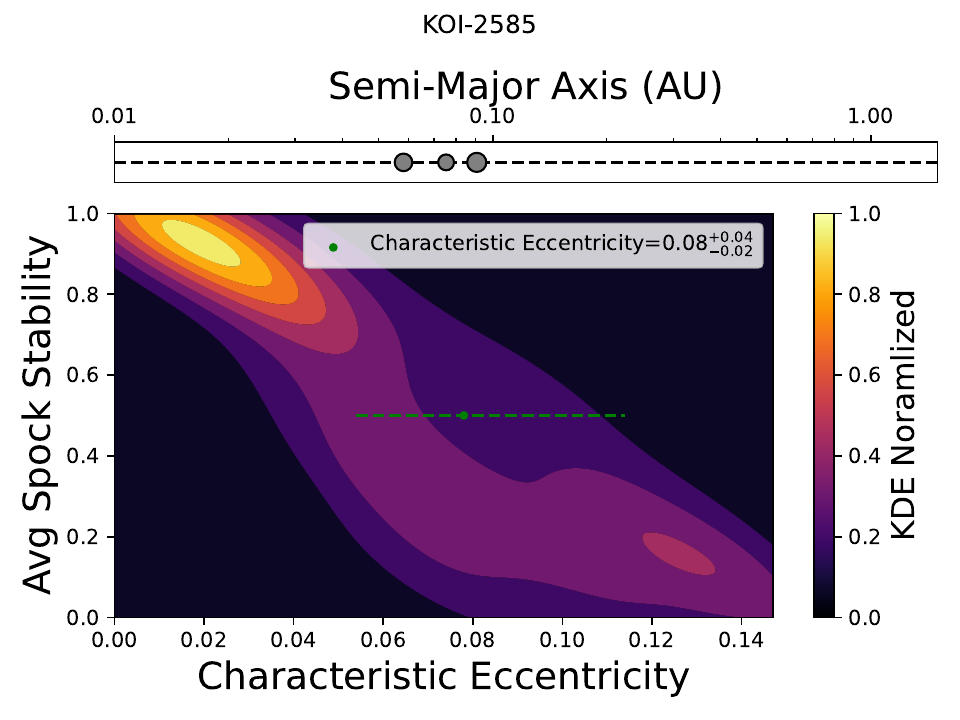}
    \end{subfigure}
    \begin{subfigure}{}
        \includegraphics[width=0.48\textwidth]{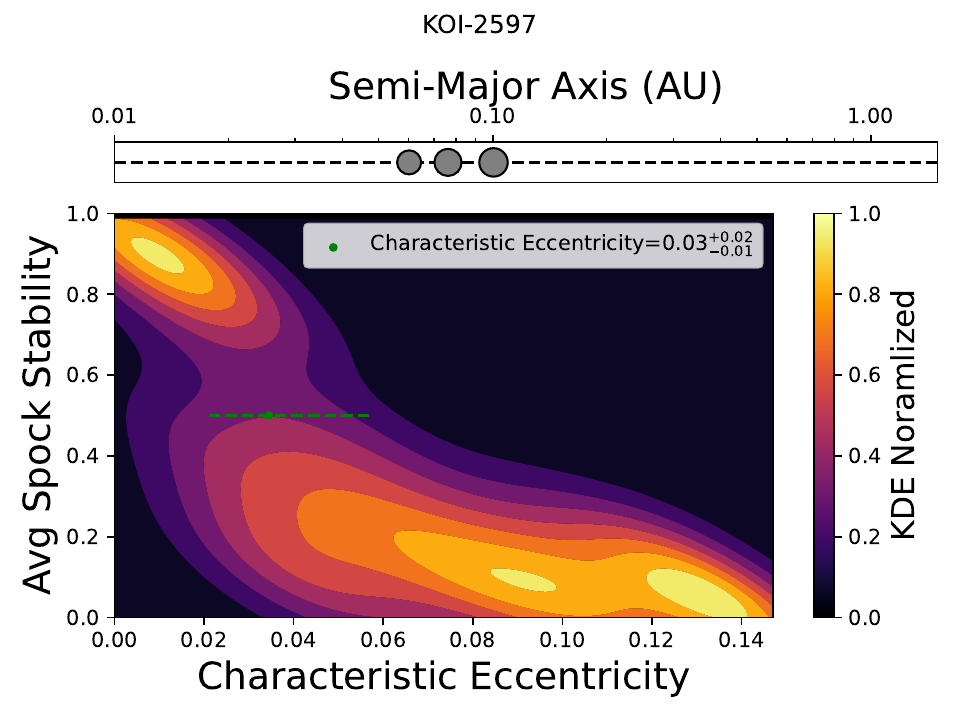}
    \end{subfigure}
    \begin{subfigure}{}
        \includegraphics[width=0.48\textwidth]{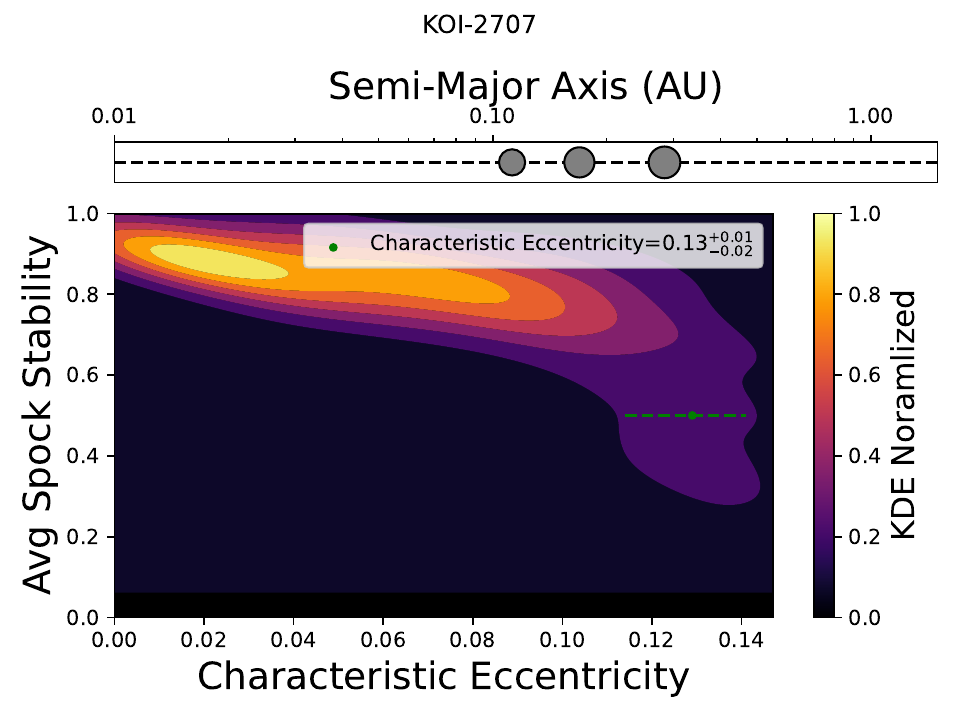}
    \end{subfigure}
\end{figure*}
\begin{figure*}
    \begin{subfigure}{}
        \includegraphics[width=0.48\textwidth]{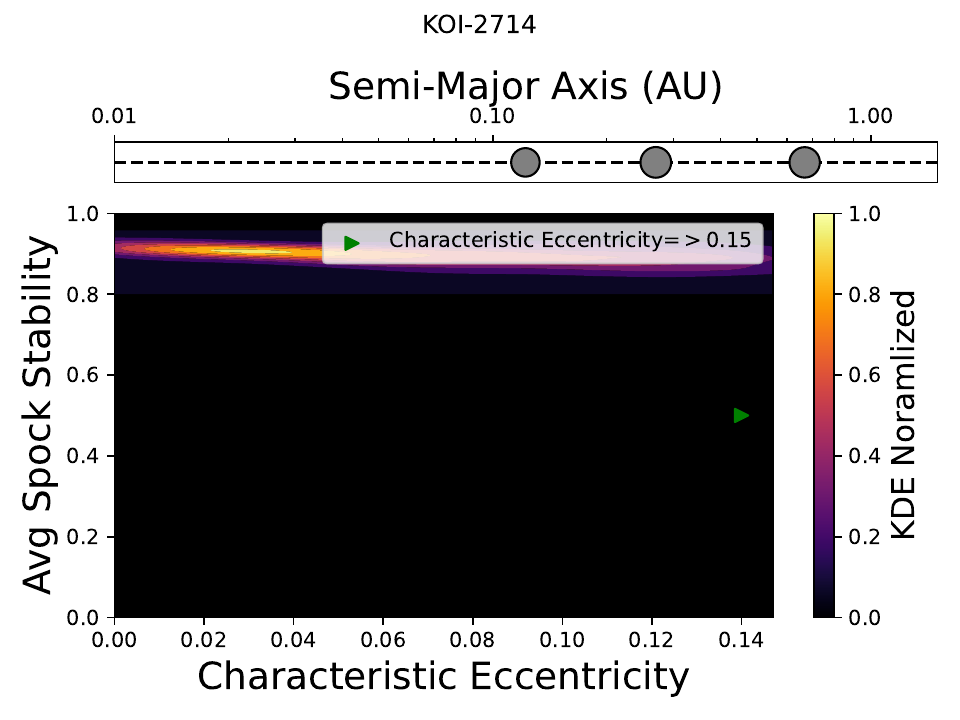}
    \end{subfigure}
    \begin{subfigure}{}
        \includegraphics[width=0.48\textwidth]{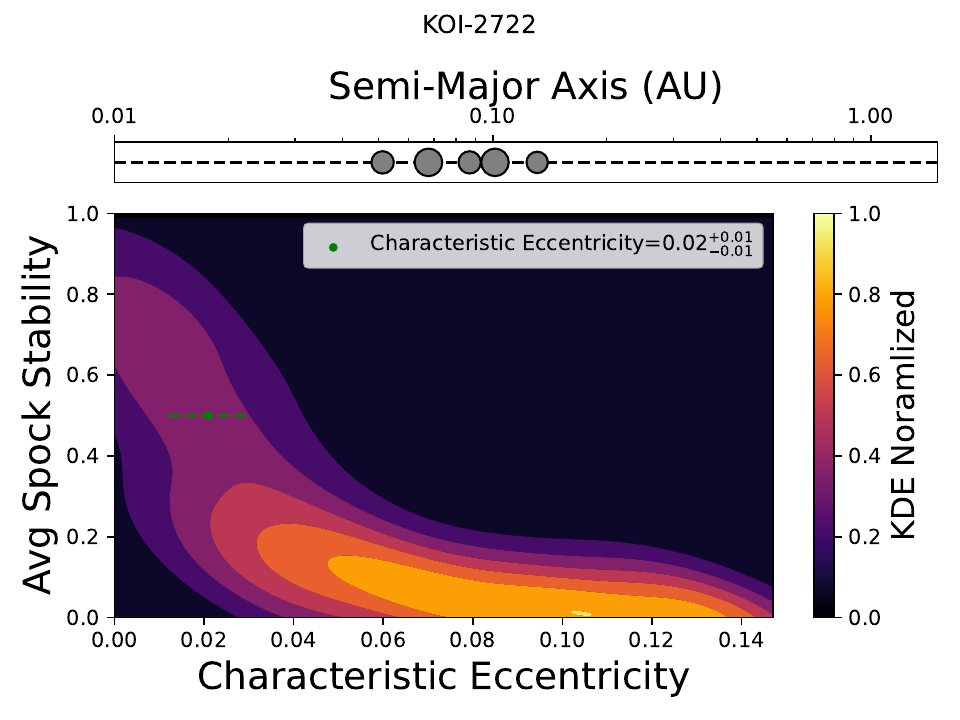}
    \end{subfigure}
    \begin{subfigure}{}
        \includegraphics[width=0.48\textwidth]{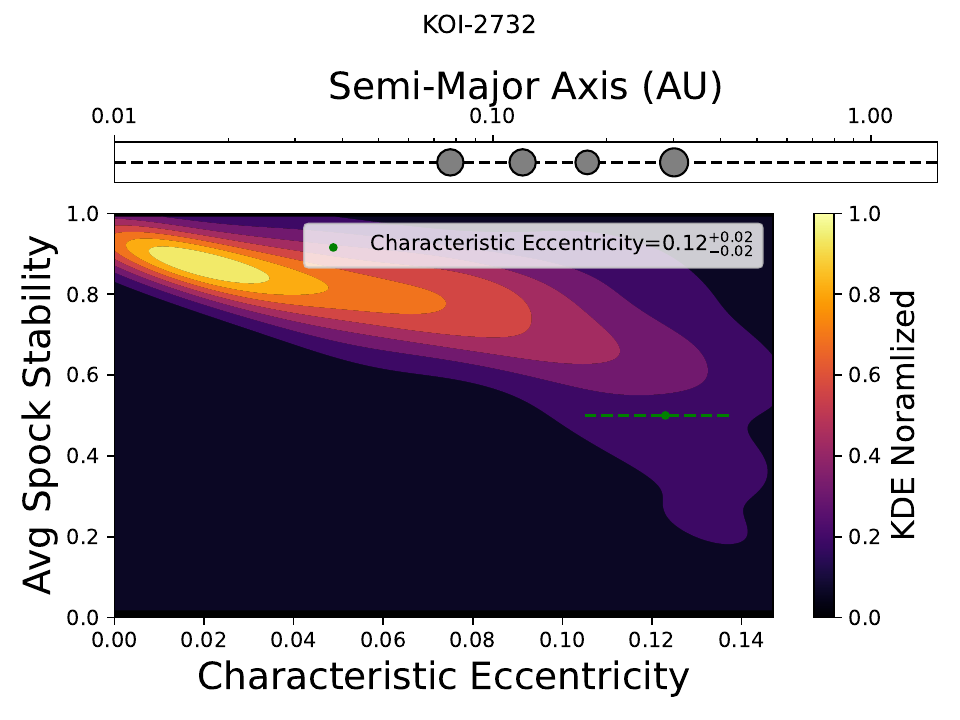}
    \end{subfigure}
    \begin{subfigure}{}
        \includegraphics[width=0.48\textwidth]{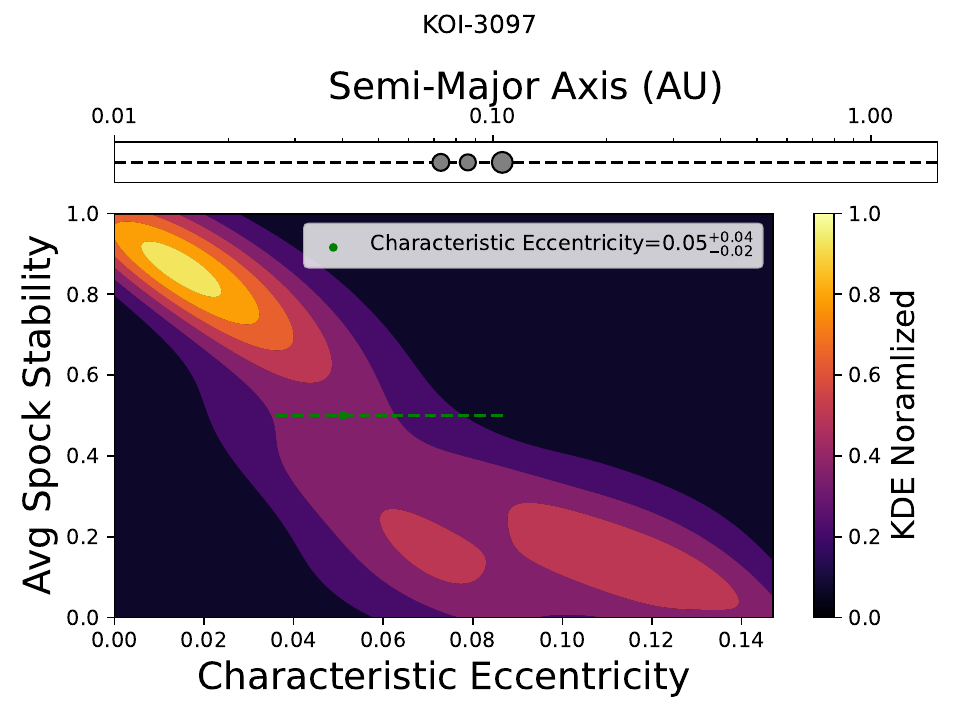}
    \end{subfigure}
    \begin{subfigure}{}
        \includegraphics[width=0.48\textwidth]{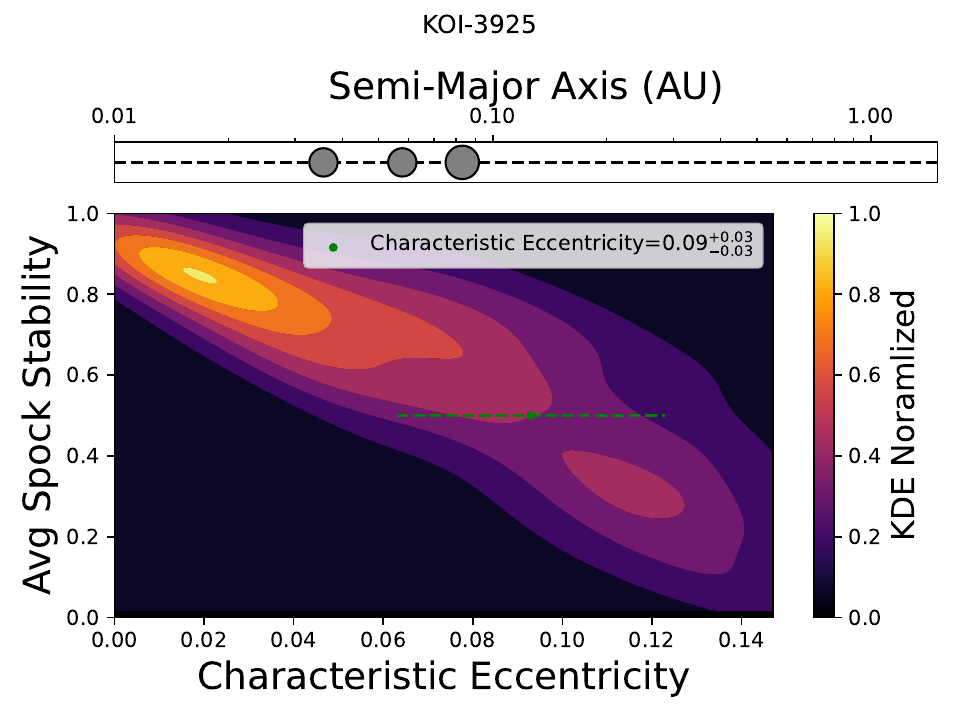}
    \end{subfigure}
    \begin{subfigure}{}
        \includegraphics[width=0.48\textwidth]{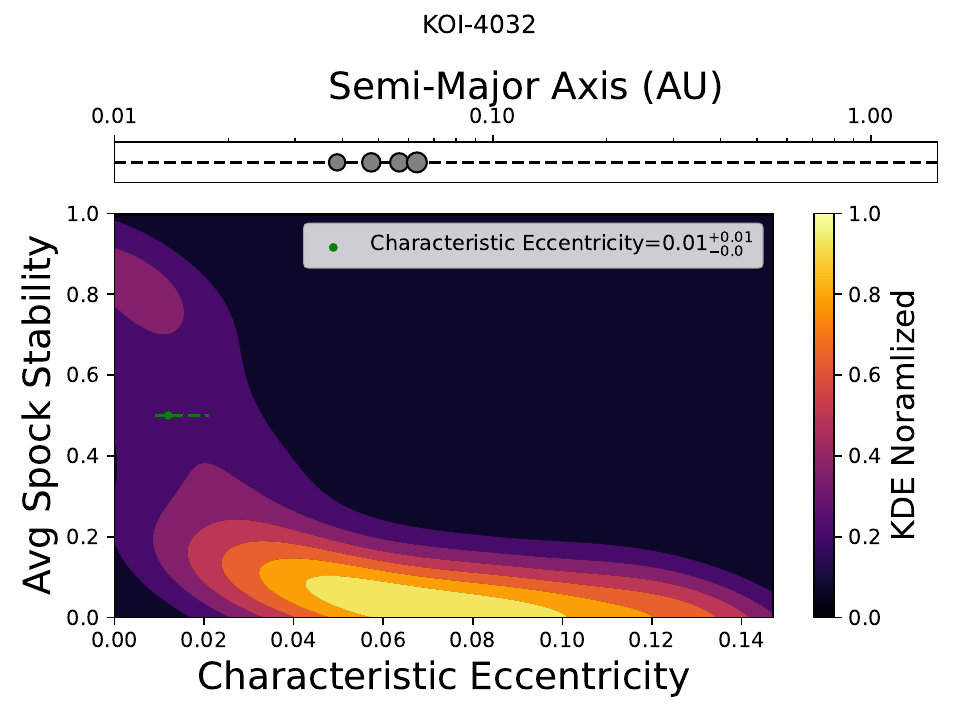}
    \end{subfigure}
    \caption{Gallery of KDE plots for 126 SPOCK compatible CKS systems. The tested eccentricity is on the x and the corresponding SPOCK stability is on the y. The KDE is normalized with 1 being the highest density of points for the system and 0 being the lowest density of points for the system. The characteristic eccentricity for a specific system that leads to a SPOCK stability of 0.5, or 50$\%$ is marked as a red point on the graph and listed in the table below. Systems that did not reach a SPOCK stability of 50$\%$ have a lower limit for their characteristic eccentricity of 0.15. On these graphs a a red arrow pointing to the right denotes the lower limit.  As an be seen, most plots follow roughly an s shape with an area of stability, and area of instability and a transition region. However as can be seen there is variance in the sharpness and location of the transition region between systems.}
\end{figure*}
\clearpage
\subsection{Table of all System Values}
\begin{longtable}[hbt!]{||c || c || c|c|c|c|c|c|c||} 
     \hline
     \textbf{System Name} & \textbf{$e_{system}$} & \textbf{$e_{1}$} & \textbf{$e_{2}$} & \textbf{$e_{3}$} & \textbf{$e_{4}$} & \textbf{$e_{5}$} & \textbf{$e_{6}$} & \textbf{$e_{7}$}
    \endhead
    \hline\hline
    KOI-70 & 0.08 & $0.08$ & $0.07$ & $0.08$ & $0.10$ & $0.097$ & $N/A$ & $N/A$  \\ 
    \hline
    KOI-82 & 0.02 & 0.02 & 0.02 & $0.02$ & $0.02$ & $0.031$ & $N/A$ & $N/A$ \\
    \hline
    KOI-$85$ & 0.1 &  $0.08$ & $0.10$ & $0.06$ & $0.18$ & $N/A$ & $N/A$ & $N/A$ \\
    \hline
    KOI-94 & 0.1 &  $0.11$ & $0.11$ & $0.10$ & $0.12$ & $N/A$ & $N/A$ & $N/A$ \\
    \hline
    KOI-$111$ & 0.12 &  $0.16$ & $0.13$ & $0.15$ & $N/A$ & $N/A$ & $N/A$ & $N/A$ \\
    \hline
    KOI-116 & 0.11 &  $0.11$ & $0.15$ & $0.12$ & $0.13$ & $N/A$ & $N/A$ & $N/A$ \\ 
    \hline
    KOI-117 & 0.07 &  $0.06$ & $0.07$ & $0.06$ & $0.07$ & $N/A$ & $N/A$ & $N/A$ \\ 
    \hline
    KOI-137 & 0.11 & $0.14$ & $0.13$ & $0.13$ & $N/A$ & $N/A$ & $N/A$ & $N/A$  \\ 
    \hline
    KOI-$148$ & 0.13 &  $0.17$ & $0.11$ & $0.15$ & $N/A$ & $N/A$ & $N/A$ & $N/A$ \\
    \hline
    KOI-152 & 0.08 & $0.10$ & $0.09$ & $0.08$ & $0.08$ & $N/A$ & $N/A$ & $N/A$ \\  
    \hline
    KOI-156 & 0.04 & $0.04$ & $0.03$ & $0.03$ & $N/A$ & $N/A$ & $N/A$ & $N/A$ \\ 
    \hline
    KOI-157 & 0.02 & $0.02$ & $0.02$ & $0.02$ & $0.01$ & $0.01$ & $0.02$ & $N/A$ \\ 
    \hline
    KOI-168 & 0.03 & $0.03$ & $0.03$ & $0.03$ & $N/A$ & $N/A$ & $N/A$ & $N/A$ \\ 
    \hline
    KOI-191 & $>$0.15 &  $N/A$ & $N/A$ & $N/A$ & $N/A$ & $N/A$ & $N/A$ & $N/A$ \\
    \hline
    KOI-232 & 0.11 & $0.14$ & $0.12$ & $0.13$ & $0.16$ & $N/A$ & $N/A$ & $N/A$ \\ 
    \hline
    KOI-241 & $>$0.15 &  $N/A$ & $N/A$ & $N/A$ & $N/A$ & $N/A$ & $N/A$ & $N/A$ \\
    \hline
    KOI-260 & $>$0.15 &  $N/A$ & $N/A$ & $N/A$ & $N/A$ & $N/A$ & $N/A$ & $N/A$ \\
    \hline
    KOI-$271$ & 0.12 & $0.16$ & $0.11$ & $0.15$ & $N/A$ & $N/A$ & $N/A$ & $N/A$ \\ 
    \hline
    KOI-$279$ & $0.12$ &  $0.15$ & $0.11$ & $0.14$ & $0.14$ & $N/A$ & $N/A$ & $N/A$ \\
    \hline
    KOI-282 & $>$0.15 &  $N/A$ & $N/A$ & $N/A$ & $N/A$ & $N/A$ & $N/A$ & $N/A$ \\
    \hline
    KOI-285 & 0.11 & $0.15$ & $0.10$ & $0.12$ & $N/A$ & $N/A$ & $N/A$ & $N/A$ \\ 
    \hline
    KOI-316 & $>$0.15 &  $N/A$ & $N/A$ & $N/A$ & $N/A$ & $N/A$ & $N/A$ & $N/A$ \\
    \hline
    KOI-339 & $>$0.15 &  $N/A$ & $N/A$ & $N/A$ & $N/A$ & $N/A$ & $N/A$ & $N/A$ \\
    \hline
    KOI-343 & $>$0.15 &  $N/A$ & $N/A$ & $N/A$ & $N/A$ & $N/A$ & $N/A$ & $N/A$ \\
    \hline
    KOI-351 & 0.02 & $0.02$ & $0.03$ & $0.02$ & $0.03$ & $0.03$ & $0.02$ & $0.02$ \\
    \hline
    KOI-$377$ & 0.1 &  $0.17$ & $0.16$ & $0.16$ & $N/A$ & $N/A$ & $N/A$ & $N/A$ \\
    \hline
    KOI-401 & 0.13 & $0.14$ & $0.15$ & $0.12$ & $N/A$ & $N/A$ & $N/A$ & $N/A$ \\ 
    \hline
    KOI-408 & 0.11 & $0.12$ & $0.11$ & $0.08$ & $0.11$ & $N/A$ & $N/A$ & $N/A$ \\  
    \hline
    KOI-$416$ & 0.14 &  $0.11$ & $0.20$ & $0.18$ & $N/A$ & $N/A$ & $N/A$ & $N/A$ \\
    \hline
    KOI-435 & 0.1 & $0.10$ & $0.12$ & $0.10$ & $0.11$ & $0.104$ & $0.133$ & $N/A$ \\ 
    \hline
    KOI-474 & $>$0.15 &  $N/A$ & $N/A$ & $N/A$ & $N/A$ & $N/A$ & $N/A$ & $N/A$ \\
    \hline
    KOI-481 & $>$0.15 &  $N/A$ & $N/A$ & $N/A$ & $N/A$ & $N/A$ & $N/A$ & $N/A$ \\
    \hline
    KOI-490 & 0.1 & $0.10$ & $0.09$ & $0.12$ & $0.095$ & $N/A$ & $N/A$ & $N/A$ \\ 
    \hline
    KOI-500 & 0.02 & $0.03$ & $0.03$ & $0.02$ & $0.02$ & $0.025$ & $N/A$ & $N/A$ \\ 
    \hline
    KOI-505 & 0.04 & $0.06$ & $0.05$ & $0.05$ & $0.04$ & $0.05$ & $N/A$ & $N/A$ \\ 
    \hline
    KOI-509 & $>$0.15 &  $N/A$ & $N/A$ & $N/A$ & $N/A$ & $N/A$ & $N/A$ & $N/A$ \\
    \hline
    KOI-510 & 0.13 &  $0.15$ & $0.15$ & $0.16$ & $0.18$ & $N/A$ & $N/A$ & $N/A$ \\
    \hline
    KOI-518 & $>$0.15 &  $N/A$ & $N/A$ & $N/A$ & $N/A$ & $N/A$ & $N/A$ & $N/A$ \\
    \hline
    KOI-520 & 0.08 & $0.09$ & $0.08$ & $0.09$ & $0.08$ & $N/A$ & $N/A$ & $N/A$ \\
    \hline
    KOI-528 & 0.14 & $0.14$ & $0.16$ & $0.15$ & $N/A$ & $N/A$ & $N/A$ & $N/A$ \\
    \hline
    KOI-$564$ & 0.14 &  $N/A$ & $N/A$ & $N/A$ & $N/A$ & $N/A$ & $N/A$ & $N/A$ \\
    \hline
    KOI-567 & 0.05 & $0.05$ & $0.06$ & $0.06$ & $N/A$ & $N/A$ & $N/A$ & $N/A$ \\ 
    \hline
    KOI-582 & 0.07 & $0.07$ & $0.06$ & $0.07$ & $N/A$ & $N/A$ & $N/A$ & $N/A$ \\ 
    \hline
    KOI-$584$ & 0.09 & $0.12$ & $0.11$ & $0.11$ & $N/A$ & $N/A$ & $N/A$ & $N/A$ \\ 
    \hline
    KOI-$597$ & 0.1 &  $0.11$ & $0.09$ & $0.08$ & $N/A$ & $N/A$ & $N/A$ & $N/A$ \\
    \hline
    KOI-623 & 0.1 & $0.11$ & $0.10$ & $0.08$ & $0.12$ & $N/A$ & $N/A$ & $N/A$ \\ 
    \hline
    KOI-$624$ & $>$0.15 &  $N/A$ & $N/A$ & $N/A$ & $N/A$ & $N/A$ & $N/A$ & $N/A$ \\
    \hline
    KOI-$658$ & 0.1 & $0.13$ & $0.13$ & $0.15$ & $N/A$ & $N/A$ & $N/A$ & $N/A$ \\ 
    \hline
    KOI-664 & 0.12 & $0.13$ & $0.10$ & $0.12$ & $N/A$ & $N/A$ & $N/A$ & $N/A$ \\ 
    \hline
    KOI-665 & 0.12 & $0.13$ & $0.13$ & $0.11$ & $N/A$ & $N/A$ & $N/A$ & $N/A$ \\
    \hline
    KOI-671 & 0.04 & $0.05$ & $0.03$ & $0.03$ & $0.05$ & $N/A$ & $N/A$ & $N/A$ \\
    \hline
    KOI-700 & 0.07 & $0.08$ & $0.09$ & $0.10$ & $0.11$ & $N/A$ & $N/A$ & $N/A$ \\ 
    \hline
    KOI-701 & 0.08 & $0.11$ & $0.08$ & $0.07$ & $0.08$ & $0.082$ & $N/A$ & $N/A$ \\ 
    \hline
    KOI-707 & 0.03 & $0.03$ & $0.04$ & $0.03$ & $0.02$ & $0.043$ & $N/A$ & $N/A$ \\ 
    \hline
    KOI-710 & 0.07 & $0.05$ & $0.05$ & $0.08$ & $N/A$ & $N/A$ & $N/A$ & $N/A$ \\ 
    \hline
    KOI-$711$ & $>$0.15 &  $N/A$ & $N/A$ & $N/A$ & $N/A$ & $N/A$ & $N/A$ & $N/A$ \\
    \hline
    KOI-718 & $>$0.15 &  $N/A$ & $N/A$ & $N/A$ & $N/A$ & $N/A$ & $N/A$ & $N/A$ \\
    \hline
    KOI-$719^{}$ & 0.13 &  $N/A$ & $N/A$ & $N/A$ & $N/A$ & $N/A$ & $N/A$ & $N/A$ \\
    \hline
    KOI-720 & 0.08 & $0.11$ & $0.08$ & $0.09$ & $0.12$ & $N/A$ & $N/A$ & $N/A$ \\ 
    \hline
    KOI-$732$ & 0.13 &  $0.13$ & $0.17$ & $0.17$ & $N/A$ & $N/A$ & $N/A$ & $N/A$ \\
    \hline
    KOI-733 & 0.08 & $0.07$ & $0.07$ & $0.08$ & $0.08$ & $N/A$ & $N/A$ & $N/A$ \\ 
    \hline
    KOI-749 & 0.03 & $0.02$ & $0.02$ & $0.02$ & $N/A$ & $N/A$ & $N/A$ & $N/A$ \\ 
    \hline
    KOI-756 & 0.11 & $0.13$ & $0.08$ & $0.15$ & $N/A$ & $N/A$ & $N/A$ & $N/A$ \\ 
    \hline
    KOI-$757$ & 0.14 &  $0.22$ & $0.14$ & $0.12$ & $N/A$ & $N/A$ & $N/A$ & $N/A$ \\
    \hline
    KOI-806 & 0.03 & $0.03$ & $0.02$ & $0.02$ & $N/A$ & $N/A$ & $N/A$ & $N/A$ \\ 
    \hline
    KOI-834 & 0.08 & $0.10$ & $0.11$ & $0.10$ & $0.07$ & $0.10$ & $N/A$ & $N/A$ \\ 
    \hline
    KOI-841 & 0.09 & $0.13$ & $0.10$ & $0.10$ & $0.12$ & $N/A$ & $N/A$ & $N/A$ \\ 
    \hline
    KOI-$864$ & 0.12 &  $0.15$ & $0.16$ & $0.14$ & $N/A$ & $N/A$ & $N/A$ & $N/A$ \\
    \hline
    KOI-869 & 0.12 &  $0.14$ & $0.14$ & $0.13$ & $0.13$ & $N/A$ & $N/A$ & $N/A$ \\
    \hline
    KOI-880 & $>$0.15 &  $N/A$ & $N/A$ & $N/A$ & $N/A$ & $N/A$ & $N/A$ & $N/A$ \\
    \hline
    KOI-$884$ & 0.14 &  $0.17$ & $0.20$ & $0.14$ & $N/A$ & $N/A$ & $N/A$ & $N/A$ \\
    \hline
    KOI-$896$ & 0.09 & $0.12$ & $0.10$ & $0.08$ & $N/A$ & $N/A$ & $N/A$ & $N/A$ \\ 
    \hline
    KOI-$906$ & 0.12 & $0.11$ & $0.12$ & $0.13$ & $N/A$ & $0N/A$ & $N/A$ & $N/A$ \\ 
    \hline
    KOI-907 & 0.14 &  $N/A$ & $N/A$ & $N/A$ & $N/A$ & $N/A$ & $N/A$ & $N/A$ \\
    \hline
    KOI-921 & 0.13 &  $0.15$ & $0.14$ & $0.16$ & $N/A$ & $N/A$ & $N/A$ & $N/A$ \\
    \hline
    KOI-934 & 0.08 & $0.09$ & $0.10$ & $0.08$ & $N/A$ & $N/A$ & $N/A$ & $N/A$ \\ 
    \hline
    KOI-935 & 0.1 & $0.12$ & $0.11$ & $0.10$ & $0.11$ & $N/A$ & $N/A$ & $N/A$ \\ 
    \hline
    KOI-939 & 0.1 & $0.14$ & $0.09$ & $0.10$ & $0.13$ & $N/A$ & $N/A$ & $N/A$ \\ 
    \hline
    KOI-941 & $>$0.15 &  $N/A$ & $N/A$ & $N/A$ & $N/A$ & $N/A$ & $N/A$ & $N/A$ \\
    \hline
    KOI-1015 & 0.13 &  $0.20$ & $0.13$ & $0.14$ & $N/A$ & $N/A$ & $N/A$ & $N/A$ \\
    \hline
    KOI-1052 & 0.12 &  $0.14$ & $0.17$ & $0.12$ & $0.12$ & $N/A$ & $N/A$ & $N/A$ \\
    \hline
    KOI-1060 & 0.07 & $0.07$ & $0.07$ & $0.07$ & $0.065$ & $N/A$ & $N/A$ & $N/A$ \\ 
    \hline
    KOI-1127 & 0.1 & $0.12$ & $0.08$ & $0.08$ & $N/A$ & $N/A$ & $N/A$ & $N/A$ \\ 
    \hline
    KOI-1151 & 0.03 & $0.03$ & $0.04$ & $0.03$ & $0.03$ & $0.03$ & $N/A$ & $N/A$ \\ 
    \hline
    KOI-1161 & 0.11 & $0.13$ & $0.10$ & $0.12$ & $N/A$ & $N/A$ & $N/A$ & $N/A$ \\ 
    \hline
    KOI-1198 & 0.11 & $0.11$ & $0.09$ & $0.15$ & $N/A$ & $N/A$ & $N/A$ & $N/A$ \\ 
    \hline
    KOI-1203 & 0.1 & $0.13$ & $0.10$ & $0.11$ & $N/A$ & $N/A$ & $N/A$ & $N/A$ \\ 
    \hline
    KOI-1278 & 0.06 & $0.07$ & $0.05$ & $0.07$ & $0.07$ & $N/A$ & $N/A$ & $N/A$ \\ 
    \hline
    KOI-1306 & 0.1 & $0.12$ & $0.12$ & $0.13$ & $0.16$ & $N/A$ & $N/A$ & $N/A$ \\ 
    \hline
    KOI-1360 & $>$0.15 &  $N/A$ &$N/A$ & $N/A$ & $N/A$ & $N/A$ & $N/A$ & $N/A$ \\
    \hline
    KOI-1364 & 0.06 & $0.05$ & $0.05$ & $0.04$ & $0.05$ & $0.05$ & $N/A$ & $N/A$ \\ 
    \hline
    KOI-1432 & 0.14 &  $0.16$ & $0.17$ & $0.14$ & $0.10$ & $N/A$ & $N/A$ & $N/A$ \\
    \hline
    KOI-1436 & 0.14 &  $N/A$ & $N/A$ & $N/A$ & $N/A$ & $N/A$ & $N/A$ & $N/A$ \\
    \hline
    KOI-1557 & 0.08 & $0.12$ & $0.08$ & $0.08$ & $0.07$ & $N/A$ & $N/A$ & $N/A$ \\ 
    \hline
    KOI-1563 & 0.03 & $0.03$ & $0.02$ & $0.02$ & $0.03$ & $N/A$ & $N/A$ & $N/A$ \\ 
    \hline
    KOI-1567 & 0.07 & $0.10$ & $0.10$ & $0.11$ & $0.11$ & $N/A$ & $N/A$ & $N/A$ \\ 
    \hline
    KOI-1598 & 0.13 &  $0.18$ & $0.16$ & $0.16$ & $N/A$ & $N/A$ & $N/A$ & $N/A$ \\
    \hline
    KOI-1608 & $>$0.15 &  $N/A$ & $N/A$ & $N/A$ & $N/A$ & $N/A$ & $N/A$ & $N/A$ \\
    \hline
    KOI-1805 & 0.12 & $0.15$ & $0.10$ & $0.16$ & $N/A$ & $N/A$ & $N/A$ & $N/A$ \\ 
    \hline
    KOI-1860 & 0.11 & $0.11$ & $0.13$ & $0.13$ & $0.14$ & $N/A$ & $N/A$ & $N/A$ \\ 
    \hline
    KOI-$1905$ & 0.14 &  $0.15$ & $0.15$ & $0.18$ & $N/A$ & $N/A$ & $N/A$ & $N/A$ \\
    \hline
    KOI-$1909$ & 0.14 &  $N/A$ & $N/A$ & $N/A$ & $N/A$ & $N/A$ & $N/A$ & $N/A$ \\
    \hline
    KOI-1916 & $>$0.15 &  $N/A$ & $N/A$ & $N/A$ & $N/A$ & $N/A$ & $N/A$ & $N/A$ \\
    \hline
    KOI-1930 & 0.06 & $0.06$ & $0.05$ & $0.07$ & $0.06$ & $N/A$ & $N/A$ & $N/A$ \\ 
    \hline
    KOI-1931 & 0.04 & $0.03$ & $0.04$ & $0.03$ & $N/A$ & $N/A$ & $N/A$ & $N/A$ \\ 
    \hline
    KOI-$1952$ & 0.11 & $0.08$ & $0.14$ & $0.14$ & $N/A$ & $N/A$ & $N/A$ & $N/A$ \\ 
    \hline
    KOI-$1955$ & 0.14 &  $0.11$ & $0.16$ & $0.13$ & $N/A$ & $N/A$ & $N/A$ & $N/A$ \\
    \hline
    KOI-2029 & 0.06 & $0.06$ & $0.05$ & $0.06$ & $0.06$ & $N/A$ & $N/A$ & $N/A$ \\ 
    \hline
    KOI-$2037$ & 0.11 & $0.13$ & $0.11$ & $0.16$ & $N/A$ & $N/A$ & $N/A$ & $N/A$ \\ 
    \hline
    KOI-$2073$ & 0.14 &  $0.21$ & $0.19$ & $0.08$ & $0.16$ & $N/A$ & $N/A$ & $N/A$ \\
    \hline
    KOI-2135 & $>$0.15 &  $N/A$ & $N/A$ & $N/A$ & $N/A$ & $N/A$ & $N/A$ & $N/A$ \\
    \hline
    KOI-2148 & 0.11 & $0.18$ & $0.15$ & $0.10$ & $N/A$ & $N/A$ & $N/A$ & $N/A$  \\ 
    \hline
    KOI-2169 & 0.04 & $0.04$ & $0.04$ & $0.03$ & $0.04$ & $N/A$ & $N/A$ & $N/A$  \\ 
    \hline
    KOI-2194 & $>$0.15 &  $N/A$ & $N/A$ & $N/A$ & $N/A$ & $N/A$ & $N/A$ & $N/A$ \\
    \hline
    KOI-2220 & 0.05 & $0.05$ & $0.06$ & $0.04$ & $0.05$ & $0.06$ & $N/A$ & $N/A$ \\ 
    \hline
    KOI-2352 & 0.11 & $0.14$ & $0.14$ & $0.15$ & $N/A$ & $N/A$ & $N/A$ & $N/A$ \\ 
    \hline
    KOI-2433 & 0.11 & $0.17$ & $0.13$ & $0.12$ & $0.11$ & $N/A$ & $N/A$ & $N/A$ \\ 
    \hline
    KOI-2585 & 0.07 & $0.07$ & $0.06$ & $0.05$ & $N/A$ & $N/A$ & $N/A$ & $N/A$ \\
    \hline
    KOI-2597 & 0.03 & $0.04$ & $0.03$ & $0.03$ & $N/A$ & $N/A$ & $N/A$ & $N/A$ \\
    \hline
    KOI-2707 & 0.13 & $0.14$ & $0.10$ & $0.15$ & $N/A$ & $N/A$ & $N/A$ & $N/A$ \\
    \hline
    KOI-2714 & $>$0.15 &  $N/A$ & $N/A$ & $N/A$ & $N/A$ & $N/A$ & $N/A$ & $N/A$ \\
    \hline
    KOI-2722 & 0.02 & $0.02$ & $0.02$ & $0.02$ & $0.02$ & $0.02$ & $N/A$ & $N/A$ \\
    \hline
    KOI-$2732$ & 0.12 &  $0.15$ & $0.12$ & $0.14$ & $0.207$ & $N/A$ & $N/A$ & $N/A$ \\
    \hline
    KOI-3097 & 0.05 & $0.05$ & $0.04$ & $0.04$ & $N/A$ & $N/A$ & $N/A$ & $N/A$ \\
    \hline
    KOI-3925 & 0.09 & $0.12$ & $0.11$ & $0.13$ & $N/A$ & $N/A$ & $N/A$ & $N/A$ \\
    \hline
    KOI-4032 & 0.01 & $0.02$ & $0.01$ & $0.01$ & $0.01$ & $N/A$ & $N/A$ & $N/A$ \\
    \hline
    \caption{The system characteristic eccentricity ($e_{system}$) as well as the individual planet characteristic eccentricities ($e_{1}$ for planet 1, $e_{2}$ for planet 2, etc.) for the systems tested in this work. For systems where only a lower limit was obtained, only the system characteristic eccentricity is listed. These eccentricities were calculated by finding the values that would lead to a SPOCK stability of 0.5, or 50$\%$. The more specific methodology is located in the first paragraph of section 4, and in paragraph 2 of section 2. Because we expected lower characteristic eccentricity values, systems with lower planetary or system wide characteristic eccentricities are generally better constrained than systems with higher values. Systems with planetary or system wide characteristic eccentricities greater than 0.09 are less constrained. As can be seen in some of the KDE plots shown in Section 8.1, sometimes systems were still relatively stable even with high eccentricities. }
    \label{tab:my_label}
\end{longtable}
\clearpage


\bibliography{sample631}{}

\begin{thebibliography}{}
\expandafter\ifx\csname natexlab\endcsname\relax\def\natexlab#1{#1}\fi
\providecommand{\url}[1]{\href{#1}{#1}}
\providecommand{\dodoi}[1]{doi:~\href{http://doi.org/#1}{\nolinkurl{#1}}}
\providecommand{\doeprint}[1]{\href{http://ascl.net/#1}{\nolinkurl{http://ascl.net/#1}}}
\providecommand{\doarXiv}[1]{\href{https://arxiv.org/abs/#1}{\nolinkurl{https://arxiv.org/abs/#1}}}

\bibitem[{{Berger} {et~al.}(2020){Berger}, {Huber}, {Gaidos}, {van Saders}, \& {Weiss}}]{2020AJ....160..108B}
{Berger}, T.~A., {Huber}, D., {Gaidos}, E., {van Saders}, J.~L., \& {Weiss}, L.~M. 2020, \aj, 160, 108, \dodoi{10.3847/1538-3881/aba18a}

\bibitem[{{Best} {et~al.}(2024){Best}, {Sefilian}, \& {Petrovich}}]{2024ApJ...960...89B}
{Best}, M., {Sefilian}, A.~A., \& {Petrovich}, C. 2024, \apj, 960, 89, \dodoi{10.3847/1538-4357/ad0965}

\bibitem[{{Chambers} {et~al.}(1996){Chambers}, {Wetherill}, \& {Boss}}]{1996Icar..119..261C}
{Chambers}, J.~E., {Wetherill}, G.~W., \& {Boss}, A.~P. 1996, \icarus, 119, 261, \dodoi{10.1006/icar.1996.0019}

\bibitem[{{Dawson} {et~al.}(2016){Dawson}, {Lee}, \& {Chiang}}]{2016ApJ...822...54D}
{Dawson}, R.~I., {Lee}, E.~J., \& {Chiang}, E. 2016, \apj, 822, 54, \dodoi{10.3847/0004-637X/822/1/54}

\bibitem[{{Deck} {et~al.}(2013){Deck}, {Payne}, \& {Holman}}]{2013ApJ...774..129D}
{Deck}, K.~M., {Payne}, M., \& {Holman}, M.~J. 2013, \apj, 774, 129, \dodoi{10.1088/0004-637X/774/2/129}

\bibitem[{{Fabrycky} {et~al.}(2014){Fabrycky}, {Lissauer}, {Ragozzine}, {Rowe}, {Steffen}, {Agol}, {Barclay}, {Batalha}, {Borucki}, {Ciardi}, {Ford}, {Gautier}, {Geary}, {Holman}, {Jenkins}, {Li}, {Morehead}, {Morris}, {Shporer}, {Smith}, {Still}, \& {Van Cleve}}]{2014ApJ...790..146F}
{Fabrycky}, D.~C., {Lissauer}, J.~J., {Ragozzine}, D., {et~al.} 2014, \apj, 790, 146, \dodoi{10.1088/0004-637X/790/2/146}

\bibitem[{{Fang} \& {Margot}(2012)}]{2012ApJ...761...92F}
{Fang}, J., \& {Margot}, J.-L. 2012, \apj, 761, 92, \dodoi{10.1088/0004-637X/761/2/92}

\bibitem[{{Fulton} \& {Petigura}(2018)}]{2018AJ....156..264F}
{Fulton}, B.~J., \& {Petigura}, E.~A. 2018, \aj, 156, 264, \dodoi{10.3847/1538-3881/aae828}

\bibitem[{{Gilbert} \& {Fabrycky}(2020)}]{2020AJ....159..281G}
{Gilbert}, G.~J., \& {Fabrycky}, D.~C. 2020, \aj, 159, 281, \dodoi{10.3847/1538-3881/ab8e3c}

\bibitem[{{Gladman}(1993)}]{1993Icar..106..247G}
{Gladman}, B. 1993, \icarus, 106, 247, \dodoi{10.1006/icar.1993.1169}

\bibitem[{{Hadden} \& {Lithwick}(2018)}]{2018AJ....156...95H}
{Hadden}, S., \& {Lithwick}, Y. 2018, \aj, 156, 95, \dodoi{10.3847/1538-3881/aad32c}

\bibitem[{{He} {et~al.}(2020){He}, {Ford}, {Ragozzine}, \& {Carrera}}]{2020AJ....160..276H}
{He}, M.~Y., {Ford}, E.~B., {Ragozzine}, D., \& {Carrera}, D. 2020, \aj, 160, 276, \dodoi{10.3847/1538-3881/abba18}

\bibitem[{{Lammers} {et~al.}(2024){Lammers}, {Hadden}, \& {Murray}}]{2024arXiv240317928L}
{Lammers}, C., {Hadden}, S., \& {Murray}, N. 2024, arXiv e-prints, arXiv:2403.17928, \dodoi{10.48550/arXiv.2403.17928}

\bibitem[{{Laskar}(1997)}]{1997A&A...317L..75L}
{Laskar}, J. 1997, \aap, 317, L75

\bibitem[{{Laskar} \& {Petit}(2017)}]{2017A&A...605A..72L}
{Laskar}, J., \& {Petit}, A.~C. 2017, \aap, 605, A72, \dodoi{10.1051/0004-6361/201630022}

\bibitem[{{Leleu} {et~al.}(2021){Leleu}, {Alibert}, {Hara}, {Hooton}, {Wilson}, {Robutel}, {Delisle}, {Laskar}, {Hoyer}, {Lovis}, {Bryant}, {Ducrot}, {Cabrera}, {Delrez}, {Acton}, {Adibekyan}, {Allart}, {Allende Prieto}, {Alonso}, {Alves}, {Anderson}, {Angerhausen}, {Anglada Escud{\'e}}, {Asquier}, {Barrado}, {Barros}, {Baumjohann}, {Bayliss}, {Beck}, {Beck}, {Bekkelien}, {Benz}, {Billot}, {Bonfanti}, {Bonfils}, {Bouchy}, {Bourrier}, {Bou{\'e}}, {Brandeker}, {Broeg}, {Buder}, {Burdanov}, {Burleigh}, {B{\'a}rczy}, {Cameron}, {Chamberlain}, {Charnoz}, {Cooke}, {Corral Van Damme}, {Correia}, {Cristiani}, {Damasso}, {Davies}, {Deleuil}, {Demangeon}, {Demory}, {Di Marcantonio}, {Di Persio}, {Dumusque}, {Ehrenreich}, {Erikson}, {Figueira}, {Fortier}, {Fossati}, {Fridlund}, {Futyan}, {Gandolfi}, {Garc{\'\i}a Mu{\~n}oz}, {Garcia}, {Gill}, {Gillen}, {Gillon}, {Goad}, {Gonz{\'a}lez Hern{\'a}ndez}, {Guedel}, {G{\"u}nther}, {Haldemann}, {Henderson}, {Heng}, {Hogan}, {Isaak}, {Jehin}, {Jenkins}, {Jord{\'a}n}, {Kiss},
  {Kristiansen}, {Lam}, {Lavie}, {Lecavelier des Etangs}, {Lendl}, {Lillo-Box}, {Lo Curto}, {Magrin}, {Martins}, {Maxted}, {McCormac}, {Mehner}, {Micela}, {Molaro}, {Moyano}, {Murray}, {Nascimbeni}, {Nunes}, {Olofsson}, {Osborn}, {Oshagh}, {Ottensamer}, {Pagano}, {Pall{\'e}}, {Pedersen}, {Pepe}, {Persson}, {Peter}, {Piotto}, {Polenta}, {Pollacco}, {Poretti}, {Pozuelos}, {Queloz}, {Ragazzoni}, {Rando}, {Ratti}, {Rauer}, {Raynard}, {Rebolo}, {Reimers}, {Ribas}, {Santos}, {Scandariato}, {Schneider}, {Sebastian}, {Sestovic}, {Simon}, {Smith}, {Sousa}, {Sozzetti}, {Steller}, {Su{\'a}rez Mascare{\~n}o}, {Szab{\'o}}, {S{\'e}gransan}, {Thomas}, {Thompson}, {Tilbrook}, {Triaud}, {Turner}, {Udry}, {Van Grootel}, {Venus}, {Verrecchia}, {Vines}, {Walton}, {West}, {Wheatley}, {Wolter}, \& {Zapatero Osorio}}]{2021A&A...649A..26L}
{Leleu}, A., {Alibert}, Y., {Hara}, N.~C., {et~al.} 2021, \aap, 649, A26, \dodoi{10.1051/0004-6361/202039767}

\bibitem[{{Lithwick} \& {Wu}(2011)}]{2011ApJ...739...31L}
{Lithwick}, Y., \& {Wu}, Y. 2011, \apj, 739, 31, \dodoi{10.1088/0004-637X/739/1/31}

\bibitem[{{Mills} {et~al.}(2019){Mills}, {Howard}, {Petigura}, {Fulton}, {Isaacson}, \& {Weiss}}]{2019AJ....157..198M}
{Mills}, S.~M., {Howard}, A.~W., {Petigura}, E.~A., {et~al.} 2019, \aj, 157, 198, \dodoi{10.3847/1538-3881/ab1009}

\bibitem[{{Morrison} \& {Kratter}(2016)}]{2016ApJ...823..118M}
{Morrison}, S.~J., \& {Kratter}, K.~M. 2016, \apj, 823, 118, \dodoi{10.3847/0004-637X/823/2/118}

\bibitem[{{Mudryk} \& {Wu}(2006)}]{2006ApJ...639..423M}
{Mudryk}, L.~R., \& {Wu}, Y. 2006, \apj, 639, 423, \dodoi{10.1086/499347}

\bibitem[{{Murray} \& {Dermott}(1999)}]{1999ssd..book.....M}
{Murray}, C.~D., \& {Dermott}, S.~F. 1999, {Solar System Dynamics}, \dodoi{10.1017/CBO9781139174817}

\bibitem[{{Petit}(2021)}]{2021CeMDA.133...39P}
{Petit}, A.~C. 2021, Celestial Mechanics and Dynamical Astronomy, 133, 39, \dodoi{10.1007/s10569-021-10035-7}

\bibitem[{{Petit} {et~al.}(2020){Petit}, {Pichierri}, {Davies}, \& {Johansen}}]{2020A&A...641A.176P}
{Petit}, A.~C., {Pichierri}, G., {Davies}, M.~B., \& {Johansen}, A. 2020, \aap, 641, A176, \dodoi{10.1051/0004-6361/202038764}

\bibitem[{{Pu} \& {Wu}(2015)}]{2015ApJ...807...44P}
{Pu}, B., \& {Wu}, Y. 2015, \apj, 807, 44, \dodoi{10.1088/0004-637X/807/1/44}

\bibitem[{{Quillen}(2011)}]{2011MNRAS.418.1043Q}
{Quillen}, A.~C. 2011, \mnras, 418, 1043, \dodoi{10.1111/j.1365-2966.2011.19555.x}

\bibitem[{{Sidlichovsky}(1990)}]{1990CeMDA..49..177S}
{Sidlichovsky}, M. 1990, Celestial Mechanics and Dynamical Astronomy, 49, 177, \dodoi{10.1007/BF00050713}

\bibitem[{{Tamayo} {et~al.}(2021){Tamayo}, {Murray}, {Tremaine}, \& {Winn}}]{2021AJ....162..220T}
{Tamayo}, D., {Murray}, N., {Tremaine}, S., \& {Winn}, J. 2021, \aj, 162, 220, \dodoi{10.3847/1538-3881/ac1c6a}

\bibitem[{{Tamayo} {et~al.}(2020){Tamayo}, {Cranmer}, {Hadden}, {Rein}, {Battaglia}, {Obertas}, {Armitage}, {Ho}, {Spergel}, {Gilbertson}, {Hussain}, {Silburt}, {Jontof-Hutter}, \& {Menou}}]{2020PNAS..11718194T}
{Tamayo}, D., {Cranmer}, M., {Hadden}, S., {et~al.} 2020, Proceedings of the National Academy of Science, 117, 18194, \dodoi{10.1073/pnas.2001258117}

\bibitem[{{Van Eylen} \& {Albrecht}(2015)}]{2015ApJ...808..126V}
{Van Eylen}, V., \& {Albrecht}, S. 2015, \apj, 808, 126, \dodoi{10.1088/0004-637X/808/2/126}

\bibitem[{{Volk} \& {Malhotra}(2020)}]{2020AJ....160...98V}
{Volk}, K., \& {Malhotra}, R. 2020, \aj, 160, 98, \dodoi{10.3847/1538-3881/aba0b0}

\bibitem[{{Volk} \& {Malhotra}(2024)}]{2024AJ....167..271V}
---. 2024, \aj, 167, 271, \dodoi{10.3847/1538-3881/ad3de5}

\bibitem[{{Weiss} \& {Marcy}(2014)}]{2014ApJ...783L...6W}
{Weiss}, L.~M., \& {Marcy}, G.~W. 2014, \apjl, 783, L6, \dodoi{10.1088/2041-8205/783/1/L6}

\bibitem[{{Weiss} {et~al.}(2023){Weiss}, {Millholland}, {Petigura}, {Adams}, {Batygin}, {Block}, \& {Mordasini}}]{2023ASPC..534..863W}
{Weiss}, L.~M., {Millholland}, S.~C., {Petigura}, E.~A., {et~al.} 2023, in Astronomical Society of the Pacific Conference Series, Vol. 534, Protostars and Planets VII, ed. S.~{Inutsuka}, Y.~{Aikawa}, T.~{Muto}, K.~{Tomida}, \& M.~{Tamura}, 863

\bibitem[{{Weiss} {et~al.}(2018){Weiss}, {Marcy}, {Petigura}, {Fulton}, {Howard}, {Winn}, {Isaacson}, {Morton}, {Hirsch}, {Sinukoff}, {Cumming}, {Hebb}, \& {Cargile}}]{2018AJ....155...48W}
{Weiss}, L.~M., {Marcy}, G.~W., {Petigura}, E.~A., {et~al.} 2018, \aj, 155, 48, \dodoi{10.3847/1538-3881/aa9ff6}

\bibitem[{{Wu} {et~al.}(2024){Wu}, {Malhotra}, \& {Lithwick}}]{2024ApJ...971....5W}
{Wu}, Y., {Malhotra}, R., \& {Lithwick}, Y. 2024, \apj, 971, 5, \dodoi{10.3847/1538-4357/ad5a09}

\bibitem[{{Xie} {et~al.}(2016){Xie}, {Dong}, {Zhu}, {Huber}, {Zheng}, {De Cat}, {Fu}, {Liu}, {Luo}, {Wu}, {Zhang}, {Zhang}, {Zhou}, {Cao}, {Hou}, {Wang}, \& {Zhang}}]{2016PNAS..11311431X}
{Xie}, J.-W., {Dong}, S., {Zhu}, Z., {et~al.} 2016, Proceedings of the National Academy of Science, 113, 11431, \dodoi{10.1073/pnas.1604692113}

\bibitem[{{Yee} {et~al.}(2021){Yee}, {Tamayo}, {Hadden}, \& {Winn}}]{2021AJ....162...55Y}
{Yee}, S.~W., {Tamayo}, D., {Hadden}, S., \& {Winn}, J.~N. 2021, \aj, 162, 55, \dodoi{10.3847/1538-3881/ac00a9}

\bibitem[{{Zhou} \& {Sun}(2007)}]{2007IJMPB..21.3981Z}
{Zhou}, J.-L., \& {Sun}, Y.-S. 2007, International Journal of Modern Physics B, 21, 3981, \dodoi{10.1142/S0217979207045050}

\end{thebibliography}
\bibliographystyle{aasjournal}



\end{document}